\documentclass[twocolumn,letterpaper,aps,prd,longbibliography,superscriptaddress,nofootinbib,floatfix]{revtex4-1}

\usepackage{graphicx}	
\usepackage{amsmath}
\usepackage{pifont}
\usepackage{multirow}
\usepackage{xspace}	

\newcommand{\dphi}{\mbox{$\Delta\phi$}\xspace}
\newcommand{\mumu}{\mbox{$\mu\mu$}\xspace}
\newcommand{\lsmumu}{\mbox{$\mu^{\pm}\mu^{\pm}$}\xspace}
\newcommand{\ulmumu}{\mbox{$\mu^{+}\mu^{-}$}\xspace}
\newcommand{\ee}{\mbox{$e^{+}e^{-}$}\xspace}
\newcommand{\pt}{\mbox{$p_T$}\xspace}
\newcommand{\gevcc}{\mbox{GeV/$c^{2}$}\xspace}
\newcommand{\gevc}{\mbox{GeV/$c$}\xspace}

\newcommand{\pA}{\mbox{$p$$+A$}\xspace}
\newcommand{\dA}{\mbox{$d$$+$$A$}\xspace}
\newcommand{\pp}{\mbox{$p$+$p$}\xspace}

\newcommand{\cc}{$c\bar{c}$\xspace}
\newcommand{\bb}{$b\bar{b}$\xspace}
\newcommand{\z}{$z$\xspace}
\newcommand{\powheg}{{\sc powheg}\xspace}
\newcommand{\pythia}{{\sc pythia}\xspace}
\newcommand{\mcnlo}{{\sc mcnlo}\xspace}

\newcommand{\sigmabb}{\mbox{$\sigma_{b\bar{b}}$}\xspace}

\begin{document}

\title{Measurements of $\mu\mu$ pairs from open heavy flavor and Drell-Yan 
in $p$$+$$p$ collisions at $\sqrt{s}=200$ GeV}

\newcommand{\abilene}{Abilene Christian University, Abilene, Texas 79699, USA}
\newcommand{\augie}{Department of Physics, Augustana University, Sioux Falls, South Dakota 57197, USA}
\newcommand{\banaras}{Department of Physics, Banaras Hindu University, Varanasi 221005, India}
\newcommand{\barc}{Bhabha Atomic Research Centre, Bombay 400 085, India}
\newcommand{\baruch}{Baruch College, City University of New York, New York, New York, 10010 USA}
\newcommand{\bnlcoll}{Collider-Accelerator Department, Brookhaven National Laboratory, Upton, New York 11973-5000, USA}
\newcommand{\bnlphys}{Physics Department, Brookhaven National Laboratory, Upton, New York 11973-5000, USA}
\newcommand{\caucr}{University of California-Riverside, Riverside, California 92521, USA}
\newcommand{\charlesczech}{Charles University, Ovocn\'{y} trh 5, Praha 1, 116 36, Prague, Czech Republic}
\newcommand{\chonbuk}{Chonbuk National University, Jeonju, 561-756, Korea}
\newcommand{\cns}{Center for Nuclear Study, Graduate School of Science, University of Tokyo, 7-3-1 Hongo, Bunkyo, Tokyo 113-0033, Japan}
\newcommand{\colorado}{University of Colorado, Boulder, Colorado 80309, USA}
\newcommand{\columbia}{Columbia University, New York, New York 10027 and Nevis Laboratories, Irvington, New York 10533, USA}
\newcommand{\czechtech}{Czech Technical University, Zikova 4, 166 36 Prague 6, Czech Republic}
\newcommand{\debrecen}{Debrecen University, H-4010 Debrecen, Egyetem t{\'e}r 1, Hungary}
\newcommand{\elte}{ELTE, E{\"o}tv{\"o}s Lor{\'a}nd University, H-1117 Budapest, P{\'a}zm{\'a}ny P.~s.~1/A, Hungary}
\newcommand{\eszterhazy}{Eszterh\'azy K\'aroly University, K\'aroly R\'obert Campus, H-3200 Gy\"ongy\"os, M\'atrai \'ut 36, Hungary}
\newcommand{\ewha}{Ewha Womans University, Seoul 120-750, Korea}
\newcommand{\fsu}{Florida State University, Tallahassee, Florida 32306, USA}
\newcommand{\gsu}{Georgia State University, Atlanta, Georgia 30303, USA}
\newcommand{\hiroshima}{Hiroshima University, Kagamiyama, Higashi-Hiroshima 739-8526, Japan}
\newcommand{\howard}{Department of Physics and Astronomy, Howard University, Washington, DC 20059, USA}
\newcommand{\ihepprot}{IHEP Protvino, State Research Center of Russian Federation, Institute for High Energy Physics, Protvino, 142281, Russia}
\newcommand{\illuiuc}{University of Illinois at Urbana-Champaign, Urbana, Illinois 61801, USA}
\newcommand{\inrras}{Institute for Nuclear Research of the Russian Academy of Sciences, prospekt 60-letiya Oktyabrya 7a, Moscow 117312, Russia}
\newcommand{\instpasczech}{Institute of Physics, Academy of Sciences of the Czech Republic, Na Slovance 2, 182 21 Prague 8, Czech Republic}
\newcommand{\isu}{Iowa State University, Ames, Iowa 50011, USA}
\newcommand{\jaea}{Advanced Science Research Center, Japan Atomic Energy Agency, 2-4 Shirakata Shirane, Tokai-mura, Naka-gun, Ibaraki-ken 319-1195, Japan}
\newcommand{\kek}{KEK, High Energy Accelerator Research Organization, Tsukuba, Ibaraki 305-0801, Japan}
\newcommand{\korea}{Korea University, Seoul, 02841}
\newcommand{\kurchatov}{National Research Center ``Kurchatov Institute", Moscow, 123098 Russia}
\newcommand{\kyoto}{Kyoto University, Kyoto 606-8502, Japan}
\newcommand{\lawllnl}{Lawrence Livermore National Laboratory, Livermore, California 94550, USA}
\newcommand{\losalamos}{Los Alamos National Laboratory, Los Alamos, New Mexico 87545, USA}
\newcommand{\lund}{Department of Physics, Lund University, Box 118, SE-221 00 Lund, Sweden}
\newcommand{\lyon}{IPNL, CNRS/IN2P3, Univ Lyon, Université Lyon 1, F-69622, Villeurbanne, France}
\newcommand{\maryland}{University of Maryland, College Park, Maryland 20742, USA}
\newcommand{\mass}{Department of Physics, University of Massachusetts, Amherst, Massachusetts 01003-9337, USA}
\newcommand{\michigan}{Department of Physics, University of Michigan, Ann Arbor, Michigan 48109-1040, USA}
\newcommand{\muhlenberg}{Muhlenberg College, Allentown, Pennsylvania 18104-5586, USA}
\newcommand{\nara}{Nara Women's University, Kita-uoya Nishi-machi Nara 630-8506, Japan}
\newcommand{\natmephi}{National Research Nuclear University, MEPhI, Moscow Engineering Physics Institute, Moscow, 115409, Russia}
\newcommand{\newmex}{University of New Mexico, Albuquerque, New Mexico 87131, USA}
\newcommand{\nmsu}{New Mexico State University, Las Cruces, New Mexico 88003, USA}
\newcommand{\ohio}{Department of Physics and Astronomy, Ohio University, Athens, Ohio 45701, USA}
\newcommand{\ornl}{Oak Ridge National Laboratory, Oak Ridge, Tennessee 37831, USA}
\newcommand{\orsay}{IPN-Orsay, Univ.~Paris-Sud, CNRS/IN2P3, Universit\'e Paris-Saclay, BP1, F-91406, Orsay, France}
\newcommand{\peking}{Peking University, Beijing 100871, People's Republic of China}
\newcommand{\pnpi}{PNPI, Petersburg Nuclear Physics Institute, Gatchina, Leningrad region, 188300, Russia}
\newcommand{\riken}{RIKEN Nishina Center for Accelerator-Based Science, Wako, Saitama 351-0198, Japan}
\newcommand{\rikjrbrc}{RIKEN BNL Research Center, Brookhaven National Laboratory, Upton, New York 11973-5000, USA}
\newcommand{\rikkyo}{Physics Department, Rikkyo University, 3-34-1 Nishi-Ikebukuro, Toshima, Tokyo 171-8501, Japan}
\newcommand{\saispbstu}{Saint Petersburg State Polytechnic University, St.~Petersburg, 195251 Russia}
\newcommand{\seoulnat}{Department of Physics and Astronomy, Seoul National University, Seoul 151-742, Korea}
\newcommand{\stonybrkc}{Chemistry Department, Stony Brook University, SUNY, Stony Brook, New York 11794-3400, USA}
\newcommand{\stonycrkp}{Department of Physics and Astronomy, Stony Brook University, SUNY, Stony Brook, New York 11794-3800, USA}
\newcommand{\tenn}{University of Tennessee, Knoxville, Tennessee 37996, USA}
\newcommand{\titech}{Department of Physics, Tokyo Institute of Technology, Oh-okayama, Meguro, Tokyo 152-8551, Japan}
\newcommand{\tsukuba}{Tomonaga Center for the History of the Universe, University of Tsukuba, Tsukuba, Ibaraki 305, Japan}
\newcommand{\vandy}{Vanderbilt University, Nashville, Tennessee 37235, USA}
\newcommand{\weizmann}{Weizmann Institute, Rehovot 76100, Israel}
\newcommand{\wigner}{Institute for Particle and Nuclear Physics, Wigner Research Centre for Physics, Hungarian Academy of Sciences (Wigner RCP, RMKI) H-1525 Budapest 114, POBox 49, Budapest, Hungary}
\newcommand{\yonsei}{Yonsei University, IPAP, Seoul 120-749, Korea}
\newcommand{\zagreb}{Department of Physics, Faculty of Science, University of Zagreb, Bijeni\v{c}ka c.~32 HR-10002 Zagreb, Croatia}
\affiliation{\abilene}
\affiliation{\augie}
\affiliation{\banaras}
\affiliation{\barc}
\affiliation{\baruch}
\affiliation{\bnlcoll}
\affiliation{\bnlphys}
\affiliation{\caucr}
\affiliation{\charlesczech}
\affiliation{\chonbuk}
\affiliation{\cns}
\affiliation{\colorado}
\affiliation{\columbia}
\affiliation{\czechtech}
\affiliation{\debrecen}
\affiliation{\elte}
\affiliation{\eszterhazy}
\affiliation{\ewha}
\affiliation{\fsu}
\affiliation{\gsu}
\affiliation{\hiroshima}
\affiliation{\howard}
\affiliation{\ihepprot}
\affiliation{\illuiuc}
\affiliation{\inrras}
\affiliation{\instpasczech}
\affiliation{\isu}
\affiliation{\jaea}
\affiliation{\kek}
\affiliation{\korea}
\affiliation{\kurchatov}
\affiliation{\kyoto}
\affiliation{\lawllnl}
\affiliation{\losalamos}
\affiliation{\lund}
\affiliation{\lyon}
\affiliation{\maryland}
\affiliation{\mass}
\affiliation{\michigan}
\affiliation{\muhlenberg}
\affiliation{\nara}
\affiliation{\natmephi}
\affiliation{\newmex}
\affiliation{\nmsu}
\affiliation{\ohio}
\affiliation{\ornl}
\affiliation{\orsay}
\affiliation{\peking}
\affiliation{\pnpi}
\affiliation{\riken}
\affiliation{\rikjrbrc}
\affiliation{\rikkyo}
\affiliation{\saispbstu}
\affiliation{\seoulnat}
\affiliation{\stonybrkc}
\affiliation{\stonycrkp}
\affiliation{\tenn}
\affiliation{\titech}
\affiliation{\tsukuba}
\affiliation{\vandy}
\affiliation{\weizmann}
\affiliation{\wigner}
\affiliation{\yonsei}
\affiliation{\zagreb}
\author{C.~Aidala} \affiliation{\michigan} 
\author{Y.~Akiba} \email[PHENIX Spokesperson: ]{akiba@rcf.rhic.bnl.gov} \affiliation{\riken} \affiliation{\rikjrbrc} 
\author{M.~Alfred} \affiliation{\howard} 
\author{V.~Andrieux} \affiliation{\michigan} 
\author{N.~Apadula} \affiliation{\isu} 
\author{H.~Asano} \affiliation{\kyoto} \affiliation{\riken} 
\author{B.~Azmoun} \affiliation{\bnlphys} 
\author{V.~Babintsev} \affiliation{\ihepprot} 
\author{A.~Bagoly} \affiliation{\elte} 
\author{N.S.~Bandara} \affiliation{\mass} 
% ...
\collaboration{PHENIX Collaboration} \noaffiliation

\date{\today}


\begin{abstract}


PHENIX reports differential cross sections of $\mu\mu$ pairs from 
semileptonic heavy-flavor decays and the Drell-Yan production mechanism 
measured in $p$$+$$p$ collisions at $\sqrt{s}=200$ GeV at forward and backward 
rapidity ($1.2<|\eta|<2.2$). The $\mu\mu$ pairs from $c\bar{c}$, 
$b\bar{b}$, and Drell-Yan are separated using a template fit to 
unlike- and like-sign muon pair spectra in mass and $p_T$. The azimuthal 
opening angle correlation between the muons from $c\bar{c}$ and 
$b\bar{b}$ decays and the pair-$p_T$ distributions are compared to 
distributions generated using {\sc pythia} and {\sc powheg} models, 
which both include next-to-leading order processes. The measured 
distributions for pairs from $c\bar{c}$ are consistent with 
{\sc pythia} calculations. The $c\bar{c}$ data presents narrower 
azimuthal correlations and softer $p_T$ distributions compared to 
distributions generated from {\sc powheg}. The $b\bar{b}$ data are well 
described by both models. The extrapolated total cross section for 
bottom production is 
$3.75{\pm}0.24({\rm stat}){\pm}^{0.35}_{0.50}({\rm syst}){\pm}0.45({\rm global})$[$\mu$b], 
which is consistent with previous measurements at the Relativistic Heavy Ion 
Collider in the same system at the same collision energy, and is approximately 
a factor of two higher than the central value calculated with theoretical models. 
The measured Drell-Yan cross section is in good agreement with 
next-to-leading-order quantum-chromodynamics calculations.

\end{abstract}

\maketitle

\section{Introduction}

Lepton pair spectra are a classic tool to study particle production in 
collisions of hadronic beams. Famous discoveries using lepton pairs 
include the Drell-Yan mechanism for lepton pair production 
\cite{Christenson:1970um} and the $J/\psi$ meson \cite{Aubert:1974js}.

In this paper, we focus on the contribution of \cc and \bb decays to the 
lepton pair continuum above a mass of 1 \gevcc. In recent years, 
measurements of \cc and \bb via the lepton pair continuum have been 
reported for various collisions systems at the Relativistic Heavy Ion 
Collider (RHIC) by the 
PHENIX~\cite{ppg085,Adare:2009qk,Adare:2014iwg,Adare:2015ila,Adare:2017caq} 
and STAR~\cite{Adamczyk:2013caa} Collaborations. So far these 
measurements have been limited to $e^+e^-$ pairs at midrapidity. Now 
PHENIX adds a new measurement of the \mumu pair continuum at forward 
rapidity obtained in \pp collisions at $\sqrt{s}=200$ GeV. With these 
data the contributions from \cc and \bb decays and the Drell-Yan 
production mechanism can be separated and used to determine their 
differential cross sections as function of pair mass, \pt and opening 
angle.

Measurements of \cc and \bb in \pp collisions are important to further 
our understanding of the \cc and \bb production process, which despite 
considerable experimental and theoretical effort remains incomplete. 
Significant differences persist between data and 
perturbative-quantum-chromodynamics (pQCD) based model 
calculations~\cite{Cacciari:1998it, Sjostrand:2006za, Norrbin:2000zc, 
Frixione:2007nw, Frixione:2003ei, Vogt:2007aw}. Single \pt spectra of 
charm and bottom mesons, as well as their decay leptons have been 
measured over a wide range of beam energies and rapidity. For charm 
production, precise measurements at RHIC \cite{Adare:2010de, 
Aidala:2017pum,Xie:2017nal}, Tevatron~\cite{Acosta:2003ax} and the Large 
Hadron Collider 
(LHC)~\cite{Acharya:2017jgo,Aaij:2015bpa,Aad:2015zix,Sirunyan:2017xss} 
indicate that pQCD calculations underestimate the charm cross section, 
even when contributions beyond leading order are taken into account 
\cite{Sjostrand:2006za, Frixione:2003ei, Frixione:2007nw, 
Cacciari:1998it}. For bottom production, the case is less clear. At 
RHIC, the bottom cross section has been measured via various channels by 
PHENIX~\cite{Aidala:2017yte, Adare:2009ic, Adare:2017caq} and 
STAR~\cite{Aggarwal:2010xp}. The measured bottom cross sections also 
tend to be above pQCD predictions, albeit with relatively large 
uncertainties.  At higher energies, the bottom cross sections measured 
by D0 at $\sqrt{s}=1.8$ TeV~\cite{Abbott:1999se}, ALICE at 
$\sqrt{s}=2.76$ and $7$ TeV~\cite{Abelev:2014hla}, and ATLAS at 
$\sqrt{s}=7$ TeV~\cite{Aad:2012jga} again tend to be above pQCD 
predictions, while similar measurements from CDF at $\sqrt{s}=1.8$ 
TeV~\cite{Acosta:2004yw}, CMS at $\sqrt{s}=7$ 
TeV~\cite{Chatrchyan:2011pw} and LHCb at $\sqrt{s}=7$ and $13$ 
TeV~\cite{Aaij:2016avz} do not demonstrate significant deviations from 
pQCD.

Studying the angular correlation between the heavy flavor quarks, or 
their decay products, provides additional constraints on theoretical 
models and may help to disentangle different heavy flavor production 
mechanisms. Measurements at the Tevatron~\cite{Acosta:2004nj} and 
LHC~\cite{Khachatryan:2011wq, Aaij:2012dz} can be reasonably well 
described by next-to-leading-order (NLO) pQCD calculations. At RHIC, 
dilepton measurements at midrapidity \cite{ppg085, Adare:2014iwg, 
Adare:2017caq} can also be reproduced by different pQCD models in the 
measured phase space, but extrapolations beyond the measured range are 
model dependent, in particular for \cc production.

Besides the interest in the production mechanism itself, a solid 
understanding of \cc and \bb production in \pp collision is needed as a 
baseline for measurements involving nuclear beams, where deviations from 
the \pp baseline are often interpreted as evidence for hot or cold 
nuclear matter effects. In collisions with nuclei, modifications to the 
parton distribution functions, typically expressed as shadowing or 
anti-shadowing, may need to be taken into account. Also modifications in 
the final state, incorporated through changes to the fragmentation 
functions may need to be considered. It is broadly expected that in 
asymmetric collision systems like \pA or \dA, deviations from the \pp 
baseline indicate such cold nuclear matter effects. Uncertainties on \cc 
and \bb production in \pp limit the precision on the quantification of 
cold nuclear matter effects. For example, previous dilepton correlation 
studies indicated a significant modification of heavy flavor yields at 
forward-midrapidity in $d$$+$Au collisions \cite{Adare:2013xlp}, but not 
at mid-midrapidity \cite{Adare:2017caq}. In addition, in heavy-ion 
collisions the charm contribution is an important background to 
possible thermal dilepton radiation from the Quark Gluon Plasma 
\cite{Adare:2009qk, Adare:2015ila, Adamczyk:2013caa}. Current 
uncertainties in our understanding of \cc and \bb production prohibit 
this measurement at RHIC energies.

\begin{figure}[h]
\includegraphics[width=0.49\linewidth]{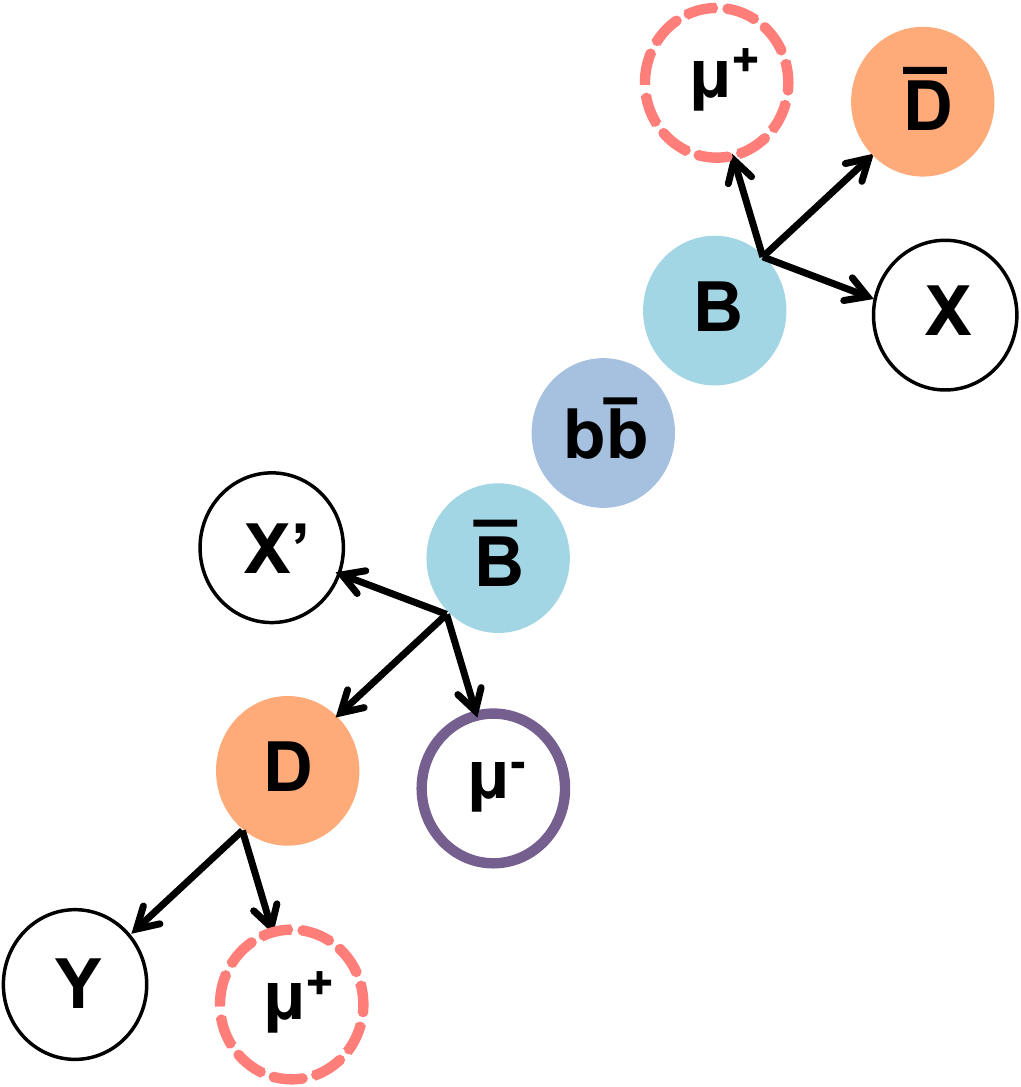}
\includegraphics[width=0.49\linewidth]{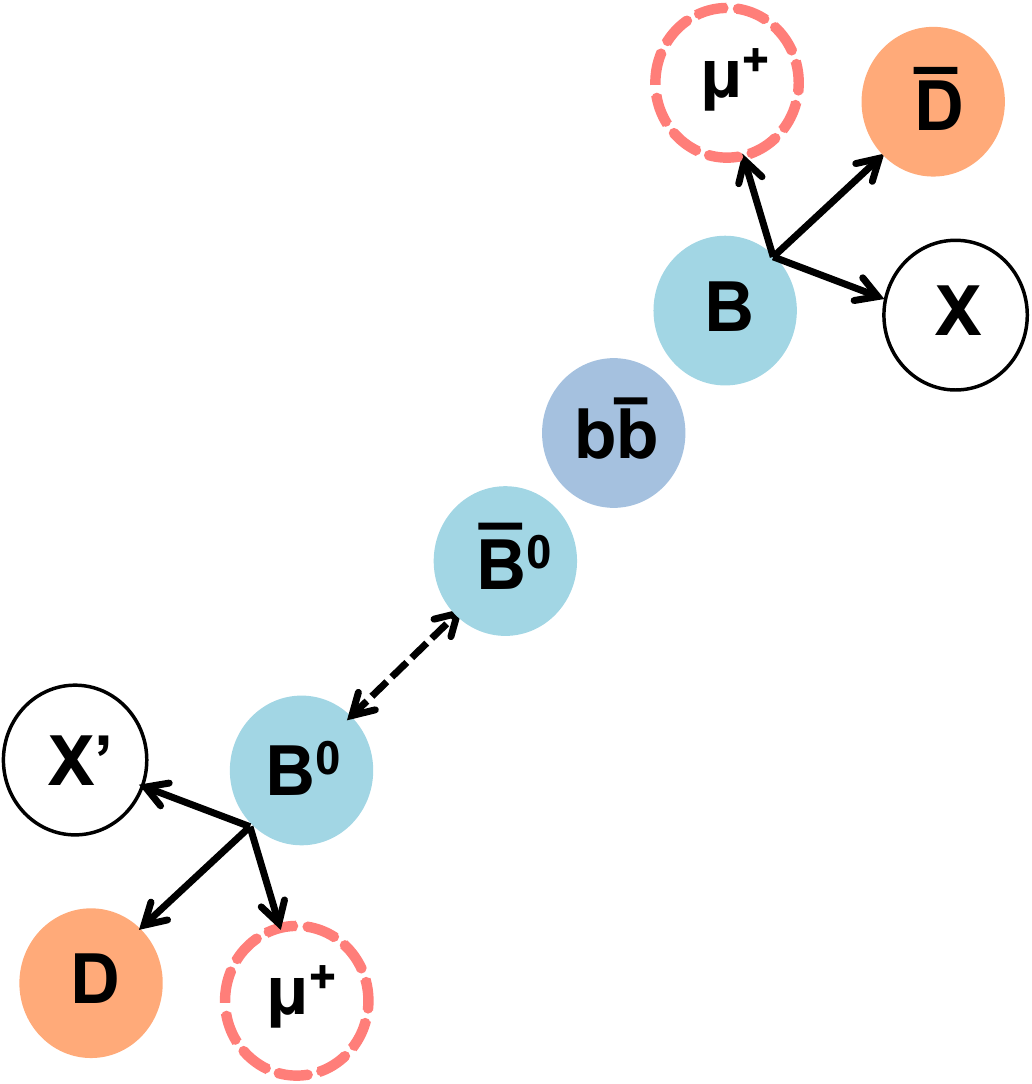}
\caption{Like-sign muon pairs from bottom may arise from a combination 
of $B \rightarrow \mu$ and $B\rightarrow D\rightarrow \mu$ decay chains 
or from decays following $B^{0}\bar{B^{0}}$ oscillations . 
\label{Fig:likesignbb} }
\end{figure}

In this study we make use of the fact that muon pairs from \cc and \bb 
decays and from Drell-Yan production contribute with different strength 
to the muon pair continuum in different phase-space regions for \ulmumu 
and \lsmumu charge combinations. Neither \cc decays nor Drell-Yan 
production contribute to \lsmumu pairs. In contrast, \bb decays do. As 
illustrated in Fig.~\ref{Fig:likesignbb}, \lsmumu muon pairs from bottom 
arises from two separate mechanisms, (i) from a combination of $B 
\rightarrow \mu$ and $B\rightarrow D\rightarrow \mu$ decay 
chains~\cite{Patrignani:2016xqp} or (ii) from decays following 
$B^{0}\bar{B^{0}}$ oscillations~\cite{Glashow:1961tr}. These two 
contributions dominate the high mass \lsmumu spectrum, which allows a 
precise measurement of the bottom cross section.

At midrapidity the $e^{+}e^{-}$ pair continuum is dominated by pairs 
from heavy flavor decays in the measurable range from 1 to 15 
\gevcc~\cite{Adare:2017caq}, and thus having established the \bb 
contribution would be sufficient to extract the \cc cross section. 
However, at forward rapidity, \ulmumu pairs from Drell-Yan can not be 
neglected. The Drell-Yan process involves quark-antiquark 
annihilation~\cite{Drell:1970wh}, whereas heavy flavor production is 
dominated by gluon fusion~\cite{Norrbin:2000zc}. Due to the relative 
large Bjorken-$x$ of valence quarks compared to gluons, at forward 
rapidity the \mumu pair yield above a mass of 6 \gevcc is dominated by 
pairs from the Drell-Yan process. Thus, the Drell-Yan contribution can 
be determined from \ulmumu pairs at high masses.

Once the contributions from \bb decays and Drell-Yan production are 
constrained, the yield from \cc can be measured in the mass range from 1 
to 3 \gevcc, where it is significant, but only one of multiple 
contributions to the total yield in the mass range.

The paper is organized as follows: Sec.~\ref{Sec:phenix} outlines the 
experimental apparatus and the relevant triggers. 
Sec.~\ref{Sec:analysis} describes the procedure to extract muon pairs 
from the data. The expected \mumu pair sources are discussed in 
Sec.~\ref{Sec:expectedsources}. The Monte Carlo simulation used to 
generate templates for \mumu pair spectra from the expected sources, 
which can be compared to the data, are presented in 
Sec.~\ref{Sec:simframe}. In Sec.~\ref{Sec:fitting} we document the 
iterative template fitting method used to determine \cc, \bb and 
Drell-Yan cross sections. Sec.~\ref{Sec:syserror} discusses the sources 
of systematic uncertainties. The results are presented in 
Sec.~\ref{Sec:results} and finally we summarize our findings in 
Sec.~\ref{Sec:summary}.

\section{Experimental Setup}
\label{Sec:phenix}

\begin{figure}[h]
\includegraphics[width=0.98\linewidth]{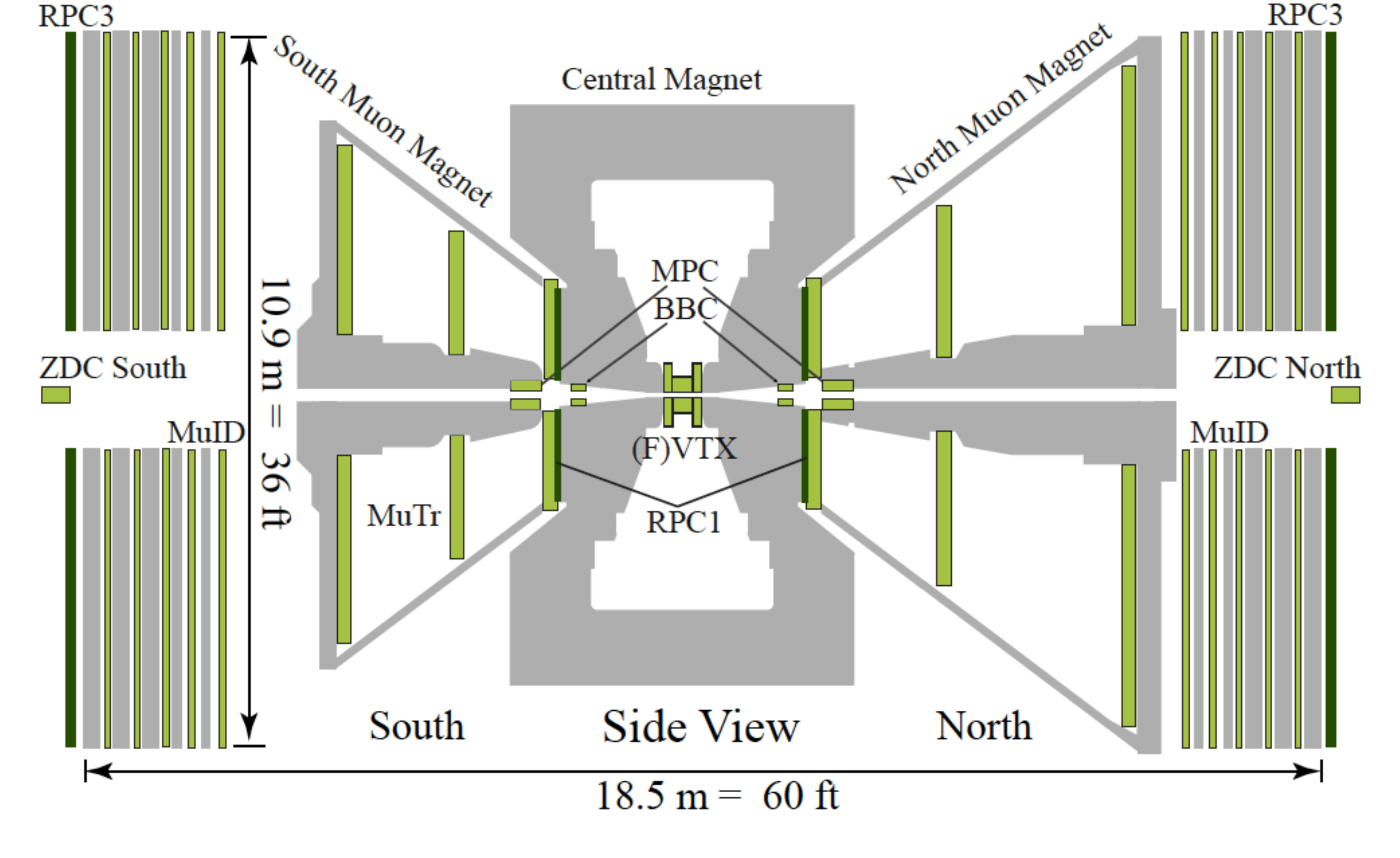}
\caption{\label{Fig:PHENIX}
Side view of the PHENIX detector in the 2015 run.}
\end{figure}

The PHENIX detector comprises two central arms at midrapidity and two 
muon arms at forward and backward rapidity~\cite{Adcox:2003zm}. The 
configuration of the experiment used for data taking with \pp collisions 
in 2015 is shown in Fig.~\ref{Fig:PHENIX}. Two muon spectrometers cover 
$\Delta\phi = 2\pi$ in azimuth and $-2.2<\eta<-1.2$ (south arm) and 
$1.2<\eta<2.4$ (north arm) in pseudorapidity. The central arms are not 
used in this analysis.

Each muon arm comprises a forward-silicon vertex tracker (FVTX), 
followed by a hadron absorber with a muon spectrometer behind it. The 
spectrometer is composed of a charged particle tracker (MuTr) inside a 
magnet and a muon identification system (MuID). The FVTX allows for 
precision tracking, but has limited acceptance and is thus not used in 
this analysis.

The hadron absorber is composed of layers of copper, iron, and stainless 
steel, corresponding to a total of 7.2 interaction lengths 
($\lambda_{I}$). The absorber suppresses muons from pion and kaon decays 
by about a factor of 1000, as it absorbs most pions and kaons before 
they decay. A small fraction of pions and kaons decays before they reach 
the absorber, which starts about 40 cm away from the nominal interaction 
point.

The MuTr has three stations of cathode strip chambers and provides a 
momentum measurement for the charged particles remaining after the 
absorber. The MuID is comprised of five alternating planes of steel 
absorbers [$4.8~(5.4)~\lambda_{I}$ for south (north) arm] and Iarocci 
tubes (gap 0--gap 4). The MuID provides identification of 
charged-particle trajectories based on the penetration depth. Only muons 
with momentum larger than 3 GeV/$c$ can penetrate all layers of 
absorbers. Signals in multiple MuID planes are combined to MuID tracks, 
which are used in the PHENIX trigger system to preselect events 
containing muon candidates. The trigger used to select the event sample 
for this analysis is a pair trigger (MuIDLL1-2D). For muon pairs with 
tracks that do not overlap in the MuID the MuIDLL1-2D is fired if both 
tracks independently fulfill the single track trigger requirement 
(MuDLL1-1D), which requires that the MuID track has at least one hit in 
the last two planes. A more detailed description of the PHENIX muon arms 
can be found in Ref.~\cite{Akikawa:2003zs}.

The beam-beam counters (BBC)~\cite{Allen:2003zt} comprise two arrays of 
64 quartz \v{C}erenkov detectors located at $z=\pm144~{\rm cm}$ from the 
nominal interaction point. Each BBC covers the full azimuth and the 
pseudorapidity range $3.1<|\eta|<3.9$. The BBCs are used to determine 
the collision-vertex position along the beam axis ($z_{\rm vtx}$) with a 
resolution of roughly 2~cm in \pp collisions. The BBCs information also 
provides a minimum-bias (MB) trigger, which requires a coincidence 
between both sides with at least one hit on each side. The cross section 
of inelastic \pp collisions at $\sqrt{s}=200$~GeV measured by the BBC, 
which is determined via the van der Meer scan 
technique~\cite{Drees:2003zza} ($\sigma^{p+p}_{\rm BBC}$), is 
$23.0{\pm}2.2$ mb.

\section{Data Analysis}
\label{Sec:analysis}

\subsection{Data set and event selection}

The data set analyzed here was taken with \pp collisions at 
$\sqrt{s}=200$ GeV in 2015. The data were selected with the \mumu pair 
trigger (MuIDLL1-2D) in coincidence with the MB trigger. Each event in 
the sample has a reconstructed vertex within $z=\pm30$ cm of the nominal 
collision point. The data sample corresponds to $1.2 \times 10^{12}$ 
MB events or to an integrated luminosity of 
$\int\mathcal{L}dt=51$~pb$^{-1}$.

\subsection{Track reconstruction \label{Sec:trackreco}}

\begin{figure}[h]
\includegraphics[width=0.95\linewidth]{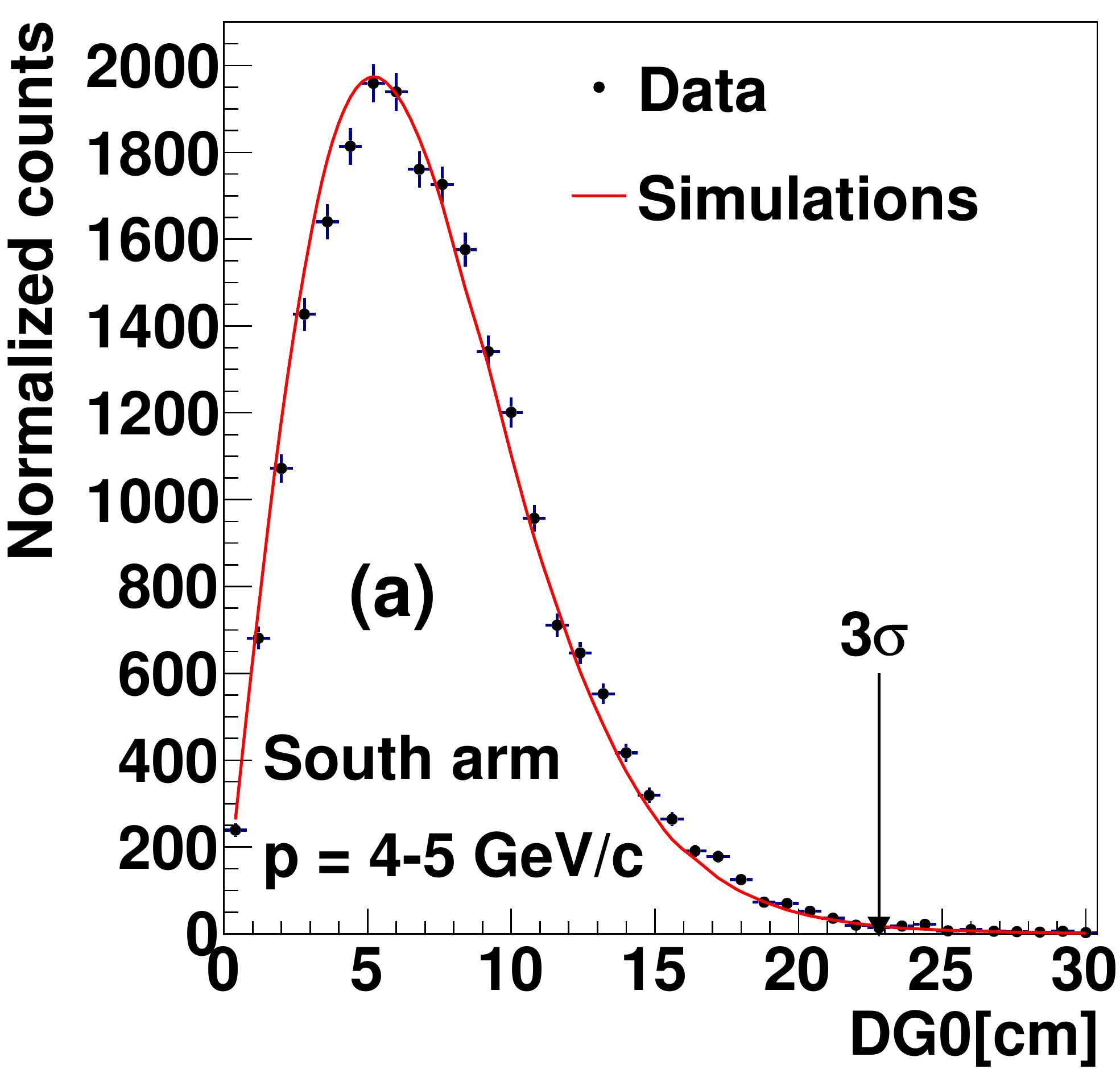}
\includegraphics[width=0.95\linewidth]{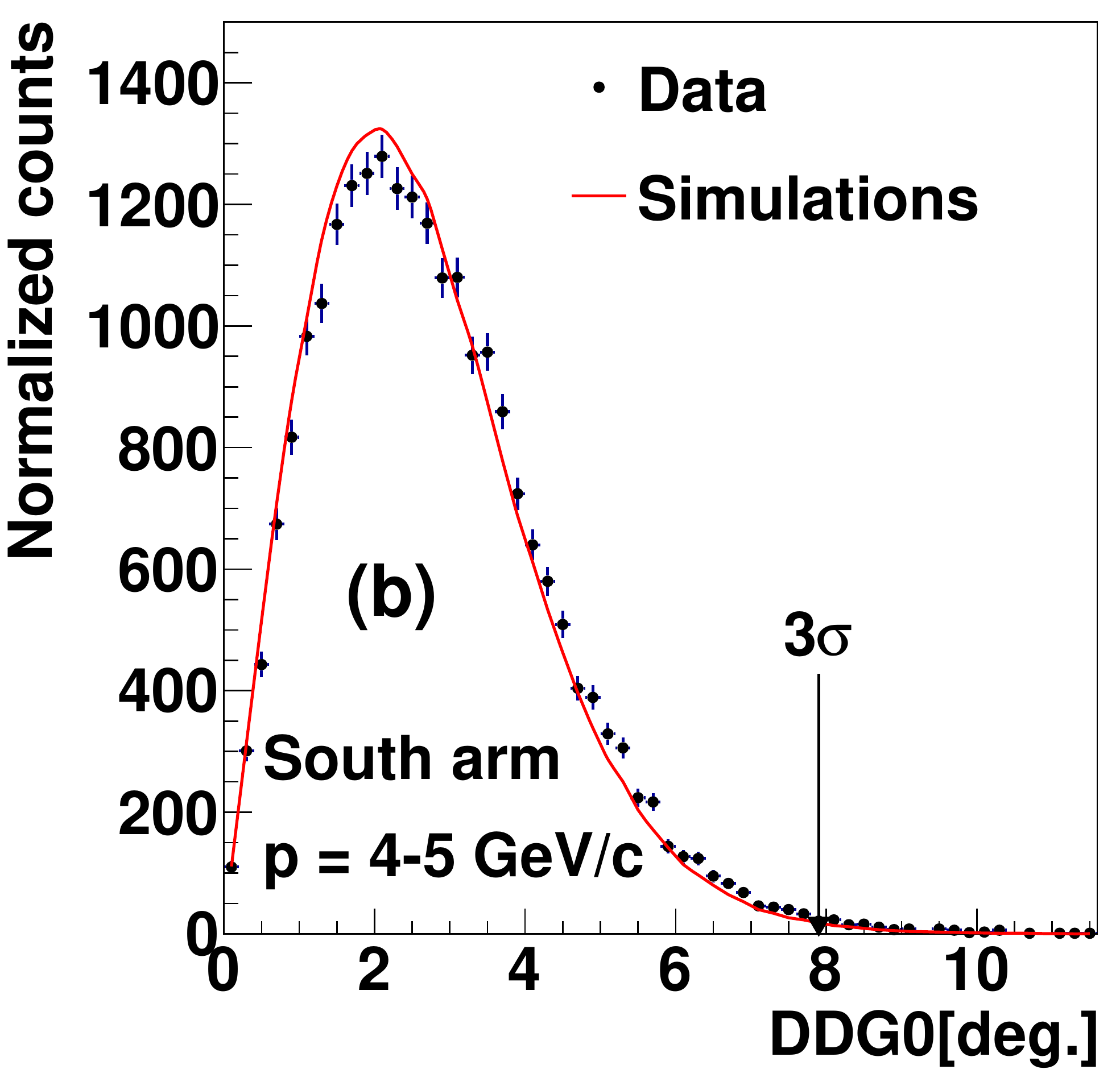}
\caption{Matching of MuTr to MuID tracklets in distance (DG0) and angle 
(DDG0) for tracks from pairs in the $J/\psi$ mass region. Data and 
simulations are compared. The $3\sigma$ cut applied in the data analysis 
is indicated. \label{Fig:dg0ddg0} }
\end{figure}

Each reconstructed muon track comprises a combination of a reconstructed 
tracklet in the MuTr and in the MuID. A number of quality cuts are 
applied to reduce the number of background muons from light hadron 
decays. They are summarized in Tab.~\ref{Tab:singlecuts}. The tracklet 
in the MuTr must have a minimum of 11 hits and a $\chi^2/NDF$ smaller 
than 15 (20) for the south (north) arm. The MuID tracklet has to 
penetrate to the last gap and must have at least 5 associated hits. MuID 
tracklets with $\chi^2/NDF$ larger than 5 are rejected. MuTr tracklets 
are projected to MuID gap 0. We apply cuts on the distance between the 
projection of the MuTr tracklet and the MuID tracklet (DG0) and the 
difference between the track angles (DDG0). Figure~\ref{Fig:dg0ddg0} 
depicts DG0 and DDG0 distributions for muons with momenta of 4 to 5 
\gevc from \mumu pairs in the mass region 2.8--3.4 \gevcc where \mumu 
pairs from $J/\psi$ dominate the yield. Both distributions are compared 
to tracks from simulated $J/\psi$ decays. These cut variables are well 
described by simulations. We apply a cut at 3$\sigma$ ($99.87\%$ 
efficiency) of the momentum dependent matching resolution of signal 
tracks determined from Monte Carlo simulations with {\sc 
geant4}~\cite{Agostinelli:2002hh}.

In addition to the basic track quality cuts, we enforce that the 
momentum of all reconstructed muon tracks are within $3<p$ [GeV/$c]<20$ 
and that their rapidity to be $1.2<|\eta|<2.2$. These requirements limit 
effects from detector acceptance edges. The upper limit on $p$ removes 
tracks from hadronic decays within the MuTr volume that lead to a 
mis-reconstructed momentum. We also require that all tracks satisfy the 
MuIDLL1-1D trigger condition.

While traversing the hadron absorber muons undergo multiple scattering 
and lose typically $2$ GeV of their energy before they reach the MuTr, 
where the momentum of the track is determined. Thus, the momentum needs 
to be corrected to correspond to the momentum in front of the absorber. 
The relative resolution has two main components, the intrinsic 
resolution of the MuTr and the resolution of the energy loss correction. 
Below 10 \gevc the resolution depends only moderately on rapidity or 
momentum and is approximately constant between 3.5 and 5\%. Towards 
larger momenta it gradually increases but remains below 10\% for all 
momenta considered in this analysis ($p<20$ \gevc). Multiple scattering 
in the absorber adds an uncertainty of 160 mrad on the angular 
measurement from the MuTr. This can be vastly improved with the FVTX, 
which measures the track in front of the absorber. However, as discussed 
in the following section we do not make use of this improvement in the 
current analysis.

\subsection{Muon pair selection}

All muon tracks in a given event are combined to pairs and their masses 
and momenta are calculated. The mass is calculated from a fit to the two 
tracks with the constraint that both originate at a common vertex within 
the range $\pm40$ cm around the nominal event vertex. This fitting 
procedure improves the resolution of the opening angle of the pair, 
which in turn significantly improves the mass resolution at $m<3$ \gevcc 
where the mass resolution is dominated by effects from multiple 
scattering. We achieve a mass resolution $\sigma_{m}/m\approx 
12.6\%,~7.4\%,~5.7\%$ at $m=1.02,~3.10,~9.46$ \gevcc corresponding to 
the $\phi,~J/\psi$ and $\Upsilon(1S)$ respectively, which is sufficient 
for the analysis of the \mumu pair continuum.

The mass resolution could be further improved by constraining the fit to 
the measured vertex position. However, our data set contains on average 
22\% of pileup events with two collisions recorded simultaneously. For 
these events only an average vertex position can be measured, which is 
often off by tens of centimeters from one or both of the collision 
points. This leads to \mumu pair masses reconstructed hundreds of 
MeV/c$^2$ different from the true mass and results in a mass resolution 
function with significant non-Gaussian tails. 

\begin{figure}[h]
\includegraphics[width=0.98\linewidth]{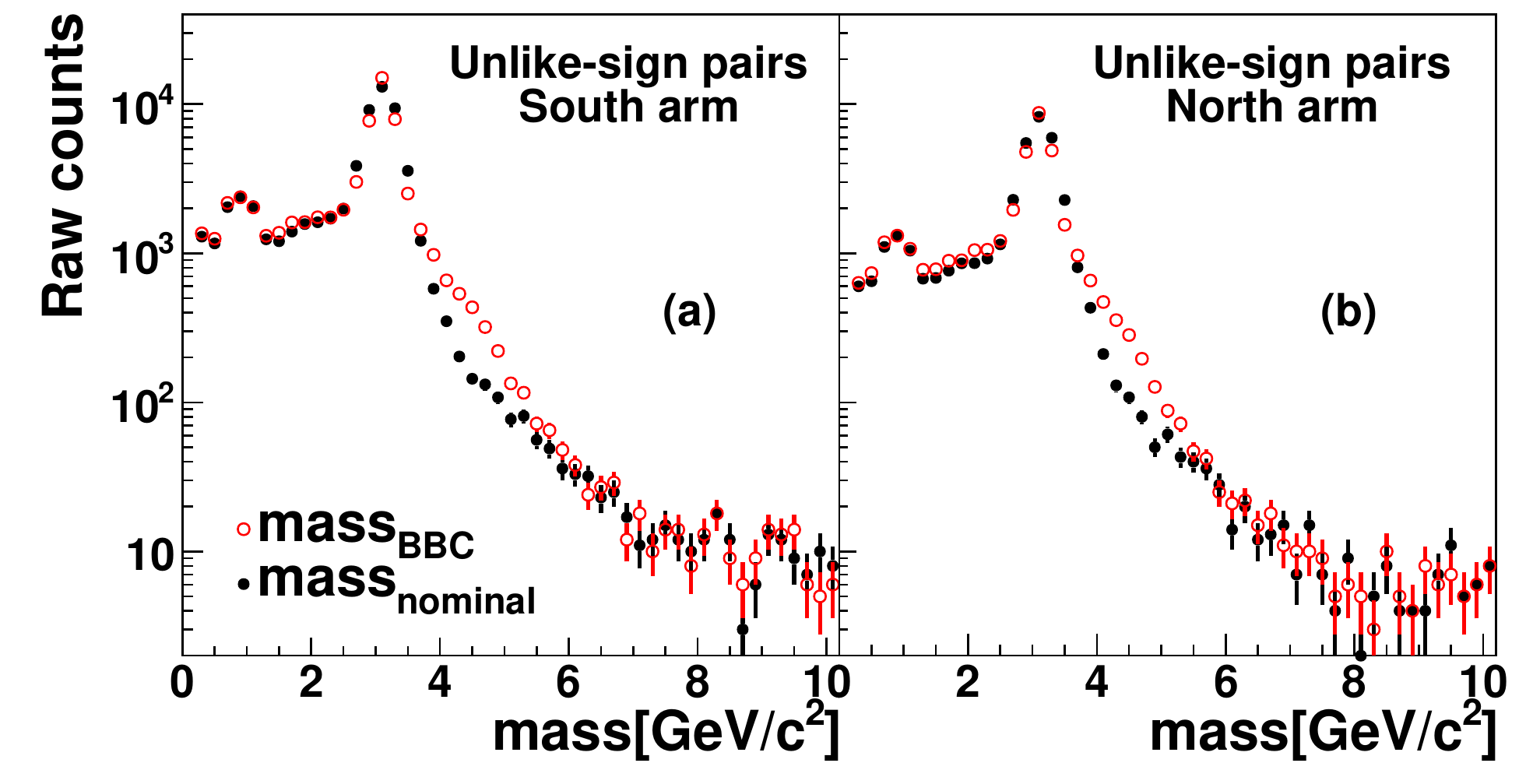}
\includegraphics[width=0.98\linewidth]{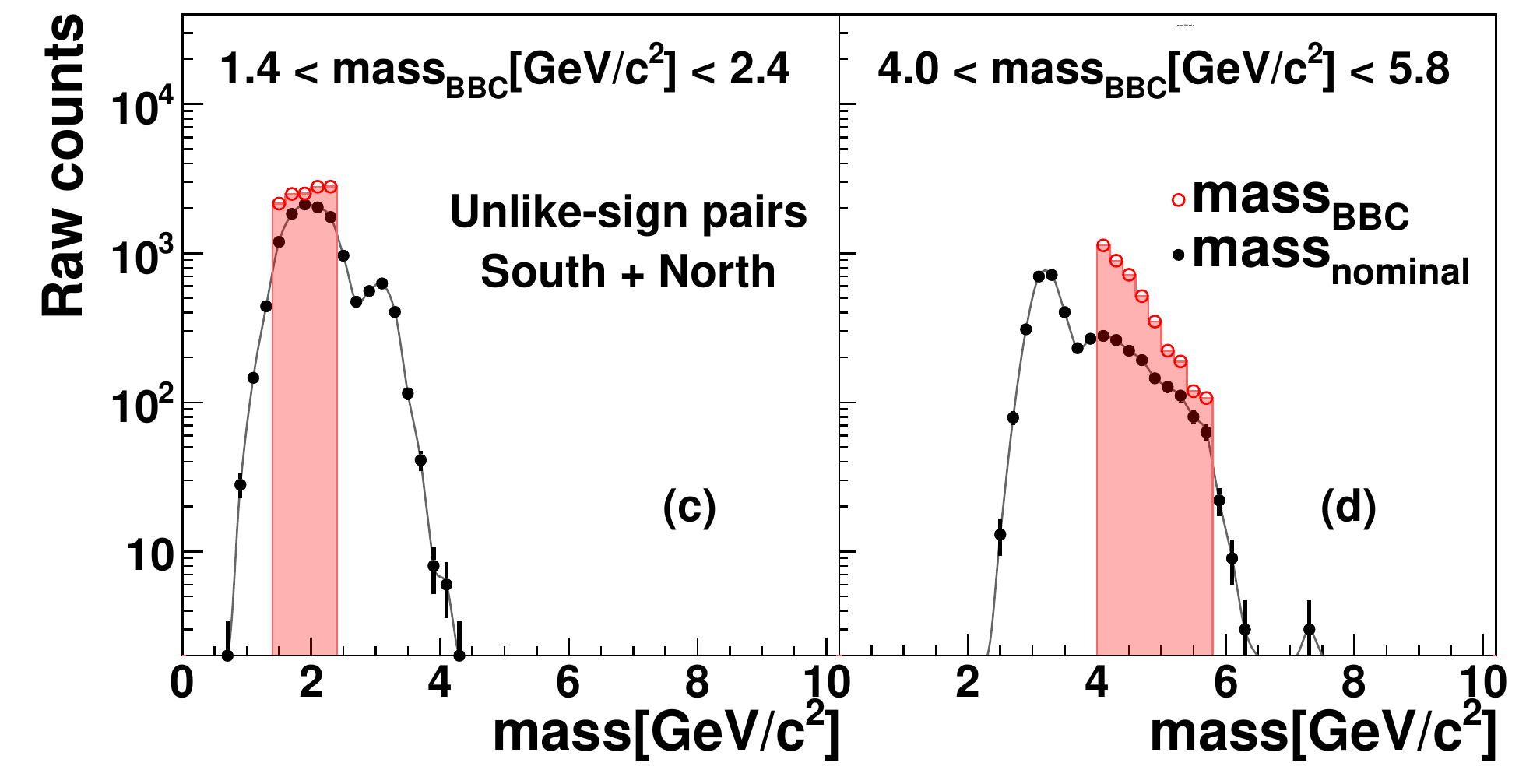}
\caption{The mass spectra from the (a) south and (b) north arms, where 
the mass is calculated with different constraints to the vertex 
position: (i) a common vertex within $\pm40$ cm around the nominal event 
vertex (mass$_{\rm nominal}$, closed circles), and (ii) the vertex 
measured by the BBC (mass$_{\rm BBC}$, open circles). mass$_{\rm BBC}$ 
and mass$_{\rm nominal}$ distributions are compared with pairs selected 
with mass$_{\rm BBC}$ (c) between 1.4 and 2.4 \gevcc, and (d) between 
4.0 and 5.8 \gevcc. \label{Fig:anamassvtx}}
\end{figure}

Figure~\ref{Fig:anamassvtx}(a) compares the mass distribution of the 
south muon arm and Fig.~\ref{Fig:anamassvtx}(b) for the north arm.  The 
mass is calculated from the fits that constrain the tracks to originate 
from a vertex located at (i) $\pm40$ cm of the nominal vertex 
(mass$_{\rm nominal}$), and (ii) $\pm2$ cm of the measured vertex using 
the BBC (mass$_{\rm BBC}$). Although the width of the $J/\psi$ is 
narrower for mass$_{\rm BBC}$ as expected, the yield at the continuum on 
either sides of the $J/\psi$ is significantly different for the two mass 
calculations. To further diagnose this issue, we selected pairs with 
mass$_{\rm BBC}$ between 1.4 and 2.4 \gevcc[panel~(c)] and between 4.0 
and 5.8 \gevcc~[panel~(d)], and compared mass$_{\rm BBC}$ and mass$_{\rm 
nominal}$ distributions. In both mass$_{\rm BBC}$ selections, a clear 
$J/\psi$ peak is observed for mass$_{\rm nominal}$, which indicates that 
the mass$_{\rm BBC}$ continuum contains a significant fraction of 
mis-reconstructed $J/\psi$ mesons, where the mis-reconstructed mass is 
due to a mis-measured vertex using the BBC in pileup events. To avoid 
this undesirable complication of the analysis of the \mumu pair 
continuum, we do not make use of the improvement of the mass resolution. 
The pileup events increase the yield of \mumu pairs per event by about 
10\%, this is taken into account in the normalization procedure.

\begin{figure*}[ht]
\includegraphics[width=0.99\linewidth]{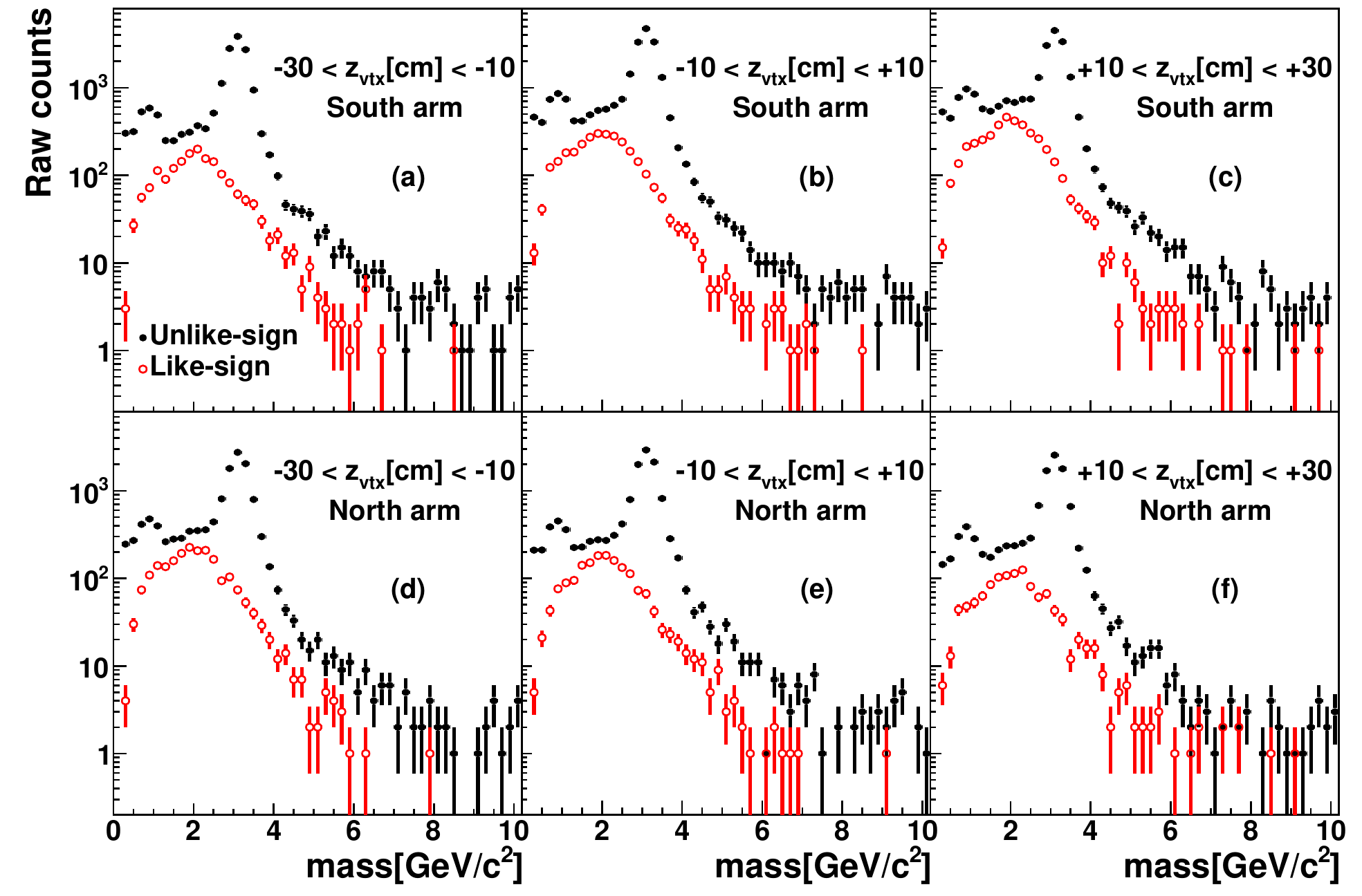}
\caption{Raw mass spectra for the south and north muon arms in different 
$z_{\rm vtx}$ slices. \label{Fig:masszslices}}
\end{figure*}

We apply additional quality cuts to the muon pairs, which are summarized 
in Table~\ref{Tab:paircuts}. The $\chi^{2}_{\rm vtx}$, computed from the 
simultaneous fit of the two muon tracks, must be less than 5. This cut 
mainly removes tracks that were either scattered by large angles in the 
absorber or that resulted from light hadron decays. We also remove pairs 
with a momentum asymmetry ($|p_{1}-p_{2}|/|p_{1}+p_{2}|$) larger than 
0.55 because these pairs are mostly from random pairs where one hadron 
has decayed into a muon inside the MuTr and is mis-reconstructed as a 
higher momentum track, thus yielding a fake high mass pair.

Finally, we impose cuts to ensure spatial separation between two tracks 
in the MuTr and MuID volumes. Specifically we require that the vertical 
and horizontal spatial separation of the two tracks at the MuID gap~0 
exceeds 20~cm. This cut removes all pairs with tracks that overlap so 
that for the remaining pairs the pair reconstruction and trigger 
efficiencies factorize into a product of single track efficiencies.

\begin{table}[h]
\caption{\label{Tab:singlecuts}Track quality cuts used in this analysis. }
\begin{ruledtabular} \begin{tabular}{ccc}
& south & north\\
\hline
Penetrate MuID last gap & & \\
MuTr $\chi^2$ & $<15$ & $<20$\\
Number of hits in MuTr & $>10$ & $>10$\\
MuID $\chi^2$ & $<5$ & $<5$\\
Number of hits in MuID & $>5$ & $>5$\\
${\rm DG0}(p)$ & $<3\sigma$ & $<3\sigma$\\
${\rm DDG0}(p)$ & $<3\sigma$ & $<3\sigma$\\
\end{tabular} \end{ruledtabular}
\end{table}

\begin{table}[h]
\caption{\label{Tab:paircuts}Pair cuts used in this analysis. }
\begin{ruledtabular} \begin{tabular}{cc}
$\chi^2_{\rm vtx}$ & $<5$\\
$|p_1-p_2|/|p_1+p_2|$ & $<0.55$\\
Muon pair do not share the same MuTr octant\\
$\Delta x$, $\Delta y$ at MuID gap 0 & $>20$ cm \\
\end{tabular} \end{ruledtabular}
\end{table}


Figure~\ref{Fig:masszslices} shows the raw mass spectra after imposing 
all single and pair cuts. Spectra are presented for \ulmumu and \lsmumu 
pairs measured for collisions in three vertex regions separately for the 
south and north arms.

The most prominent feature in the spectra is the $J/\psi$ peak at 
$\sim{3.1}$ \gevcc. For each arm the yield is independent of \z within 
10\%--20\%. Pairs in the north arm are reconstructed with about 2/3 of 
the efficiency compared to the south arm, which is mostly due to a 
larger dead area in the north MuTr, but otherwise the spectra are 
similar for mirrored \z ranges. The like-sign spectra have the lowest 
yield for the \z range closest to the absorber, negative and positive \z 
for south and north arm, respectively. The \lsmumu yield increases by 
roughly a factor of three as the collision point moves away from the 
absorber and more pions and kaons decay in flight before reaching the 
absorber.

\section{Expected pair sources
\label{Sec:expectedsources}}

To interpret the experimental data shown in Fig.~\ref{Fig:masszslices}, 
we need to compare it to the \mumu pairs from known sources, commonly 
referred to as ``cocktail". Besides our signal of interest, \mumu pairs 
from open heavy flavor (semi-leptonic decays of \cc and \bb) and 
Drell-Yan, the cocktail contains large contributions from hadron 
(pseudoscalar and vector meson) decays, and unphysical background pairs. 
The quantitative comparison is done through template \mumu pair 
distributions that are generated for the individual known sources.

The unphysical background pairs typically involve muons from the decays 
of light hadrons ($\pi^{\pm}$, $K^{\pm}$, and $K^{0}$). The production 
rates of decay muon from light hadrons overwhelm those of signal muons 
from \cc, \bb, and Drell-Yan. Therefore, in spite of the large hadron 
rejection power ($\sim1/1000$) of the muon arms, a substantial fraction 
of the reconstructed muons are from pion and kaon decays that occur 
before they reach the absorber. Because the distance to the absorber 
varies from 10 to 70 cm, depending on the \z location of the event 
vertex $z_{\rm vtx}$, the unphysical background varies significantly 
with $z_{\rm vtx}$. A smaller, but non negligible fraction of background 
tracks are hadrons that penetrate all layers of absorber and are 
therefore reconstructed as muon candidates. In addition, hadrons can 
interact strongly with the absorber to produce showers of secondary 
particles, which can also be reconstructed as muon candidates. Pairs 
including at least one of these so called {\it hadronic tracks}, i.e. a 
muon from light hadron decay, a punch-through hadron or a secondary 
particle from hadronic showers, are a large contribution to the measured 
\mumu pairs.

In the following subsections we discuss how we can generate the known 
sources of \mumu pairs, which are needed as input for the templates of 
\mumu pair spectra used in the subsequent analysis.

\subsection{Physical \mumu pair sources}

\subsubsection{Hadron decays to \mumu pairs ($h \rightarrow \mu\mu (X)$)
\label{Sec:hdecay}}

Decays from $\eta$, $\eta'$, $\omega$, $\rho$, and $\phi$ dominate the 
\ulmumu pair yield below a mass of 1 \gevcc, whereas decays from 
$J/\psi$, $\psi'$, and $\Upsilon(1S+2S+3S)$ dominate the \ulmumu pair 
yield in narrow mass regions at higher masses. We use existing data to 
constrain the input distributions for these mesons whenever possible.

\begin{figure*}[t]
\includegraphics[width=0.61\linewidth]{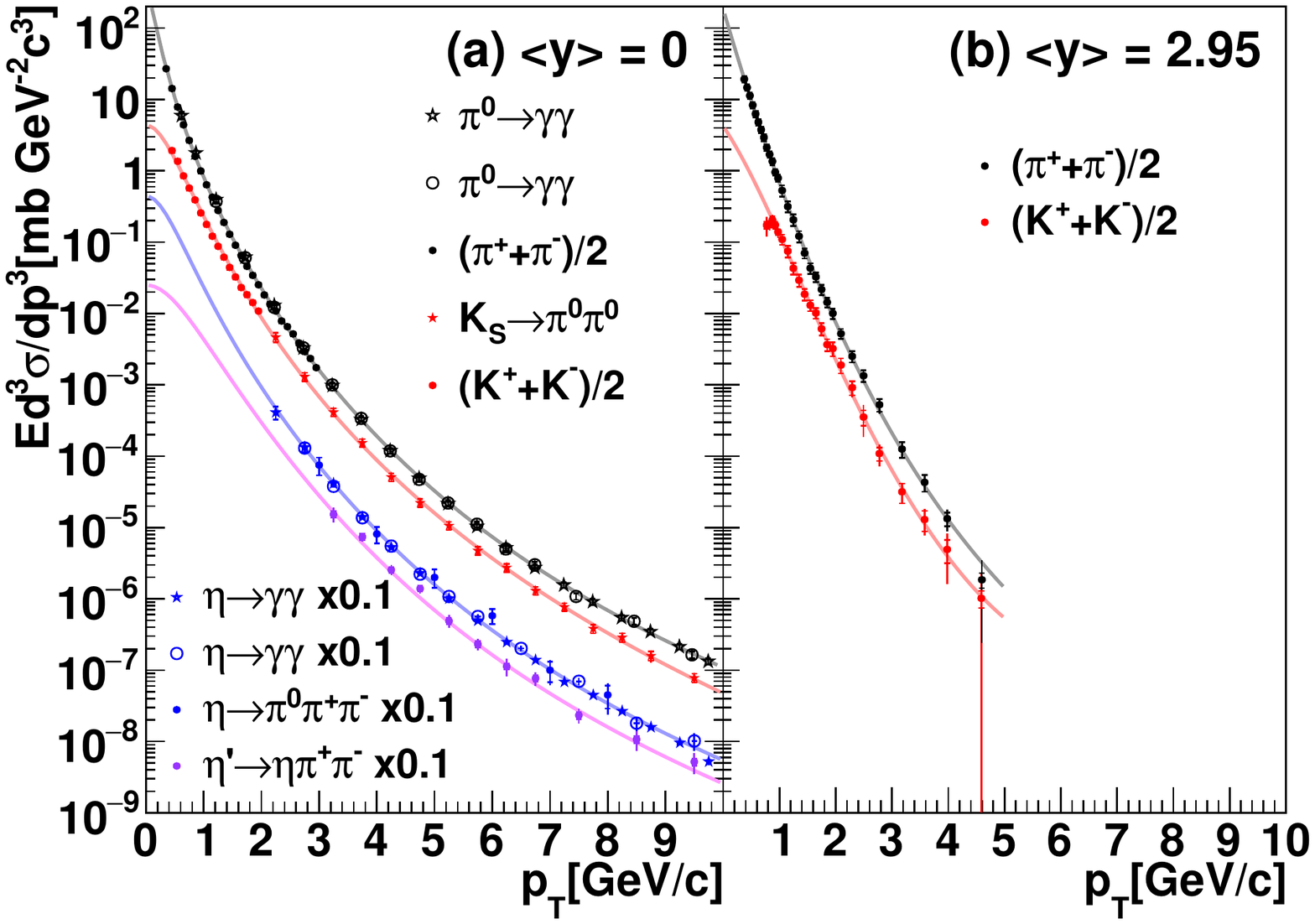}
\includegraphics[width=0.37\linewidth]{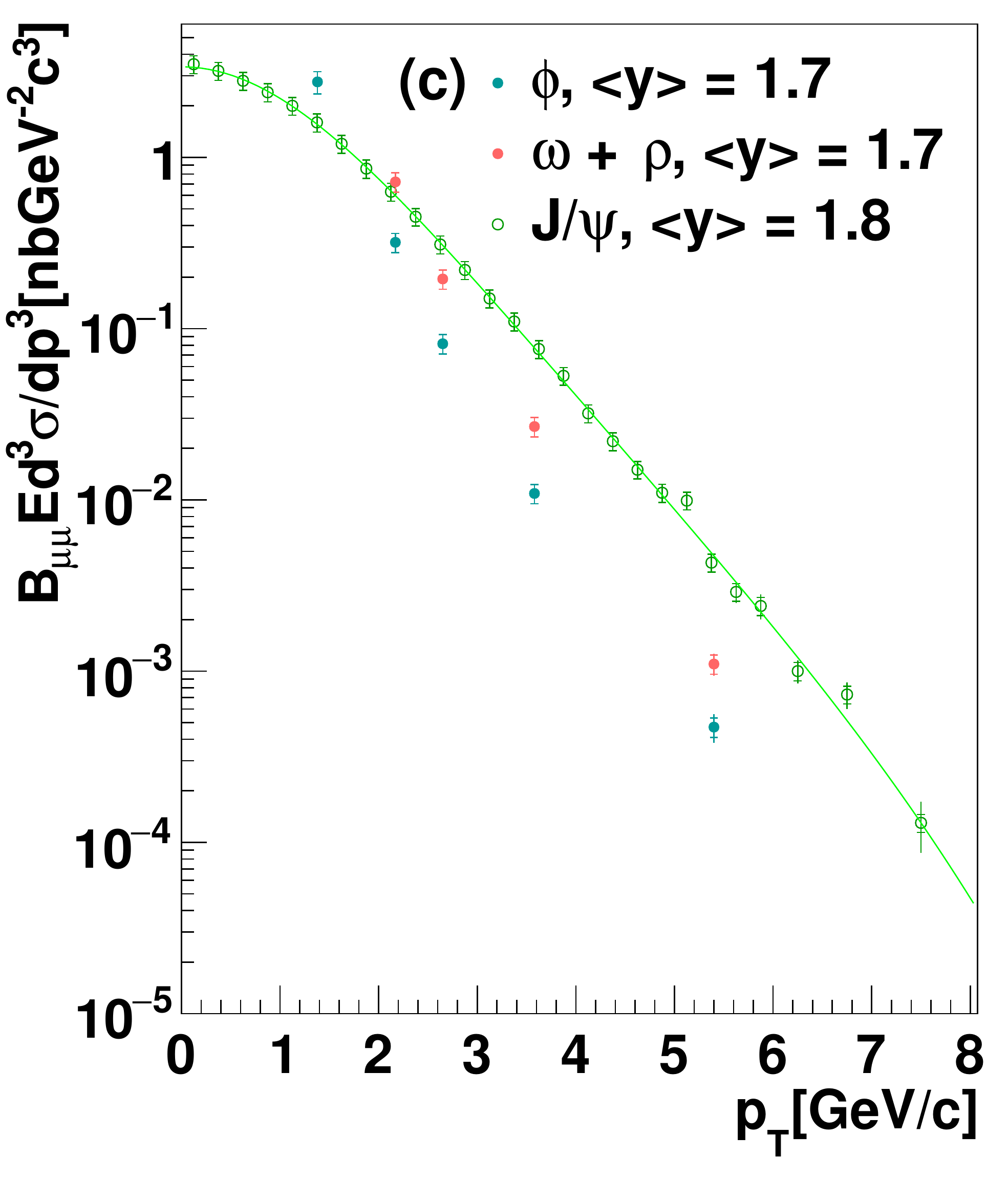}
\caption{\label{Fig:hpt} 
Compilation of meson production in \pp collisions at $\sqrt{s}=200$ GeV 
at (a) $\langle y \rangle=0$, (b) $\langle y \rangle=2.95$ and (c) 
$\langle y \rangle=1.7$--1.8. The data at $\langle y 
\rangle=0$ is taken from PHENIX: $\pi^{0}\rightarrow\gamma\gamma$ 
\cite{Adare:2007dg}(black star),\cite{Adler:2003pb}(black open circle), 
$(\pi^{+} + \pi^{-})/2$ \cite{Adare:2011vy}, 
$K_{S}\rightarrow\pi^{0}\pi^{0}$ \cite{Adare:2010fe}, $(K^{+} + 
K^{-})/2$ \cite{Adare:2011vy}, $\eta\rightarrow\gamma\gamma$ 
\cite{Adler:2006bv}(blue star),\cite{Adare:2010cy}(blue open circle), 
$\eta\rightarrow\pi^{0}\pi^{+}\pi^{-}$ \cite{Adler:2006bv}, 
$\eta'\rightarrow\eta\pi^{+}\pi^{-}$ \cite{Adare:2010fe}. The data at 
$\langle y \rangle=2.95$ is taken from BRAHMS: $(\pi^{+} + \pi^{-})/2$ 
\cite{Arsene:2007jd}, $(K^{+} + K^{-})/2$ \cite{Arsene:2007jd}. The data 
at $\langle y \rangle=1.7$--1.8 is taken from PHENIX: 
$\phi\rightarrow\mu\mu$ \cite{Adare:2014mgt}, 
$\omega+\rho\rightarrow\mu\mu$ \cite{Adare:2014mgt}, 
$J/\psi\rightarrow\mu\mu$ \cite{Adare:2011vq}. The curves are fits using 
modified Hagedorn~\cite{Adare:2017caq} or Tsallis~\cite{Tsallis:1987eu} 
functions to data.}
\end{figure*}

The mesons $\rho$, $\omega$, $\phi$, and $J/\psi$ can be generated based 
on the measured differential cross 
sections~\cite{Adare:2011vq,Adare:2014mgt} that are displayed on in 
Fig.~\ref{Fig:hpt}(c). We use the Gounaris/Sakurai parameterization to 
describe the line shape of the $\rho$ meson mass 
distribution~\cite{Gounaris:1968mw}. The $\rho$ is fixed to the $\omega$ 
with $\sigma_{\rho}/\sigma_{\omega} = 1.21 \pm 0.13$, which is 
consistent with the value found in jet 
fragmentation~\cite{Patrignani:2016xqp}. Because there is no measurement 
at forward rapidity, we constrain the $\eta$ and $\eta'$ using 
measurements at 
midrapidity~\cite{Adler:2006bv,Adare:2010fe,Adare:2010cy}, which is 
shown in Fig.~\ref{Fig:hpt}(a), and use \pythia 
v6.428~\cite{Sjostrand:2006za} to extrapolate to forward rapidity.

The \pt spectra of $\psi'$ and $\Upsilon$ are generated using \pythia 
and normalized using the measurements of $\psi'$ to $J/\psi$ 
ratio~\cite{Adare:2016psx} and 
$B_{\mu\mu}dN_{\Upsilon}/dy$~\cite{Adare:2012bv}, respectively. All 
mesons are decayed using \pythia to handle the decay kinematics.

\subsubsection{Open Heavy flavor \label{Sec:hf}}

The \mumu pairs that originate from semi-leptonic decays of heavy flavor 
hadrons, or heavy flavor pairs, are simulated using two event 
generators, \pythia and \powheg.

\begin{figure}
\includegraphics[width=1.0\linewidth]{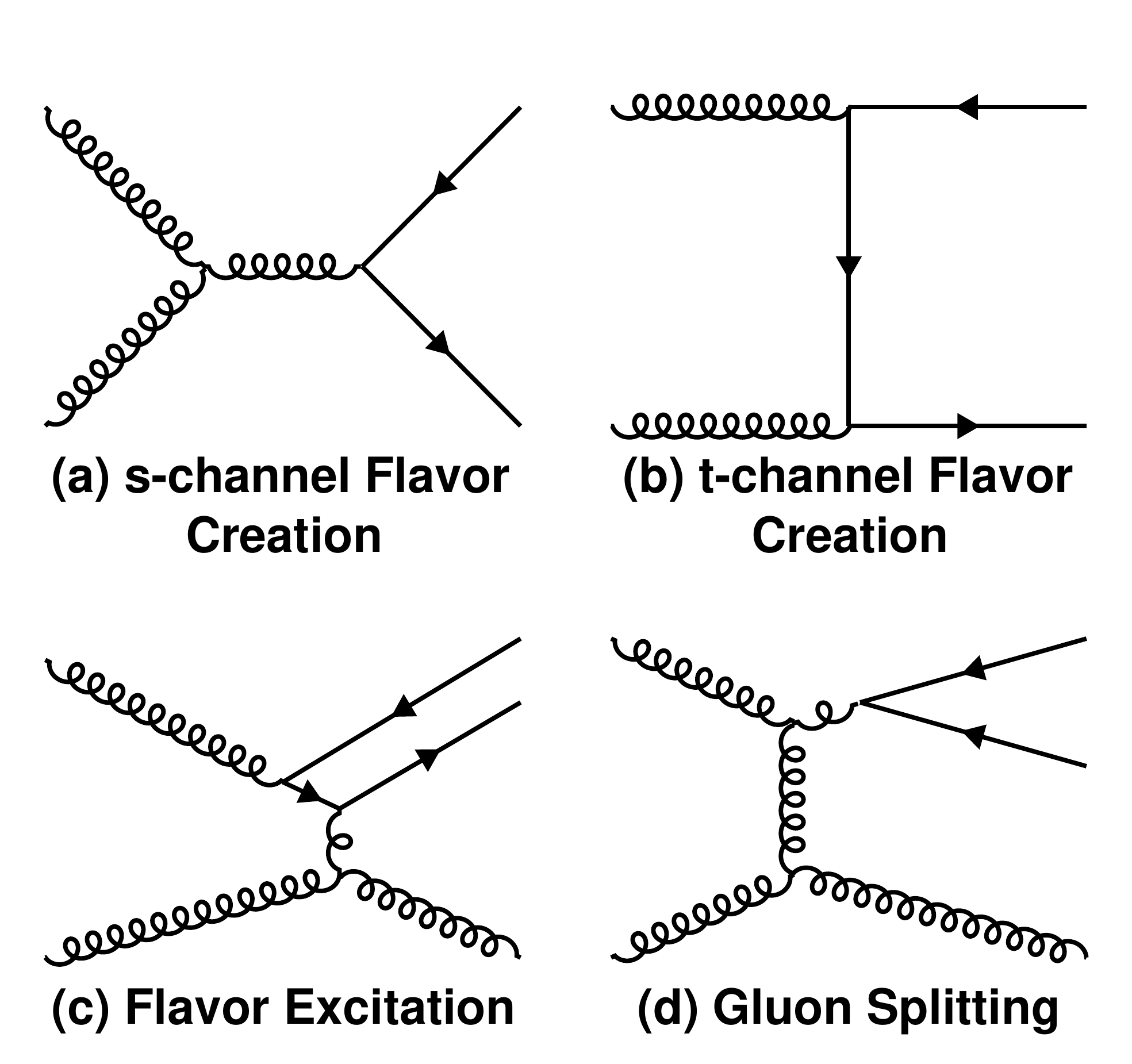}
\caption{Feynman diagrams corresponding to flavor creation (a,b), flavor 
excitation (c) and gluon splitting (d) \cite{Norrbin:2000zc, 
Ilten:2017rbd}\label{Fig:hfprocess}. }
\end{figure}

We use \pythia version v6.428~\cite{Sjostrand:2006za}. We use Tune A 
input parameters as shown in Table~\ref{Tab:mbpythia} in 
Appendix~\ref{app:simsettings}. In contrast to using the forced \cc and 
\bb production modes (MSEL4 or 5), which include only lowest-order 
process of flavor creation ($gg\to Q\bar{Q}$), we used the mode (MSEL1) 
which also simulates higher-order processes of flavor excitation ($gQ\to 
gQ$) and gluon splitting ($gg\to Q\bar{Q}g$). Figure~\ref{Fig:hfprocess} 
shows the Feynman diagrams corresponding to the different production 
processes. Leading order matrix elements are used for the initial hard 
process, and next-to-leading order corrections are implemented with a 
parton-shower approach. A classification of the three classes of 
processes can be achieved by tagging the event record which contains the 
full ancestry of any given particle; a detailed account of the 
characterization of these three classes can be found in 
Ref.~\cite{Norrbin:2000zc}.

\begin{figure*}[th]
\includegraphics[width=0.99\linewidth]{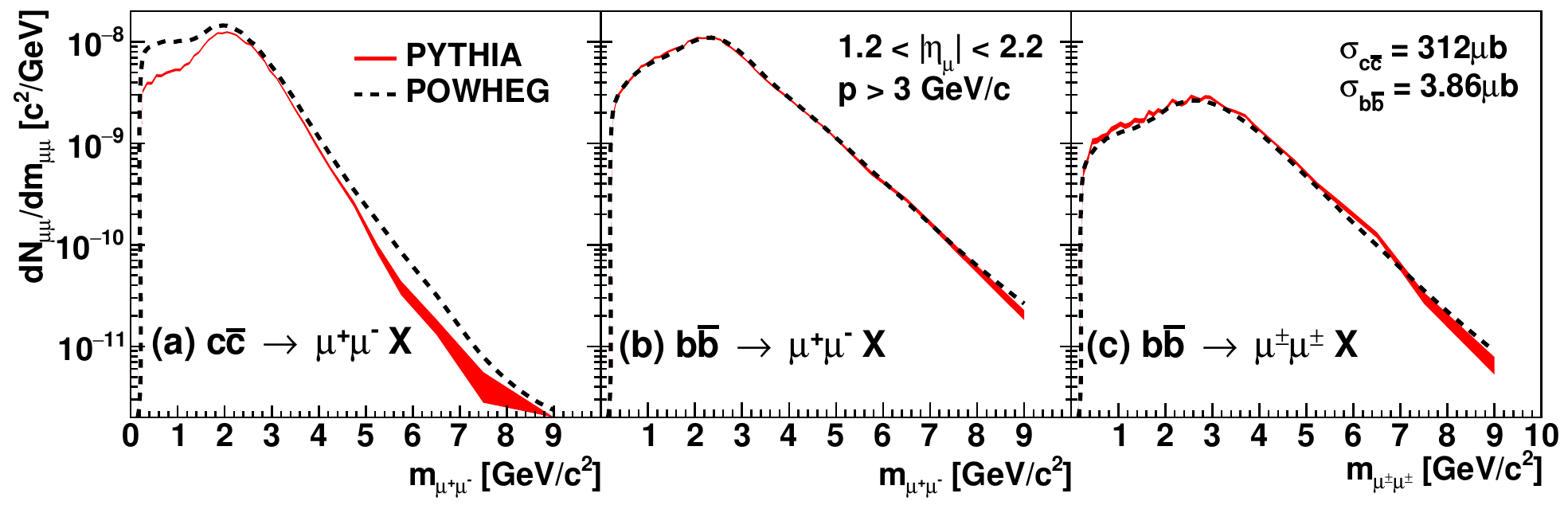}
\caption{\label{Fig:anahfmodel} Comparison of \mumu yield in the ideal 
muon arm acceptance determined using \pythia(red solid) and 
\powheg(black dotted). Both are normalized using cross 
sections($\sigma_{cc}=312\mu b$, $\sigma_{bb}=3.86\mu b$) from 
\cite{Adare:2017caq}. The width of the \pythia band represents the 
statistical uncertainty in the calculation.}
\end{figure*}

We also use \powheg version v1.0~\cite{Frixione:2007nw} interfaced with 
\pythia v8.100~\cite{Sjostrand:2007gs} to generate heavy flavor muon 
pairs. We use the default setting for \cc and \bb productions, including 
the choices for normalization and factorization scales and heavy quark 
masses. CTEQ6M is used for parton distribution functions of the proton. 
In contrast to \pythia, NLO corrections are directly implemented in the 
hard process using next-to-leading order matrix elements. As such, the 
classification of processes in \pythia is not applicable for \powheg; 
there is no trivial connection between the classes of processes in the 
\pythia formalism and the \powheg formalism.

The simulated mass spectra of pairs in the \textit{ideal muon arm 
acceptance}, which requires that each muon has a momentum $p>3$ \gevc 
and falls into the pseudorapidity range $1.2<|\eta|<2.2$, from \cc and 
\bb are shown in Fig.~\ref{Fig:anahfmodel}. Like-sign pairs from \cc is 
found to be negligible compared to \bb in the entire kinematic region 
and hence neglected for this analysis.

The \ulmumu and \lsmumu pair spectra from \bb are very similar for both 
generators; this is consistent with the findings in 
Refs.~\cite{Adare:2014iwg, Adare:2017caq} that, because of the large 
$b$-quark mass the spectra are dominated by decay kinematics rather than 
the correlation between the $b$ and $\bar{b}$ quarks. For the same 
reason variations of the scale and PDFs have a small effect on the shape 
of the mass spectra.

In contrast, we observe a significant model dependence for \ulmumu pairs 
from \cc, indicating a much larger sensitivity to the correlation 
between the $c$ and $\bar{c}$ quarks. Similar to \ee pairs 
\cite{Adare:2015ila}, this is most pronounced at low masses. This is due 
to differences in description of the correlations between the $c$ and 
$\bar{c}$ quarks; the opening angle distributions in \powheg is flatter 
which leads to higher yields at low masses. A smaller but non-negligible 
discrepancy at higher masses is also observed. Because high mass pairs 
are dominated by back-to-back pairs from leading order processes, this 
difference is likely due to a harder \pt spectrum predicted by \powheg 
compared to \pythia.

\subsubsection{Drell-Yan \label{Sec:dy}}

We use \pythia v6.428 to simulate \mumu pairs from the Drell-Yan 
production mechanism (Drell-Yan pairs). The input parameters are shown 
in Table~\ref{Tab:dysetup} in Appendix~\ref{app:simsettings}. The 
primordial $k_{T}$ is generated from a Gaussian distribution. The width 
of the distribution is $1.1$ \gevc and was determined by investigating 
the $p_{T}$ distribution of unlike-sign pairs in the mass region 
4.8--8.6 \gevcc where the yield is expected to be dominated by 
Drell-Yan~\cite{Arnaldi:2008er}. The procedure and its associated 
uncertainties will be explained in detail in Sec.~\ref{Sec:dyktsys}.

\subsection{Unphysical \mumu pair sources}
\label{Sec:corrhadrons}

Unphysical pair background is customarily subdivided into combinatorial 
and correlated pairs. Here the idea is that for combinatorial pairs the 
two tracks have no common origin and thus are uncorrelated. In contrast, 
for correlated pairs the tracks do have a common origin, for example 
they both stem from the decay chain of a heavy hadron or they were part 
of the fragmentation products of a jet or the like. 


In \pp collisions, or generally in events with a small number of 
produced particles, the distinction between combinatorial and correlated 
pairs is not well defined. A \pp collision typically produces hard 
scattered partons accompanied by an underlying event, which consists of 
initial and final state radiation, beam-beam remnants and multiple 
parton interactions. The complex event structure in a single \pp event 
forbids a clear identification of whether two particles stem from a 
common origin or not. All particles are produced from the two colliding 
protons, and thus are correlated through momentum and charge 
conservation. Therefore, the separation is more procedural and is 
defined by how the relative contributions of correlated and 
combinatorial pairs are determined. We use an approach that maximizes 
the number of pairs considered combinatorial, which will be discussed in 
detail in Sec.~\ref{Sec:normhadrons}.

The individual contributions of the unphysical pair background are 
determined using Monte-Carlo event generators. We treat pairs that are 
made from two hadronic tracks (\textit{hadron-hadron pairs}: $N_{hh}$) 
and those with one hadronic track and the other being a muon from the 
decay of a $D$, $B$, or $J/\psi$ meson (\textit{muon-hadron pairs}: 
$N_{Dh}$, $N_{Bh}$ and $N_{Jh}$) separately.

\subsubsection{Hadron-hadron pairs: $N_{hh}$ \label{sec:nhh}}

The $N_{hh}$ pairs are simulated with \pythia, using parameters listed 
in Table~\ref{Tab:mbpythia}. This Tune A setup reproduces jet-like 
hadron-hadron correlations at midrapidity in \pp collisions at 
$\sqrt{s}=200$ GeV~\cite{Adler:2005ad} reasonably well. To also 
reproduce the \pt spectra we use momentum dependent weighting to match 
the \pythia distributions to data. In the literature there are no data 
for \pt spectra of charged pions and kaons from \pp collisions at 
$\sqrt{s}=$200 GeV in the rapidity region covered by the muon arms. 
Thus, we interpolate between \pt spectra measured at 
midrapidity~\cite{Adler:2003pb,Adare:2007dg,Adare:2011vy,Adare:2010fe} 
and very forward rapidity ($2.9 < y < 3.0$)~\cite{Arsene:2007jd}. The 
data are given in Fig.~\ref{Fig:hpt}. Weighting factors are extracted 
for both rapidity ranges as a function of \pt, by taking the ratio 
between data and \pythia,

\begin{align}
w_h(y=0,p_{T})&=\frac{E\frac{d^{3}\sigma}{dp^{3}}|_{y=0,DATA}}{E\frac{d^{3}\sigma}{dp^{3}}|_{y=0,PYTHIA}},\\
w_h(y=2.95,p_{T})&=\frac{E\frac{d^{3}\sigma}{dp^{3}}|_{y=2.95,DATA}}{E\frac{d^{3}\sigma}{dp^{3}}|_{y=2.95,PYTHIA}},
\end{align}
where $h$ stands for pion or kaon. For a given \pt, we linearly interpolate
the weighting factors as a function of $y$:

\begin{align}
\nonumber
w_h(y,p_{T}) =& \frac{y}{2.95}\times[w_h(y=2.95,p_{T})-w_h(y=0,p_{T})]\\
&+ w_h(y=0,p_{T}).
\end{align}

\begin{figure}[h]
\includegraphics[width=1.0\linewidth]{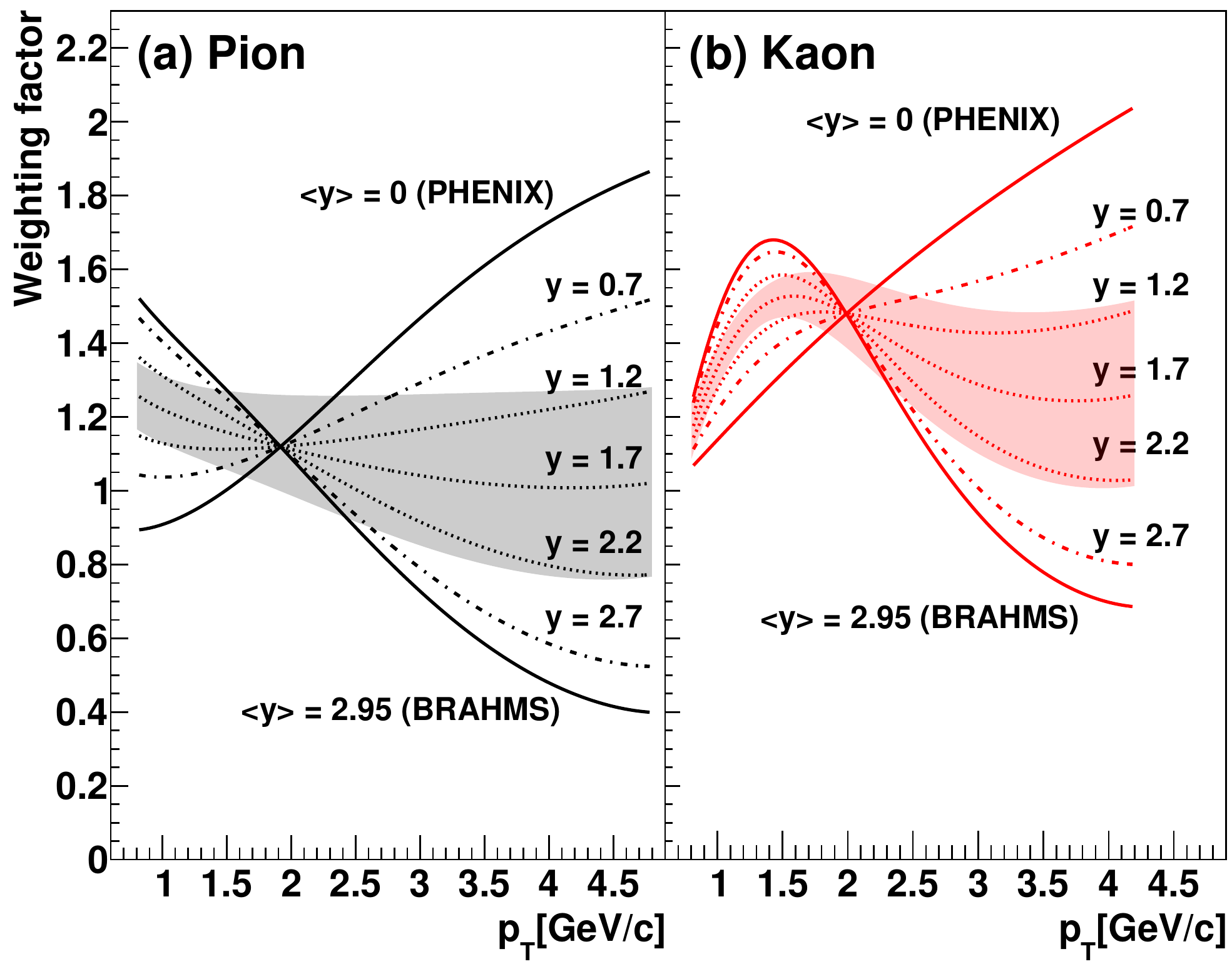}
\caption{\label{Fig:wgtfactors} Weighting factors for (a) pions and 
(b) kaons in different rapidity slices. The shaded bands indicate 
uncertainty brackets used for the investigation of systematic 
uncertainties (see Sec.~\ref{Sec:syshadronspectra}).}
\end{figure}

These weighting factors are shown in Fig.~\ref{Fig:wgtfactors}. Above 
\pt = 5 \gevc, where there are no data at forward rapidity, the weights 
are assumed to be constant. The systematic uncertainties from this 
weighting procedure are discussed in Sec.~\ref{Sec:syserror}. The 
weighting factors are applied to each input particle generated with the 
\pythia simulation.

\subsubsection{Muon-hadron pairs: $N_{Dh}$, $N_{Bh}$, and $N_{Jh}$ 
\label{sec:nmh}}

Muon-hadron pairs $N_{Dh}$ and $N_{Bh}$ as defined above are constructed 
using the same \pythia and \powheg simulations that determine the open 
heavy flavor pair input. The pion and kaon \pt spectra are tuned the 
same way as discussed above. For the muon-hadron pairs involving decays 
of the $J/\psi$ ($N_{Jh}$) we also match the \pythia $J/\psi$ momentum 
spectrum at forward rapidity to reproduce the measured $J/\psi$-hadron 
yield per MB event~\cite{Adare:2011vq} (see Fig.~\ref{Fig:hpt}).

\subsubsection{Combinatorial pair background \label{Sec:combbg}}

The combinatorial pair background is constructed via an event mixing 
technique, which combines tracks from different events of similar vertex 
position \z. This is done separately for data and the events used to 
simulate hadron-hadron pairs, and muon-hadron pairs.

To optimize the description of the pair background spectrum, we maximize 
the contribution identified as combinatorial pair background, subtract 
the combinatorial component from the simulation of hadron-hadron and 
muon-hadron pairs, and substitute the combinatorial pair background with 
the one determined from data.  The motivation of this procedure and 
the details of the normalization of individual components are discussed 
in Sec.~\ref{Sec:combbkgnorm}.

\section{Simulation framework}
\label{Sec:simframe}

To directly compare the expected sources to the data, the \mumu pairs 
from the expected sources are propagated through a Monte-Carlo 
simulation of the PHENIX detector. This simulation is designed to 
emulate in detail the detector response, and the recording and analysis 
of data taken with the PHENIX experiment. Histograms of the expected 
number of \mumu pairs are constructed in mass-\pt bins, which serve as 
templates for the subsequent fitting procedure.

The \mumu pairs from all physical sources are propagated through the 
default PHENIX simulation framework. The same approach is not practical 
for unphysical pair background from $\pi$ and $K$ decays. Because of the 
large ($\sim$1/1000) rejection power for these backgrounds, an 
undesirably large amount of simulations would be necessary to reach 
sufficient statistical accuracy. Therefore, we use a fast Monte-Carlo 
(FastMC), developed specifically for this analysis. Detailed 
descriptions of the two simulation chains can be found in 
Appendix~\ref{app:simframework}.

\section{Iterative procedure to extract charm, bottom and Drell-Yan 
cross sections
\label{Sec:fitting}}

In the previous two sections we have discussed the different expected 
sources of \mumu pairs and how template distribution of \mumu pairs are 
generated for each. In this section we compare the templates for the 
expected sources to the experimental data and determine the absolute 
contribution of each source.

After an initial normalization is chosen for each template, the key 
sources, \cc, \bb, Drell-Yan, and the hadronic pair background, are 
normalized in an iterative template fitting procedure.

\subsection{Initial normalization and data-driven tuning of 
cocktail\label{Sec:init}}

\subsubsection{Physical \mumu pair sources}

The normalization of muon pairs from hadron decays $h \rightarrow \mu\mu 
(X)$ is fixed because the cross sections of the parent mesons are set by 
experimental data as discussed in Sec.~\ref{Sec:hdecay}. The 
normalizations for each component are varied separate within 
experimental uncertainties to estimate the corresponding systematic 
uncertainties (see Sec.~\ref{Sec:syserror}).

The distributions for muon pairs from \cc, \bb, and Drell-Yan are 
normalized by the parameters $\kappa_{c\bar{c}}$, $\kappa_{b\bar{b}}$, 
and $\kappa_{\rm DY}$. These parameters will be determined via the 
iterative fitting procedure presented in this section. The initial 
values of $\kappa_{c\bar{c}}$, $\kappa_{b\bar{b}}$, and $\kappa_{\rm 
DY}$ are set based on measured data \cite{Adare:2017caq}.

\subsubsection{Correlated hadrons and combinatorial pair background}
\label{Sec:normhadrons}

The composition and normalization of the unphysical pair background 
sources is key to understanding the \mumu continuum and requires a more 
detailed discussion. In \pp collisions at $\sqrt{s}=200$ GeV, the 
multiplicity of produced particles is low, and hence there is no 
clear-cut method to differentiate between a correlated pair and a 
combinatorial pair. Great care is taken to assure that the procedure 
used to define combinatorial pairs and how their contribution is 
normalized does not affect the extraction of physical quantities.

One possibility to circumvent the distinction of correlated and 
combinatorial pairs is to generate hadron-hadron and muon-hadron pairs 
using a Monte-Carlo event generator like \pythia interfaced to the 
FastMC framework. Templates from a full event normalization include all 
background pair sources, hence the distinction between them is not 
necessary. However, in this method the extracted physical cross section 
is sensitive to how accurate \pythia describes the underlying event and 
how well {\sc geant4} treats hadronic interactions in the absorber. This 
may increase the systematic uncertainties on the extraction of the \cc, 
\bb, and Drell-Yan components.

In this analysis we use a data-driven hybrid approach, in which 

\begin{itemize}

\item the maximum possible number of combinatorial pairs is determined 
from the generated \pythia and/or \powheg events,

\item the correlated hadronic pairs are calculated by subtracting the 
combinatorial pairs determined by mixing generated events,

\item the combinatorial pairs are replaced by the combinatorial pairs 
determined from data.

\end{itemize}

Although the distinction between correlated hadronic pairs and 
combinatorial pairs depends on the choice of the normalization 
procedure, using different procedures has a negligible effect on 
extraction of physical cross sections. The separation of these two 
components is mostly important for the evaluation of systematic 
uncertainties, because the correlated hadronic pairs depend on 
simulations and the combinatorial pairs do not. Replacing the 
combinatorial pairs from the generator with mixed pairs from data should 
be regarded as a correction to the simulations to reduce systematic 
uncertainties.

\subsubsection*{Normalizing hadron-hadron and muon-hadron pairs 
\label{sec:normnhhnmh}}

The templates for hadron-hadron pairs $N_{hh}(m,p_T,z)$ are generated 
using \pythia simulations interfaced to the FastMC, as discussed above. 
Templates are determined separately for the three different \z regions 
($z'_{i}$) available in the FastMC simulations, 
$z'_{0}=(-22.5,-17.5~{\rm cm})$, $z'_{1}=(-2.5,+2.5~{\rm cm})$ and 
$z'_{2}=(+17.5,+22.5~{\rm cm})$, respectively. Only pions, kaons, and 
their decay products are considered. The momentum spectra were tuned to 
accurately describe experimental data, where available (see 
Sec.~\ref{sec:nhh}). Therefore, $N_{hh}$ contains the correct mix of 
individual hadron-hadron pair sources per event. $N_{hh}$ is initially 
normalized as a per event yield for generated MB \pp collisions.

Similarly, muon-hadron pair templates from \cc and \bb are constructed 
using \pythia and \powheg generators interfaced to the FastMC. The 
templates $N_{Dh}(m,p_T,z)$ and $N_{Bh}(m,p_T,z)$ correspond to 
muon-hadron pairs from \cc and \bb, respectively. Each is normalized per 
\cc or \bb event. Thus, they can be added to $N_{hh}$ scaled by the 
normalization factors $\kappa_{c\bar{c}}$ and $\kappa_{b\bar{b}}$, used 
for the \mumu pairs, such that $\kappa_{c\bar{c}}N_{Dh}$ and 
$\kappa_{b\bar{b}}N_{Bh}$ are the expected muon-hadron pair yields per 
MB \pp event.

For $J/\psi$, the differential cross section at forward rapidity has 
been measured~\cite{Adare:2011vq}. Analogous to the pion and kaon 
simulations, we weight the simulated $J/\psi$ momentum distribution to 
match the $J/\psi$ yield at forward rapidity. Because the simulated 
$J/\psi$ yield is normalized to the measured yield, the muon-hadron 
pair template $N_{Jh}(m,p_T,z)$ represents a yield per MB \pp event.

The full per MB \pp event hadronic pair background can thus be written 
as:

\begin{align}
\label{Eq:hbg}
N_{hbg} = \kappa_{c\bar{c}}N_{Dh}+\kappa_{b\bar{b}}N_{Bh}+N_{hh}+ N_{Jh},
\end{align}

\noindent where the templates are functions of $m$, \pt, and \z. 
Figure~\ref{Fig:hcocktail}(a) shows $N_{hbg}$ and its individual 
contributions integrated over \z and \pt as a function of mass.

\subsubsection*{Choice and normalization of the combinatorial pair background
\label{Sec:combbkgnorm}}

\begin{figure}[ht]
\includegraphics[width=0.95\linewidth]{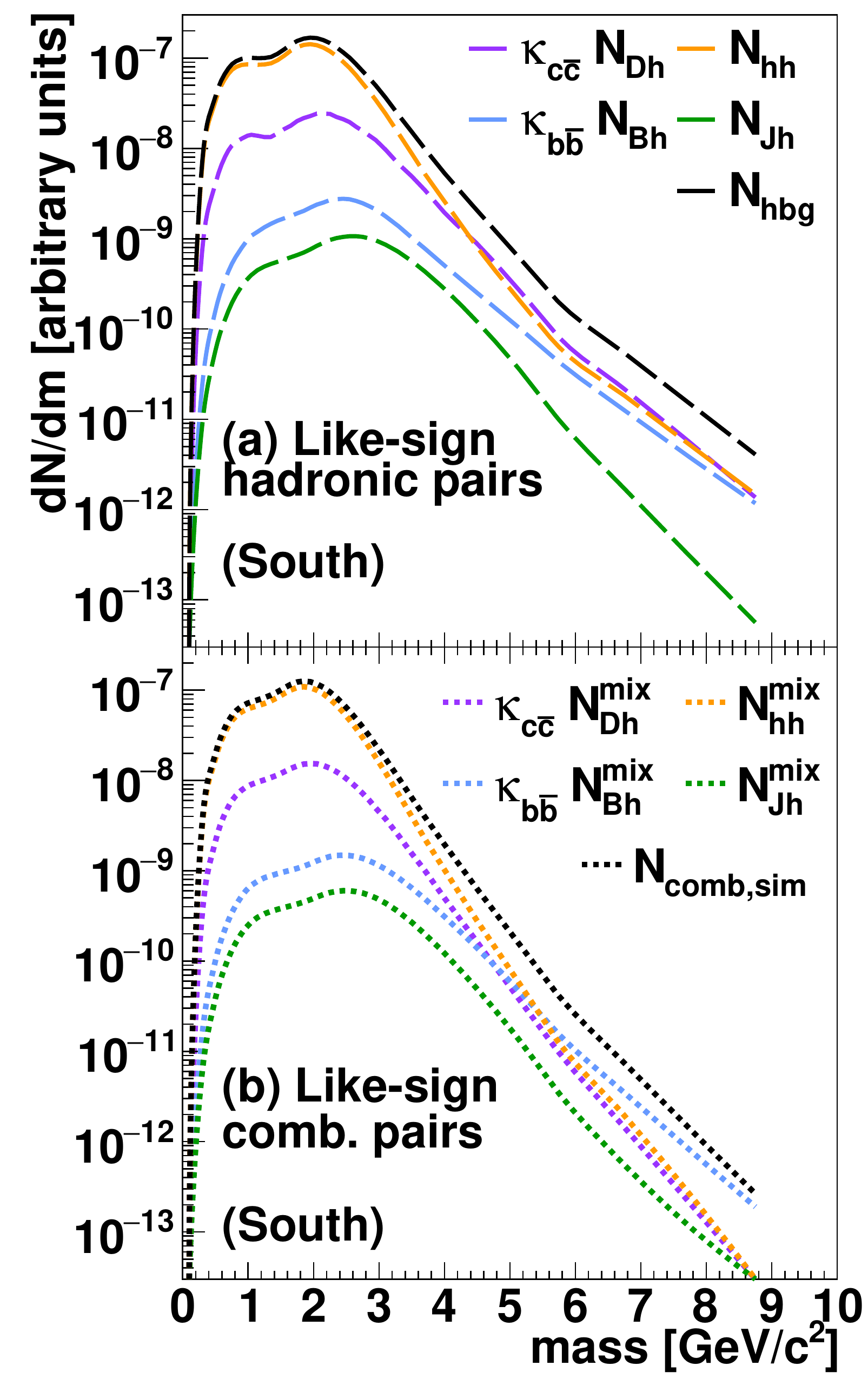}
\caption{\label{Fig:hcocktail} (a) full simulation for hadronic pairs 
and (b) combinatorial pairs for mass spectra of hadron-hadron and 
muon-hadron pairs from charm, bottom and $J/\psi$ after initial 
normalization and tuning.
}
\end{figure}

To minimize any remaining model dependence in $N_{hbg}$ used in the 
analysis, we determine the combinatorial contribution to $N_{hbg}$ from 
mixed generated events and replace it with the combinatorial pairs 
determined from data. For each simulation we determine the combinatorial 
pairs by mixing either hadron-hadron pairs or muon-hadron pairs from 
different events at the same $z'_{i}$. For a given $z'_{i}$ bin the 
combinatorial pairs are then constructed as:

\begin{align}
\label{Eq:combsim}
N_{\rm comb,sim} = \kappa_{c\bar{c}}N_{Dh}^{\rm mix}+\kappa_{b\bar{b}}N_{Bh}^{\rm mix}+N_{hh}^{\rm mix}+ N_{Jh}^{\rm mix},
\end{align}

\noindent which observes the same relative normalization of the 
individual components as in Eq.~\ref{Eq:hbg}. The contributions of each 
component to the hadronic and the combinatorial pair background, 
normalized following the above procedure are shown in 
Fig.~\ref{Fig:hcocktail}(b).

\begin{figure}[ht]
\includegraphics[width=1.0\linewidth]{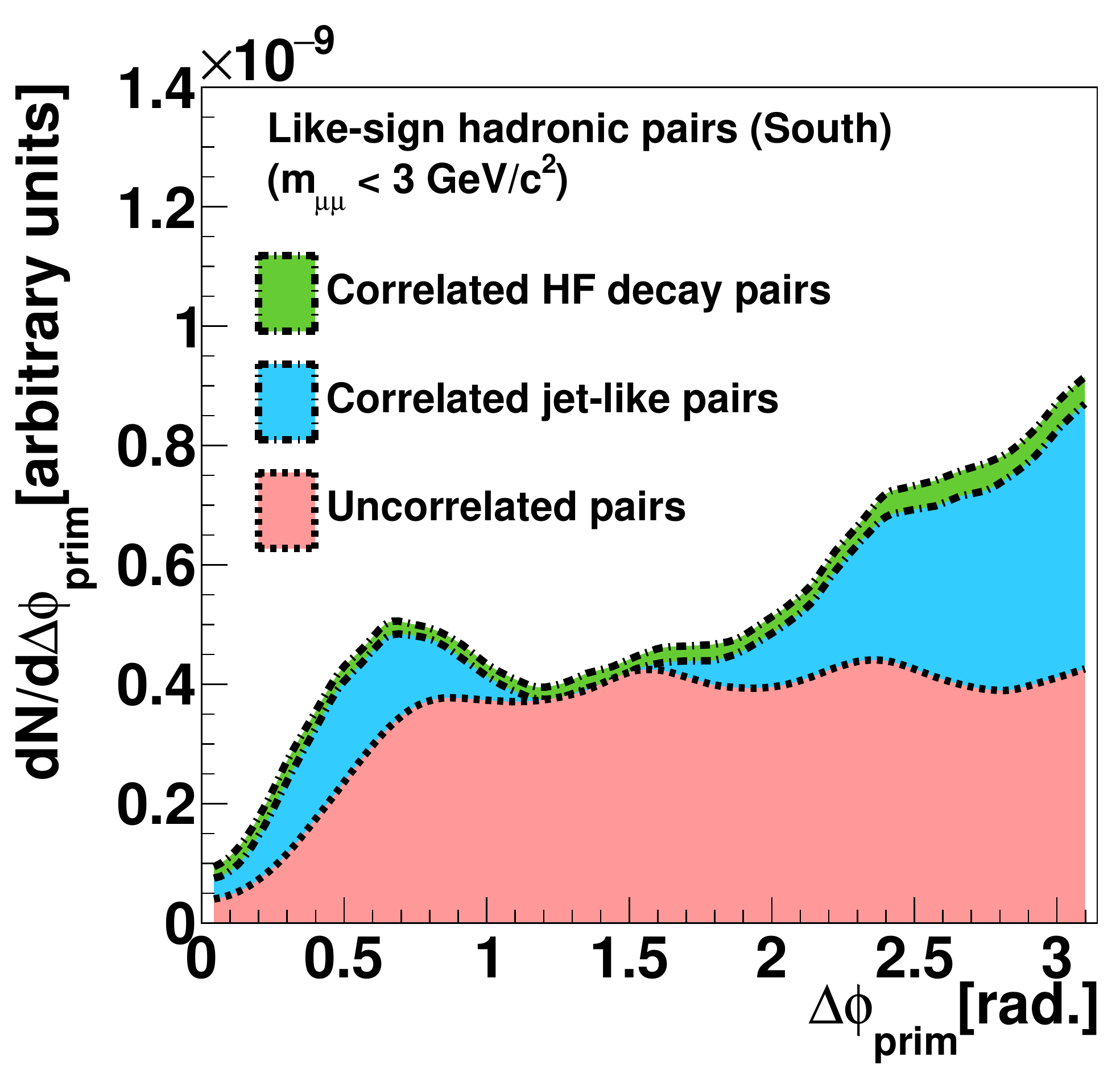}
\caption{\label{Fig:zyameg} ZYAM normalization procedure for the south 
muon arm. The normalization of the uncorrelated pairs from event mixing 
(red) is determined by enforcing the requirement that the yield of the 
uncorrelated pairs ($N_{\rm corr,sim}$) is identical to the yield of 
foreground pairs ($N_{hbg}$), excluding the pairs from heavy-flavor decay 
chains (green) at $\Delta\phi_{prim}\sim\pi/2$. The excess yield is from 
away-side and near-side jet-like correlations (blue). The periodicity of 
the distributions arises from the octant structure of the MuTr.}
\end{figure}

The normalization of the combinatorial pairs is determined statistically 
via the ZYAM (Zero Yield At Minimum) technique~\cite{Adler:2005ee} as 
described below. We use the azimuthal angle difference 
$\Delta\phi_{prim}$ of the like-sign hadronic pairs with masses less 
than 3 \gevcc. Here $\Delta\phi_{prim}$ is the difference of the 
azimuthal angles of the input particles ($\pi$, $K$, $D$, or $B$); the 
distribution is shown in Fig.~\ref{Fig:zyameg}.

\begin{figure*}[ht]
\includegraphics[width=0.99\linewidth]{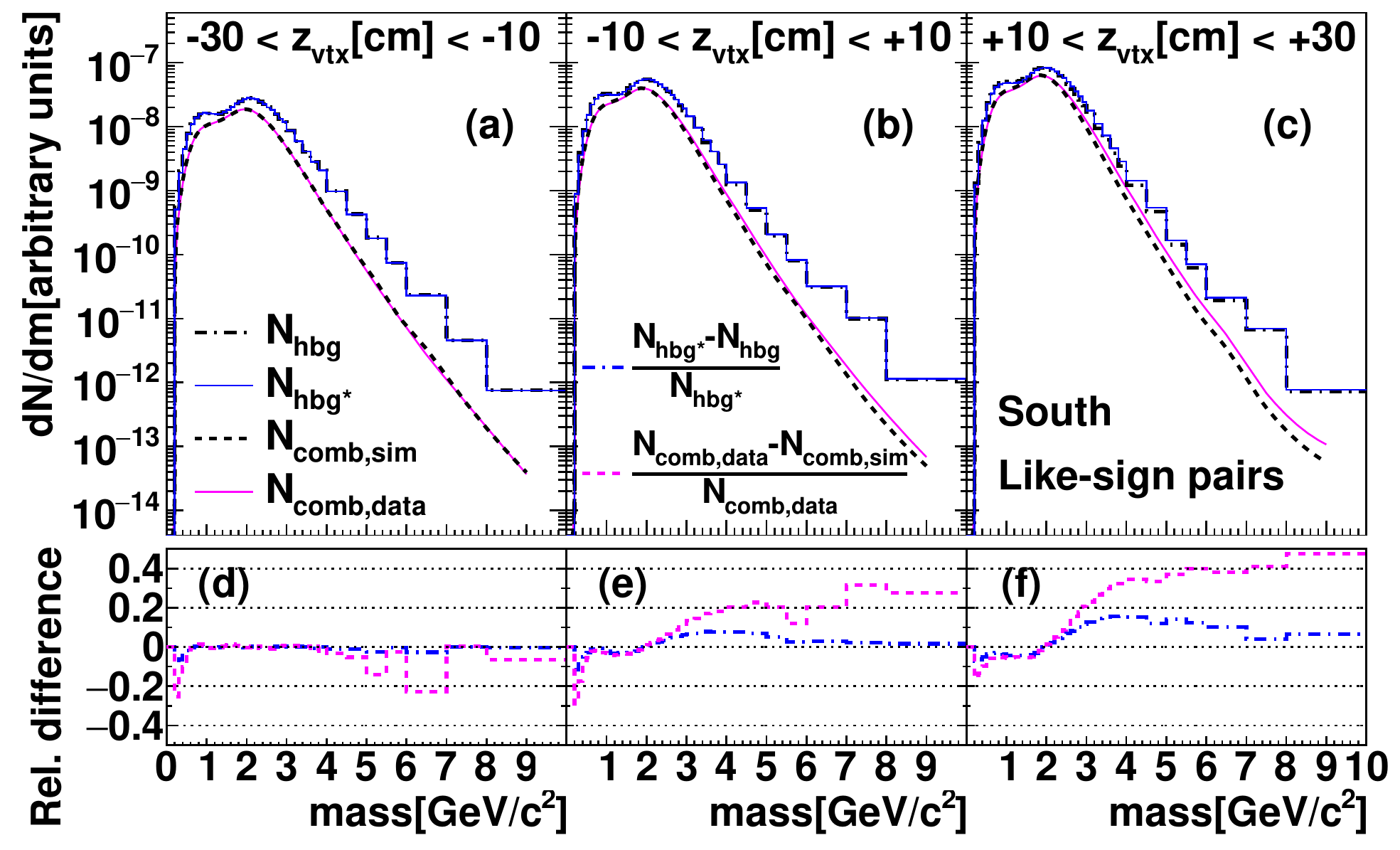}
\caption{\label{Fig:datasimeventmix} Like-sign mass spectra of the 
hadronic pair background (before and after correction by replacing with 
combinatorial pairs from data), and combinatorial background 
(simulations and data) in different $z_{\rm vtx}$ regions. Panels 
(d,e,f) show the relative difference between different mass spectra.}
\end{figure*}

First, we remove muon-hadron pairs in which both tracks originated from 
heavy flavor (\cc or \bb) pairs, because these pairs can uniquely be 
identified as correlated. For the remaining pairs we assume that 
correlations result mostly from jet-fragmentation. These should have a 
minimal contribution for $\Delta\phi_{prim} \sim \pi/2$. Thus, our ZYAM 
assumption is that the correlated yield vanishes at $\Delta\phi_{prim} = 
\pi/2$. The excess yield for $\Delta\phi_{prim}<\pi/2$ can be 
interpreted as pairs from the same jet, whereas the excess yield for 
$\Delta\phi_{prim}>\pi/2$ would correspond to \mumu pairs from 
back-to-back jets. The correlated $N_{\rm corr,sim}$ and combinatorial 
$N_{\rm comb,sim}$ contributions are now separated via the relations:

\begin{align}
\label{Eq:corrsim}
N_{\rm corr,sim}&=N_{hbg}-N_{\rm comb,sim}.
\end{align}

The separation of $N_{hbg}$ into correlated and uncorrelated components 
is done for each of the three vertex region $z'_{i}$ used in the FastMC 
simulations. In the data, mixed events are also constructed in 5~cm 
$z$-bins, but over the full range from -30~cm to 30~cm. The template 
distributions are aggregated for three broad vertex ranges, 
$z_{0}=(-30,-10~{\rm cm})$, $z_{1}=(-10,+10~{\rm cm})$ and 
$z_{2}=(+10,+30~{\rm cm})$. The normalization of the mixed events from 
the data is matched to those from the simulation by scaling 
$N_{hbg}(z'_{i})$ such that the number of combinatorial pairs of data 
and simulations are identical in the normalization mass region 
$\mathcal{M}$ ($m<3$\gevcc) for each \z bin, i.e., we require:

\begin{align}
\label{Eq:multinorm}
&\int\limits_{\mathcal{M}}N_{\rm comb,sim}(z'_{i}) = \int\limits_{\mathcal{M}}N_{\rm comb,data}(z_{i}) 
\end{align}

This rescaling is necessary because we are approximating a 
$\Delta z_{i}$ range of 20 cm from data with a $\Delta z_{i}'$ range of 
5 cm from simulations. For the two \z bins further away from the 
absorber, this approximation holds well even without rescaling because 
the multiplicity falls linearly with the distance from the absorber, and 
the center of the bin times the bin width is to first order a good 
approximation of the integral of the bin. However, for the \z bin 
closest to the absorber, this linear relation no longer holds and a 
scaling factor of $1.2$ is applied to $N_{\rm comb,sim}$ according to 
Eq.~\ref{Eq:multinorm}.

We then replace the combinatorial background from simulations by data 
for each vertex region $z_{i}$:

\begin{align}
\label{Eq:hbgstar}
N_{hbg*}(z_{i})=N_{\rm corr,sim}(z'_{i})+N_{\rm comb,data}(z_{i}).
\end{align}

The hadronic pair background in each vertex slice for the south arm, 
before and after the above replacement of the combinatorial pair 
background, is shown in Fig.~\ref{Fig:datasimeventmix}. The relative 
mass-dependent difference between the two estimates of the hadronic pair 
background ranges from $\sim 0\%$ for the $z_{\rm vtx}$ region closest 
to the absorber to a maximum of $ \sim20\%$ at $m\sim4$ \gevcc for the 
$\z_{\rm vtx}$ region furthest away from the absorber.

The same normalization is applied to unlike-sign hadronic pairs. Both 
the unlike- and like-sign hadronic pairs are scaled with a common 
normalization factor $\kappa_{h}$ to be determined in the fitting 
procedure. Finally, we define the correlated hadronic pairs, $N_{\rm 
cor}$ and combinatorial pairs, $N_{\rm comb}$ via the relations:

\begin{align}
\label{Eq:corrcomb}
\nonumber
N_{\rm cor}&=N_{\rm corr,sim},\\
N_{\rm comb}&=N_{\rm comb,data}.
\end{align}

The distinction between correlated and combinatorial hadronic pairs 
depends on the details of the normalization procedure. Different 
normalization procedures can lead to significant differences in the 
relative contributions of correlated and combinatorial components. 
However, the effect on the extraction of physical cross sections is 
small. The variations are included in the systematic uncertainties (see 
Sec.~\ref{sec:syszyam}).

\subsection{Iterative fit \label{Sec:fit}}
\subsubsection*{Fit strategy}

The absolute contribution of each of the various known sources to the 
\ulmumu and \lsmumu spectra is determined by a fitting procedure using a 
template distribution for each contribution. There are four fit 
parameters, $\kappa_{c\bar{c}}$, $\kappa_{b\bar{b}}$, $\kappa_{\rm DY}$, 
and $\kappa_{h}$, which are normalization factors for the contributions 
from \cc, \bb, Drell-Yan, and the hadronic pairs.

We adopt the following iterative fitting strategy, here parameters 
marked with a tilde correspond to fit values obtained in the previous 
step:

\renewcommand{\labelenumi}{(\roman{enumi})}
\begin{enumerate}

\item With a fixed $\kappa_{c\bar{c}}$, fit the like-sign 
spectrum with $\kappa_{b\bar{b}}$ and $\kappa_{h}$ as free parameters in 
mass-\pt-$z_{\rm vtx}$ slices in the mass range 1--10 \gevcc.

\item With the same $\kappa_{c\bar{c}}$ as in step (i) and 
$\tilde{\kappa}_{b\bar{b}}$ and $\tilde{\kappa}_{h}$ obtained in (i), 
fit mass and \pt slices in the unlike-sign mass region 4.4--8.5 \gevcc 
with $\kappa_{\rm DY}$ as a free parameter.

\item With $\tilde{\kappa}_{b\bar{b}}$ and $\tilde{\kappa}_{h}$ obtained 
in (i) and $\tilde{\kappa}_{\rm DY}$ in (ii), fit mass and \pt slices in the 
unlike-sign mass region 1.4--2.5 \gevcc with $p_{T}<2$ \gevc, with 
$\kappa_{c\bar{c}}$ as a free parameter.

\item Iterate with $\tilde{\kappa}_{c\bar{c}}$ from (iii).

\end{enumerate}

This method of fitting exploits the fact that the like-sign pairs 
contain mainly contributions from hadronic pairs and $b\bar{b}$; charm 
only contributes via muon-hadron pairs and is non-dominant while 
Drell-Yan does not contribute. Thus, the fit results in step (i) is not 
sensitive to the initial starting value of $\kappa_{c\bar{c}}$. 
The contribution of hadronic pairs to the \ulmumu and \lsmumu pairs 
increases as the distance between the event vertex $z_{\rm vtx}$ and the 
absorber becomes larger, due to enhanced probability of pions and kaons 
to decay before they hit the absorber. In contrast, the yield of \mumu 
pairs from \bb is independent of $z_{\rm vtx}$. To optimize the 
separating power between \mumu pairs from \bb and the hadronic pairs, in 
step (i) we fit like-sign pairs in mass-\pt-$z_{\rm vtx}$ slices. Step 
(i) gives strong constraints on $\kappa_{b\bar{b}}$ and $\kappa_{h}$, 
which are to first order free from systematic uncertainties on the 
$c\bar{c}$ and Drell-Yan templates. With $\kappa_{b\bar{b}}$ and 
$\kappa_{h}$ constrained, we move on to step (ii), where we fit the 
unlike-sign pairs with mass 4.4--8.5 \gevcc. This mass region is chosen 
to avoid contributions from quarkonia decays. Here, Drell-Yan and 
$b\bar{b}$ contributions are expected to dominate while contributions 
from $c\bar{c}$ and hadrons are secondary. Although Drell-Yan also 
contributes to lower masses, the sensitivity to the intrinsic $k_{T}$ 
make it unfavorable to constrain $\kappa_{\rm DY}$ in the low mass 
region. With $\kappa_{b\bar{b}}$, $\kappa_{h}$ and $\kappa_{\rm DY}$ 
constrained, we fit in the mass region 1.4--2.5 \gevcc to constrain 
$\kappa_{c\bar{c}}$. This mass region is chosen to minimize the 
contributions of decays from quarkonia and low mass mesons. In this 
step, we exclude the region with $p_{T}>2$ \gevc from the \ulmumu 
spectra from the fit, to avoid the uncertainty of the shape of Drell-Yan 
contribution in this region due to its sensitivity to $k_{T}$. We then 
repeat this fitting procedure using the fitted $\kappa_{c\bar{c}}$ value 
obtained in step (iii), and iterate until stable fit results are 
obtained. Although the fit results in step (i) is not very sensitive to 
the initial starting value of $\tilde{\kappa}_{c\bar{c}}$, the iterative 
procedure ensures consistency and robustness of the final fit results.


\subsubsection*{Fit function}
We use the log-likelihood fit which is applicable to bins having 
few (or zero) entries. For fitting the \lsmumu spectra in step (i), 
we first divide the data and simulations 
into mass, \pt and $z_{\rm vtx}$ bins. 
The parameters $\kappa_{b\bar{b}}$ and $\kappa_{h}$ are then 
varied to minimize the negative log-likelihood defined by:

\begin{equation}
\label{Eq:logll}
\begin{split}
ln \mathcal{L}(\kappa_{b\bar{b}}, \kappa_{h})&= \sum_{i}y_{i}ln 
C(i;\kappa_{b\bar{b}}, \kappa_{h})- \sum_{i}C(i;\kappa_{b\bar{b}}, 
\kappa_{h}),\\ C(i;\kappa_{b\bar{b}}, \kappa_{h})&= 
\kappa_{b\bar{b}}N_{b\bar{b}}(i)+\kappa_{h}N_{hbg*}(i;\tilde{\kappa}_{c\bar{c}},\kappa_{b\bar{b}}),
\end{split}
\end{equation}

\noindent where $y_{i}$ is the number of counts in the 
$i$\textsuperscript{th} mass-\pt-$z_{\rm vtx}$ bin and 
$C(i;\kappa_{b\bar{b}}, \kappa_{h})$ is the number of expected counts in 
the $i$\textsuperscript{th} mass-\pt-$z_{\rm vtx}$ bin from all cocktail 
components. $N_{b\bar{b}}(i)$ is the number of \mumu pairs from \bb in 
the $i^{th}$ bin per generated \bb event, 
$N_{hbg*}(i;\tilde{\kappa}_{c\bar{c}},\kappa_{b\bar{b}})$ is the sum of 
the combinatorial and correlated hadronic pairs per MB event, with fixed 
$\tilde{\kappa}_{c\bar{c}}$.

Similarly the log-likelihood function for step (ii) is defined as:

\begin{equation}
\label{Eq:logll2}
\begin{split}
ln \mathcal{L}(\kappa_{\rm DY}) = \sum_{i}y_{i}ln C(i;\kappa_{\rm DY})- \sum_{i}C(i;\kappa_{\rm DY}),\\
C(i;\kappa_{\rm DY}) = \kappa_{\rm DY}N_{\rm DY}(i) +  \tilde{\kappa}_{b\bar{b}}N_{b\bar{b}}(i) + \tilde{\kappa}_{c\bar{c}}N_{c\bar{c}}(i) \\
+\tilde{\kappa}_{h}N_{hbg*}(i;\tilde{\kappa}_{c\bar{c}},\tilde{\kappa}_{b\bar{b}}))
+N_{h\rightarrow\mu\mu(X)}(i),\\
\end{split}
\end{equation}

\noindent where $y_{i}$ is the number of counts in the 
$i$\textsuperscript{th} mass-\pt bin, $C(i;\kappa_{\rm DY})$ is the number 
of expected counts in the $i$\textsuperscript{th} mass-\pt bin from all 
cocktail components. The definitions for $N_{b\bar{b}}(i)$ is the same 
as in Eq.~\ref{Eq:logll}, while 
$N_{hbg*}(i;\tilde{\kappa}_{c\bar{c}},\tilde{\kappa}_{b\bar{b}})$ is the 
sum of the combinatorial and correlated hadronic pairs per MB event, 
with fixed $\tilde{\kappa}_{c\bar{c}}$ and fixed 
$\tilde{\kappa}_{b\bar{b}}$. $N_{c\bar{c}}(i)$ and $N_{\rm DY}(i)$ are the 
number of \mumu pairs from \cc and Drell-Yan pairs in the $i^{th}$ bin 
per generated \cc and Drell-Yan event respectively. 
$N_{h\rightarrow\mu\mu(X)}(i)$ is the number of \mumu pairs from hadron decays which 
is constrained from previous measurements.

Finally, the log-likelihood function for step (iii) is defined as:

\begin{equation}
\begin{split}
ln \mathcal{L}(\kappa_{c\bar{c}}) = \sum_{i}y_{i}ln C(i;\kappa_{c\bar{c}})- \sum_{i}C(i;\kappa_{c\bar{c}}),\\
C(i;\kappa_{c\bar{c}}) = \kappa_{c\bar{c}}N_{c\bar{c}}(i)+\tilde{\kappa}_{\rm DY}N_{\rm DY}(i) +  \tilde{\kappa}_{b\bar{b}}N_{b\bar{b}}(i)\\
+\tilde{\kappa}_{h}N_{hbg*}(i;\tilde{\kappa}_{c\bar{c}},\tilde{\kappa}_{b\bar{b}}))
+N_{h\rightarrow\mu\mu(X)}(i),\\
\end{split}
\end{equation}

\noindent where $y_{i}$ is the number of counts in the 
$i$\textsuperscript{th} mass-\pt bin, $C(i;\kappa_{c\bar{c}})$ is the 
number of expected counts in the $i$\textsuperscript{th} mass-\pt bin 
from all cocktail components. The definitions for $N_{c\bar{c}}(i)$, 
$N_{b\bar{b}}(i)$, $N_{\rm DY}(i)$, 
$N_{hbg*}(i;\tilde{\kappa}_{c\bar{c}},\tilde{\kappa}_{b\bar{b}})$, and 
$N_{h\rightarrow\mu\mu(X)}(i)$ are the same as in 
equations~(\ref{Eq:logll}) and (\ref{Eq:logll2}).

\subsubsection*{Fit results}

The three step fitting procedure is iterated until we obtain stable 
values of $\kappa_{c\bar{c}}$, $\kappa_{b\bar{b}}$, $\kappa_{\rm DY}$, 
and $\kappa_{h}$. The fitting procedure is done separately for the two 
arms. Because the contribution of charm to the like-sign spectrum is 
very small, the fit converges after two to three iterations. The fit 
results for the two arms are consistent with each other.

\begin{figure*}
\includegraphics[width=0.99\linewidth]{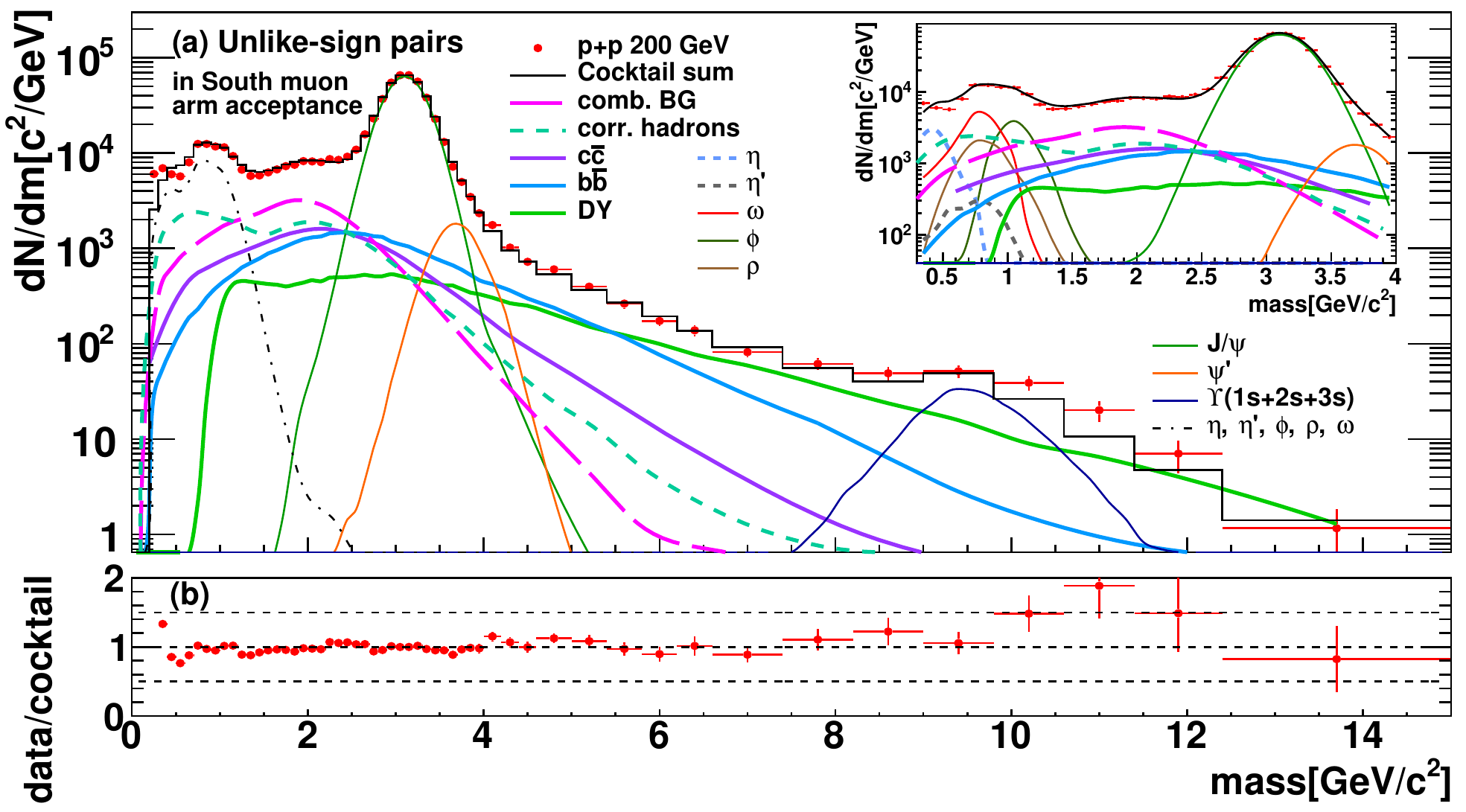}
\includegraphics[width=0.99\linewidth]{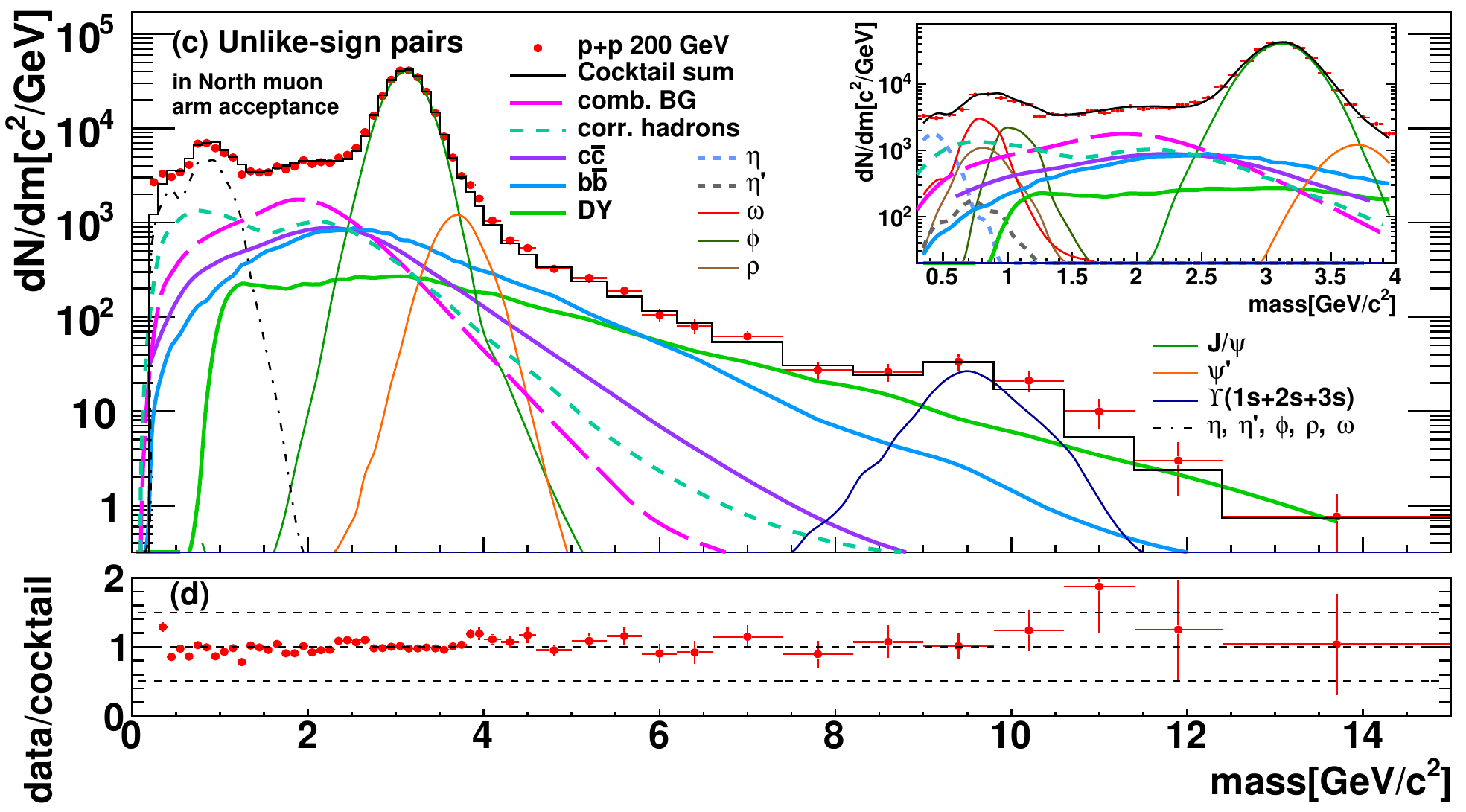}
\caption{\label{Fig:pp_unlike_mass} Inclusive \ulmumu pair mass 
distributions from \pp collisions at $\sqrt{s}=200$~GeV over the mass 
range from 0 to 15 \gevcc. The inset shows the mass region below 4 
\gevcc with more detail. Results are shown separately for the (a) south 
and (c) north muon arms. The data are compared to the cocktail of 
expected sources. Contributions from \cc and \bb are generated using 
\powheg. Panels (b) and (d) show the ratio of the data divided by the 
known sources. }
\end{figure*}

\begin{figure*}
\includegraphics[width=0.48\linewidth]{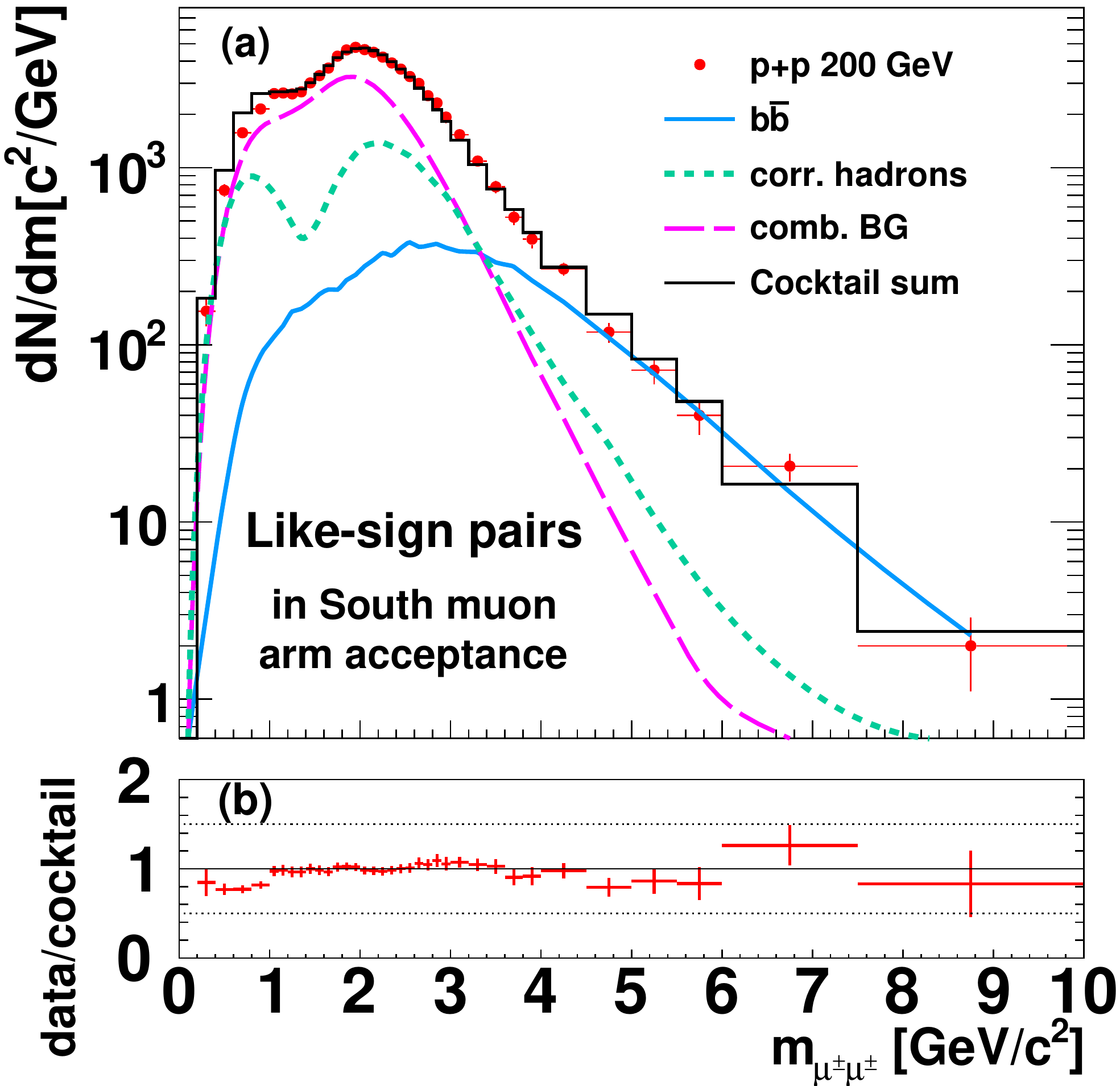}
\includegraphics[width=0.48\linewidth]{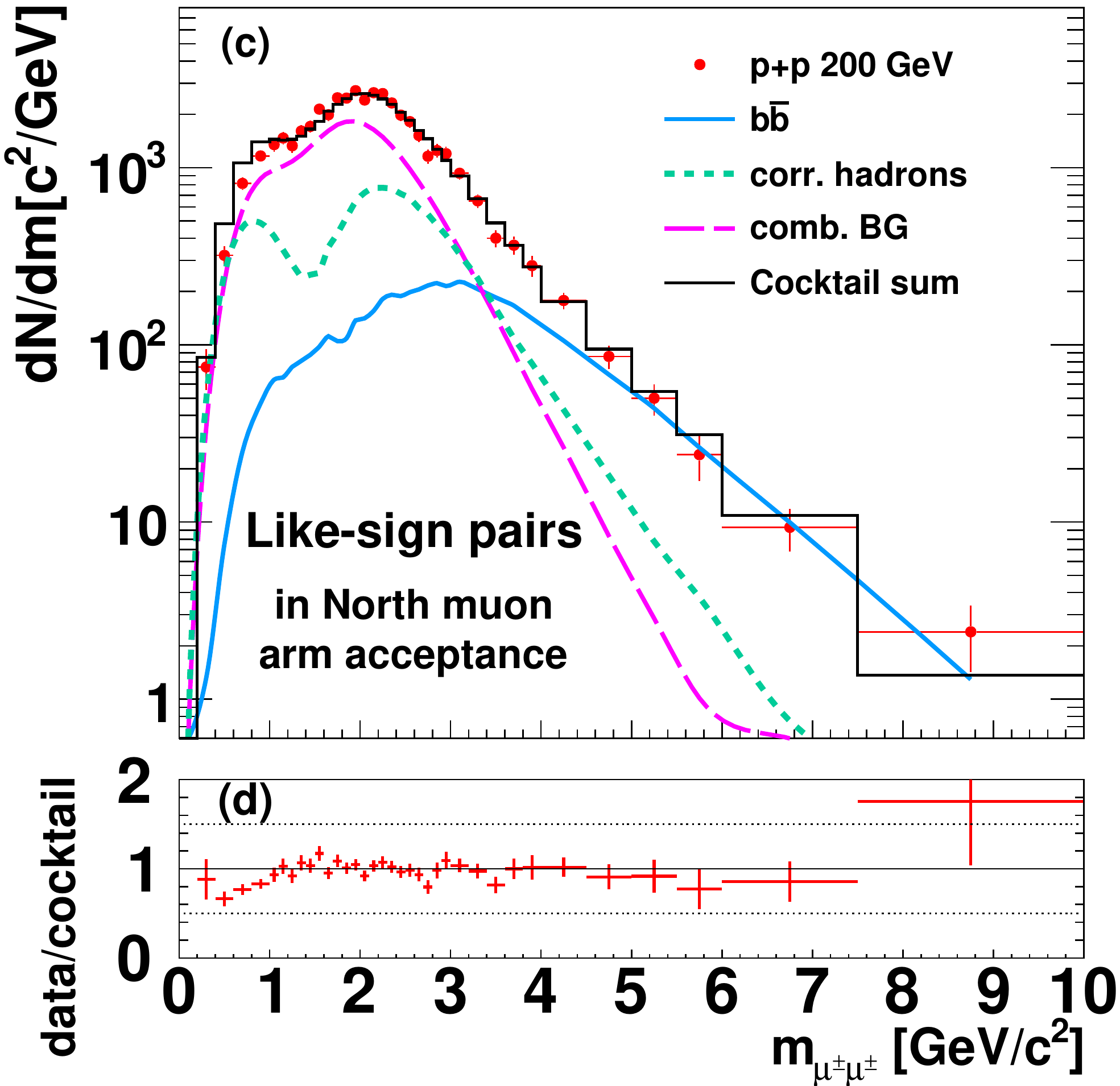}
\caption{\label{Fig:pp_like_mass} Inclusive like-sign \mumu pair yield 
from \pp collisions as a function of mass for the (a) south and (c) 
north muon arms and (b),(d) the ratio of data to expected sources. 
Contributions from \bb are generated using \powheg. }
\end{figure*}

\begin{figure*}
\includegraphics[width=0.48\linewidth]{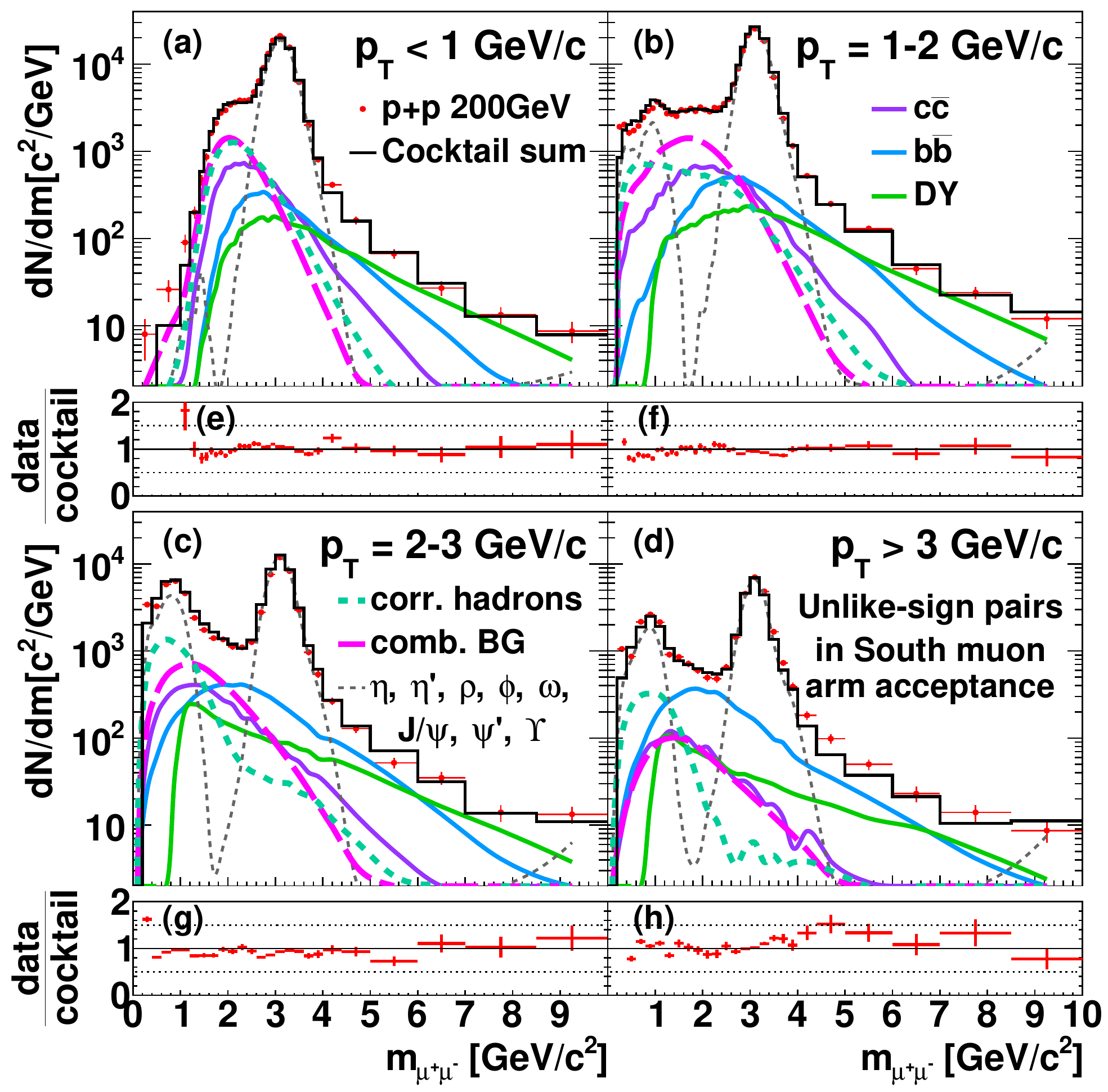}
\includegraphics[width=0.48\linewidth]{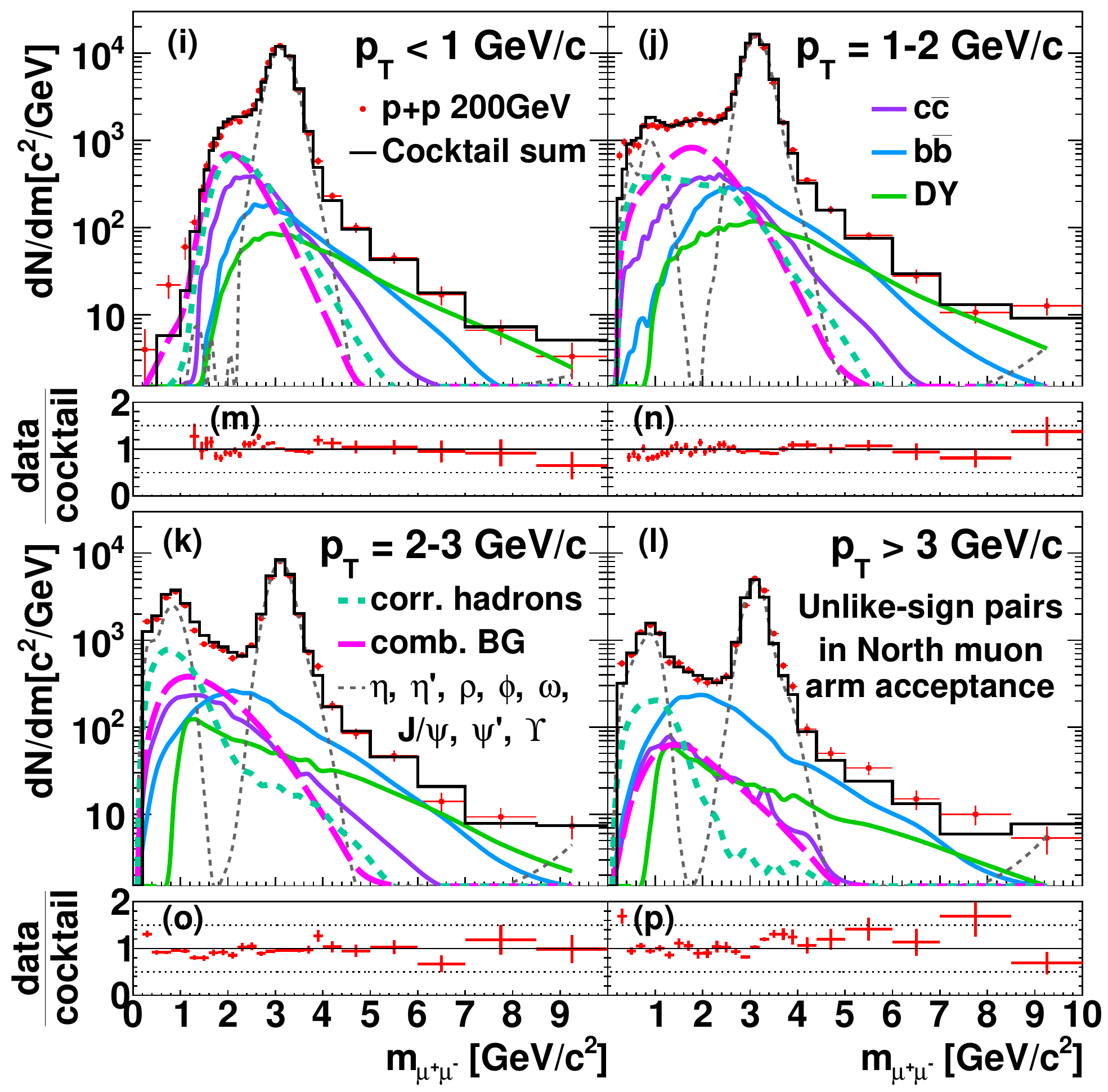}
\caption{\label{Fig:pp_unlike_masspt} Inclusive unlike-sign \mumu pair 
yield from \pp collisions at $\sqrt{s}=200$~GeV as a function of mass in 
different \pt slices for the (a,b,c,d) south and (i,j,k,l) north muon 
arms. The ratio of data to expected sources are shown in panels 
(e,f,g,h) for the south arm and (m,n,o,p) for the north arm.
}
\end{figure*}

\begin{figure*}
\includegraphics[width=0.48\linewidth]{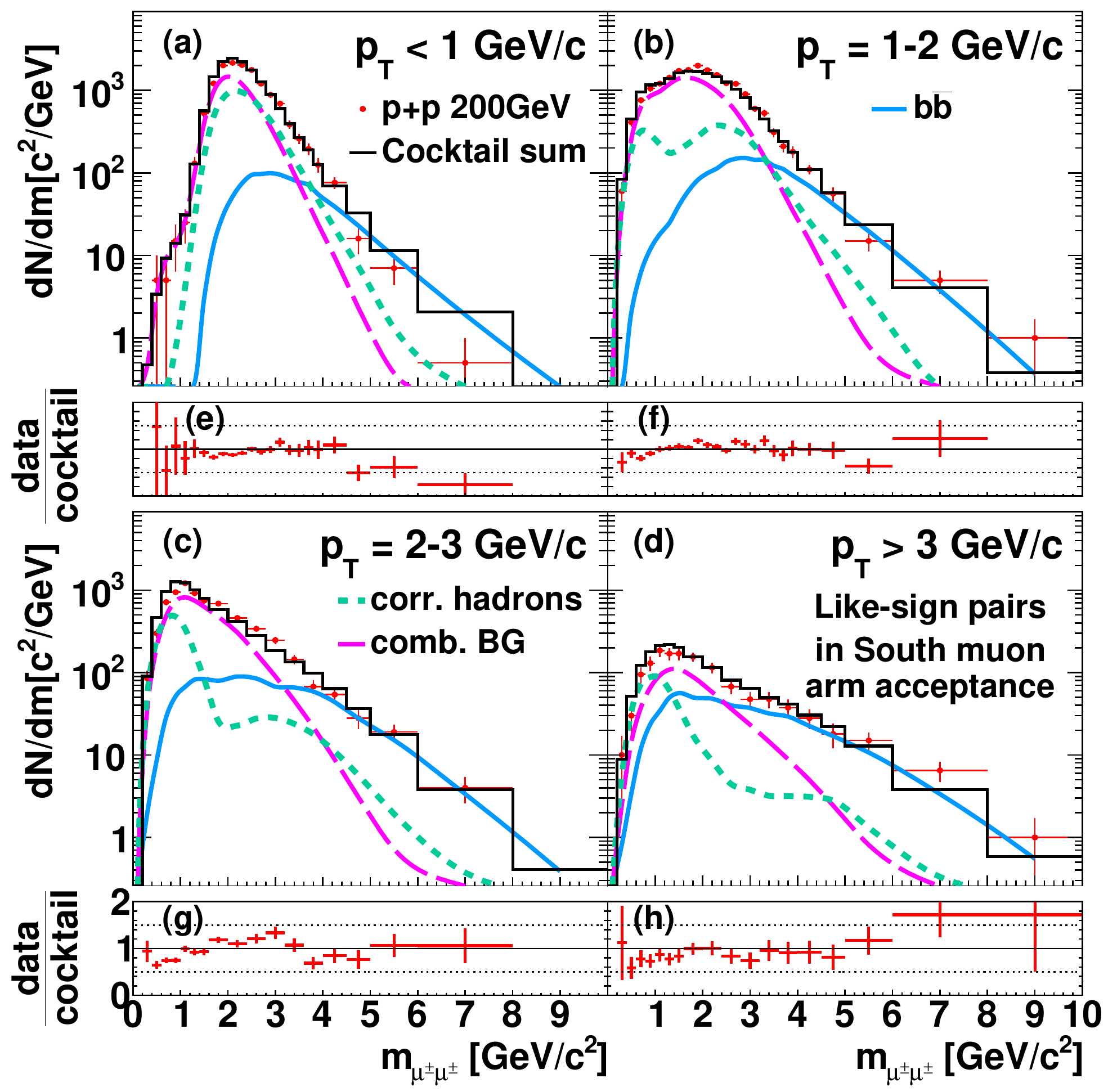}
\includegraphics[width=0.48\linewidth]{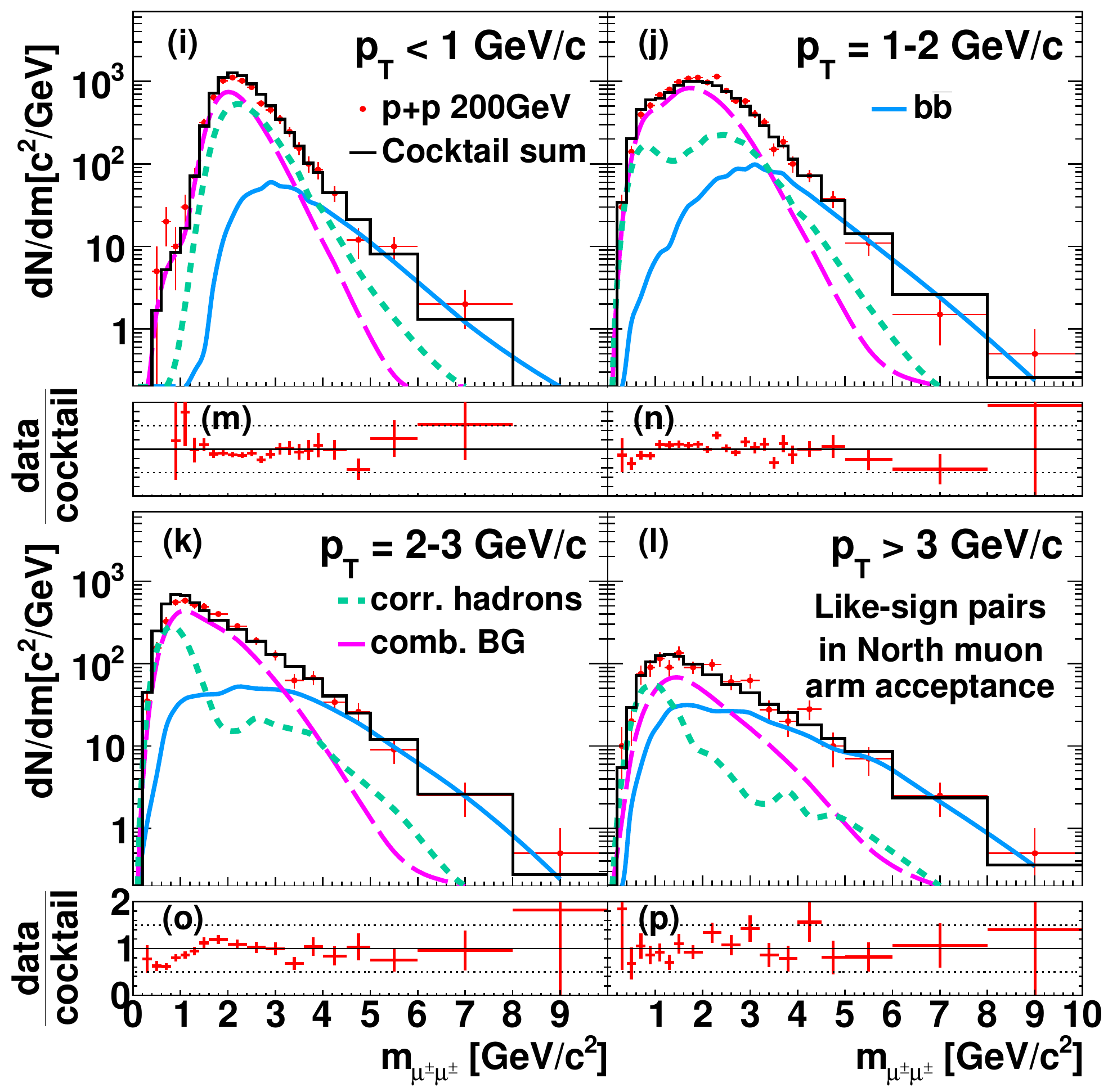}
\caption{\label{Fig:pp_like_masspt} Inclusive like-sign \mumu pair yield 
from \pp collisions at $\sqrt{s}=200$~GeV as a function of mass for the 
(a,b,c,d) south and north (i,j,k,l) muon arms. The ratio of data to 
expected sources are shown in panels (e,f,g,h) for the south arm and 
(m,n,o,p) for the north arm.}
\end{figure*}

In this section, example fit results using the following simulation 
configurations are shown: \cc and \bb generated using \powheg, Drell-Yan 
generated using \pythia with intrinsic $k_{T}=1.1$ \gevc. Variations of 
simulation settings are considered in the evaluation of systematic 
uncertainties, which will be discussed in Sec.~\ref{Sec:syserror}. Mass 
spectra of \ulmumu and \lsmumu pairs integrated over $p_{T}$ are shown 
in Figs.~\ref{Fig:pp_unlike_mass} and~\ref{Fig:pp_like_mass} 
respectively. Figs.~\ref{Fig:pp_unlike_masspt} 
and~\ref{Fig:pp_like_masspt}, give a more detailed view of \ulmumu and 
\lsmumu mass spectra in $p_{T}$ slices. The data distributions are well 
described by the cocktail simulation in both mass and \pt except for a 
small kinematic region at $m<1~\gevcc$ which is unimportant for the 
current analysis.

\subsection{Signal extraction}

Different cocktail components contribute with different strength to the 
muon pair continuum in different mass regions for \ulmumu and \lsmumu 
charge combinations. To obtain differential measurements we identify 
mass regions for the \cc, \bb, and Drell-Yan signal, where the ratio of 
the signal to all other \mumu pairs is the most favorable for that 
signal. These regions are referred to in the following as charm, bottom, 
or Drell-Yan mass region, respectively. The mass regions are:

\begin{itemize}
\item Charm: $1.5<m_{\mu^{+}\mu^{-}}<2.5$ \gevcc 
\item Bottom: $3.5<m_{\mu^{\pm}\mu^{\pm}}<10.0$ \gevcc 
\item Drell-Yan: 

\ \ \ \ \ \ \ \ \ \ \ \ \ $4.8<m_{\mu^{+}\mu^{-}}<8.2$ \gevcc 

\ \ \ \ \ \ \ \ \ \ \ \ \ and $11.2<m_{\mu^{+}\mu^{-}}<15.0$ \gevcc 

\end{itemize}

For each region we extract differential distributions by 
subtracting all other \mumu pair sources. 

\subsubsection{Azimuthal correlations and pair $p_{T}$ of \ulmumu from \cc}

\begin{figure*}[]
\includegraphics[width=0.48\linewidth]{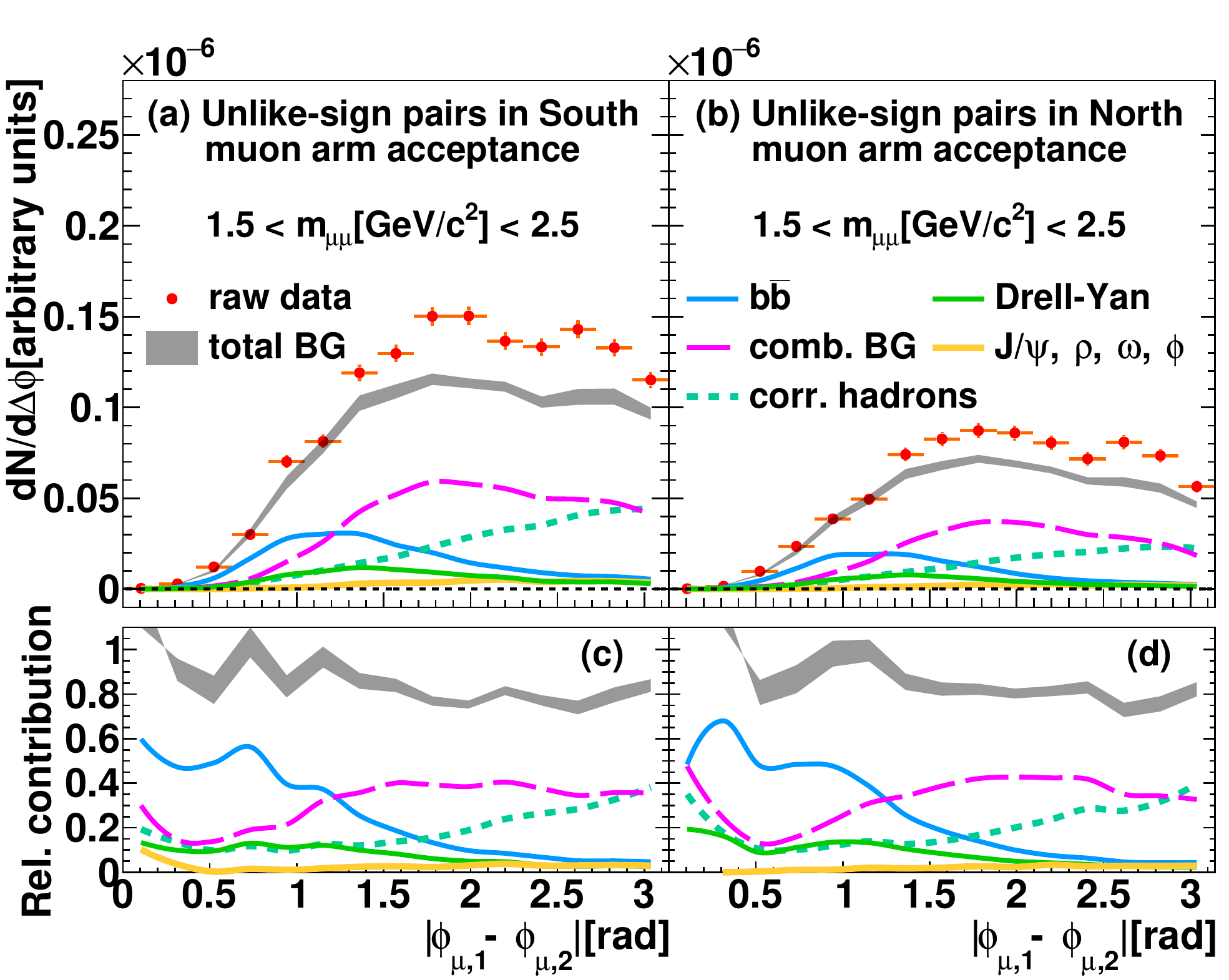}
\includegraphics[width=0.48\linewidth]{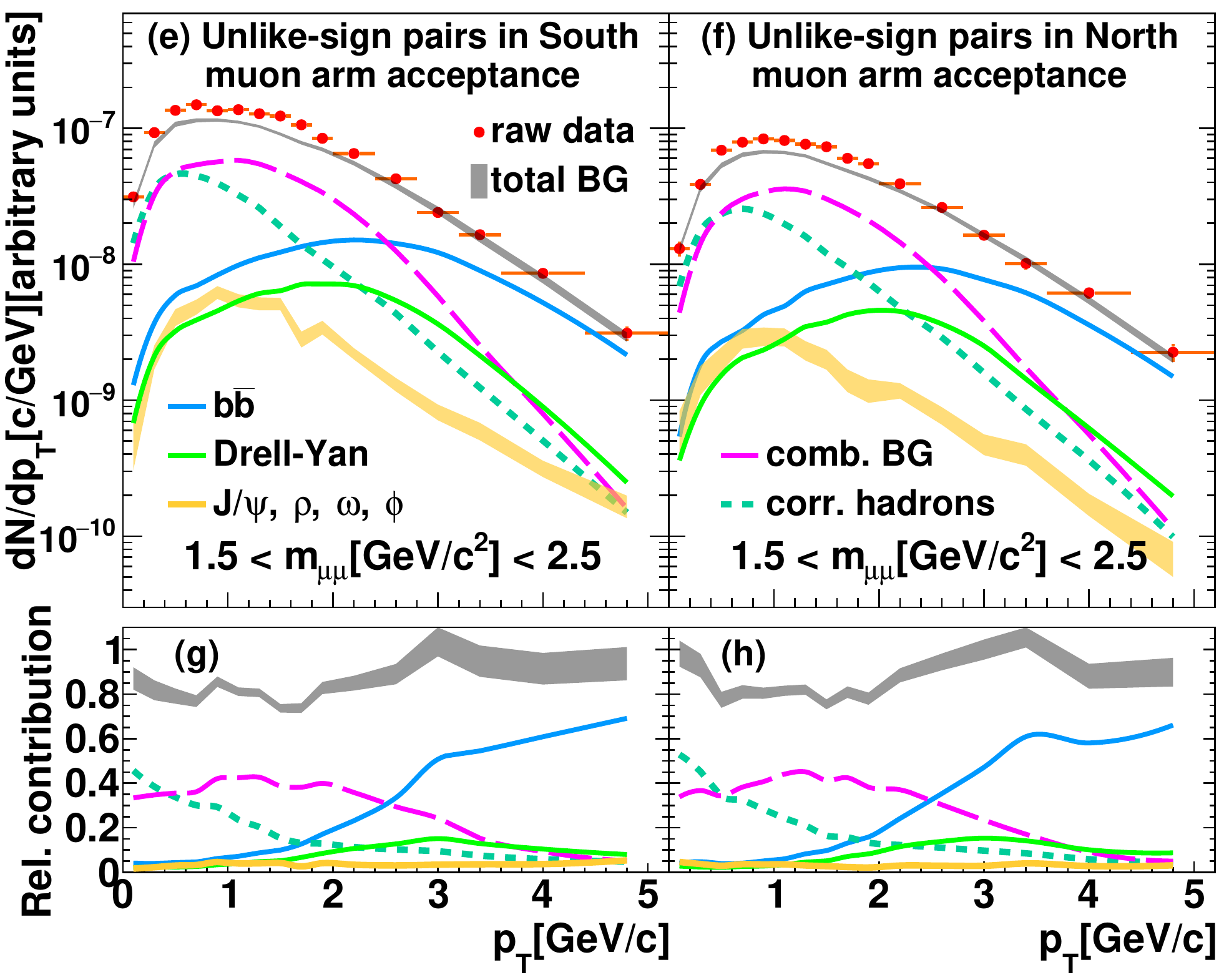}
\caption{ \label{Fig:ccbackground} The \mumu pair data in the charm mass 
region as a function of (a,b) $\Delta\phi$ or (c,d) pair $p_{T}$ are 
shown. Contributions from all known sources other than charm decays are 
also shown. Panels (c,d,g,h) give the ratio of different components to 
the total yield. Gray bands indicate the systematic uncertainty on the 
sum of all contributions.}
\end{figure*}

Figure~\ref{Fig:ccbackground} shows the number of pairs per event as a 
function of their azimuthal opening angle, $\Delta\phi$, or their pair 
transverse momentum $p_{T}$ in the charm mass region. The data are 
compared to all other sources that contribute in this region. For each 
$\Delta\phi$ or $p_{T}$ bin, the number of pairs from charm decays 
($N^{+-}_{c\bar{c}}$) is obtained as:

\begin{equation}
\begin{split}
N^{+-}_{c\bar{c}} = N_{\rm incl}^{+-}-N_{b\bar{b}}^{+-}-N_{\rm DY}^{+-}-N_{\rho,\phi,\omega}^{+-}-N_{J/\psi}^{+-}\\
-N_{\rm cor}^{+-}-N_{\rm comb}^{+-},
\label{Eq:ccbg}
\end{split}
\end{equation}

\noindent where $N_{\rm incl}^{+-}$ is the number of pairs passing all 
single and pair cuts in Tables~\ref{Tab:singlecuts} 
and~\ref{Tab:paircuts}, $N_{b\bar{b}}^{+-}$ is the estimated number of 
pairs from bottom decays, $N_{\rm DY}^{+-}$ is the estimated number of pairs 
from Drell-Yan, $N_{\rho,\phi,\omega}^{+-}$ is the estimated number of 
pairs from low mass vector meson decays, $N_{J/\psi}^{+-}$ is the 
estimated number of pairs from $J/\psi$ decays, $N_{\rm cor}^{+-}$ is 
the estimated number of pairs from correlated hadrons, and 
$N_{\rm comb}^{+-}$ is the estimated number of combinatorial pairs.

\subsubsection{Azimuthal correlations and pair \pt of \lsmumu from \bb}

\begin{figure*}
\includegraphics[width=0.48\linewidth]{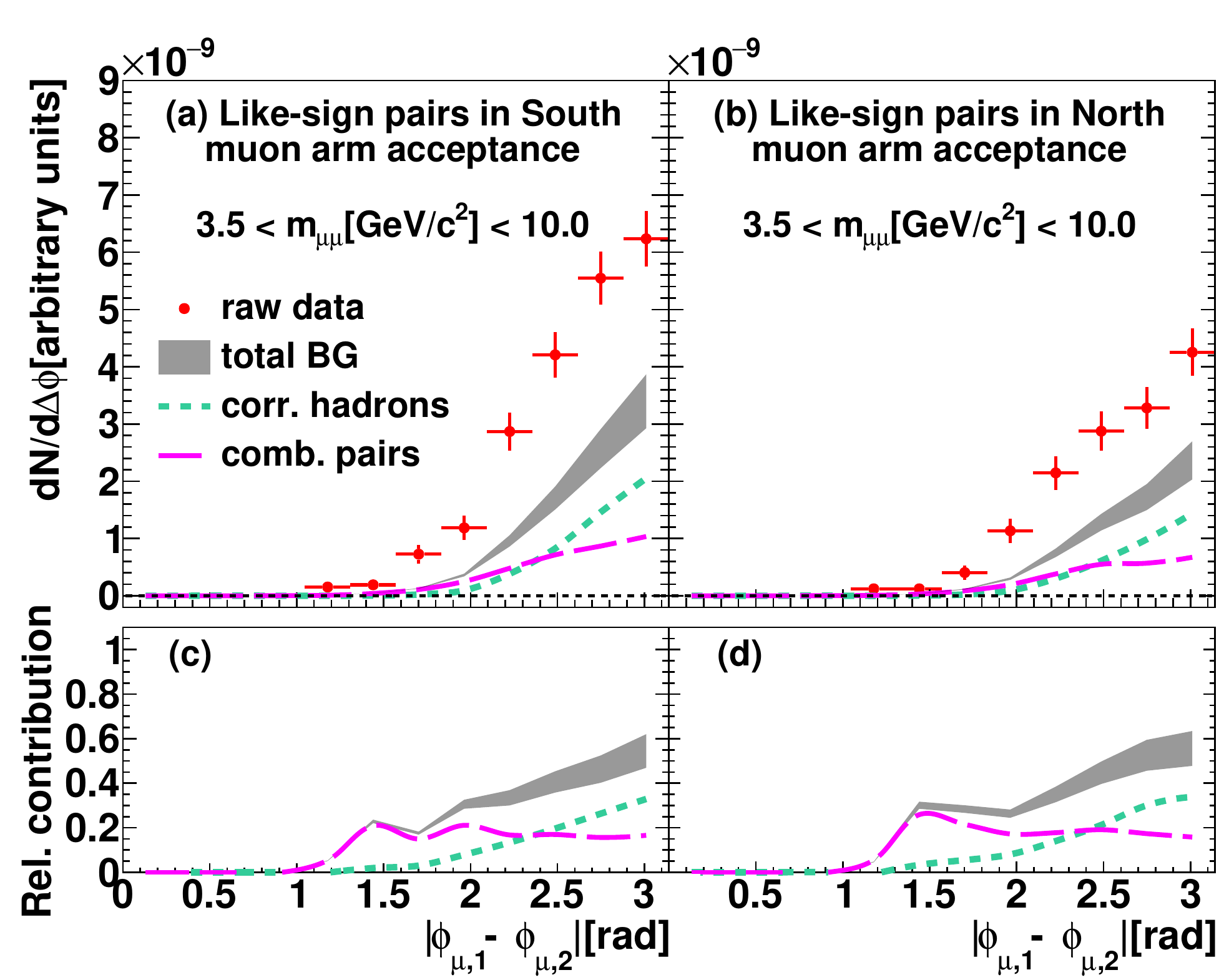}
\includegraphics[width=0.48\linewidth]{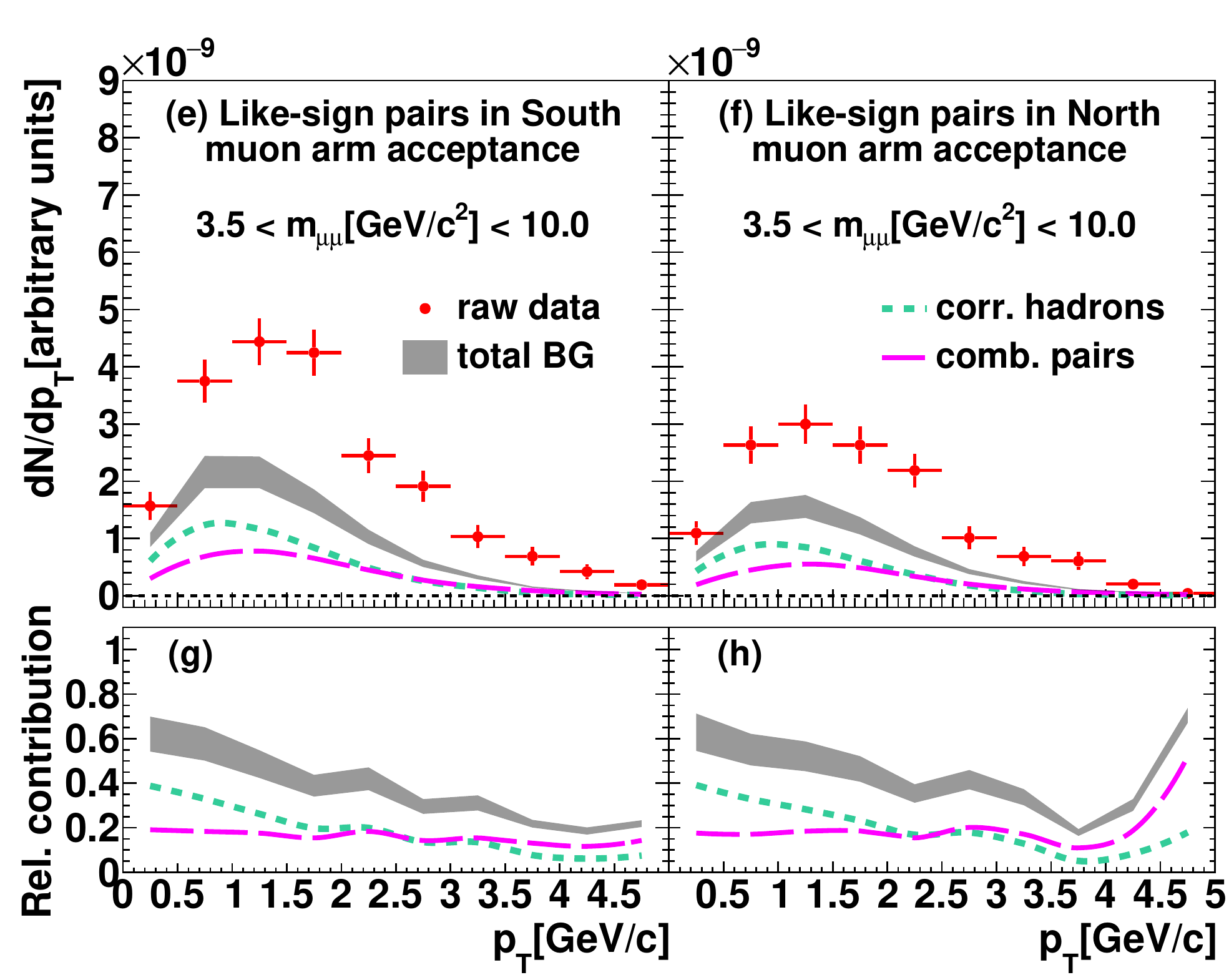}
\caption{ \label{Fig:bbbackground} The like-sign \mumu pair data in the 
bottom mass region as a function of (a,b) $\Delta\phi$ or (c,d) pair 
$p_{T}$ are shown. Contributions from all known sources other than 
bottom decays are also shown. Panels (c,d,g,h) give the ratio of 
different components to the total yield. Gray bands indicate the 
systematic uncertainty on the sum of all contributions.}
\end{figure*}

The azimuthal opening angle distribution and pair $p_{T}$ distribution 
for \lsmumu pairs from the bottom mass region is shown in 
Fig.~\ref{Fig:bbbackground}. Besides the \bb contribution there are also 
contributions from correlated and combinatorial hadronic pairs. The 
number of pairs from bottom decays ($N_{b\bar{b}}^{\pm\pm}$) is obtained 
according to the following relation:

\begin{equation}
N_{b\bar{b}}^{\pm\pm} = N_{\rm incl}^{\pm\pm}-N_{\rm cor}^{\pm\pm}-N_{\rm comb}^{\pm\pm},
\label{Eq:bbbg}
\end{equation}

\noindent where $N_{\rm incl}^{\pm\pm}$ is the number of pairs passing all 
single and pair cuts in Tables~\ref{Tab:singlecuts} 
and~\ref{Tab:paircuts}, $N_{\rm cor}^{\pm\pm}$ is the estimated number 
of pairs from correlated hadrons, and $N_{\rm comb}^{\pm\pm}$ is the 
estimated number of combinatorial pairs. We subtract the background as a 
function of \dphi or pair $p_{T}$.

\subsubsection{Pair mass and $p_{T}$ distribution of \ulmumu pairs from 
Drell-Yan}

\begin{figure*}
\includegraphics[width=0.48\linewidth]{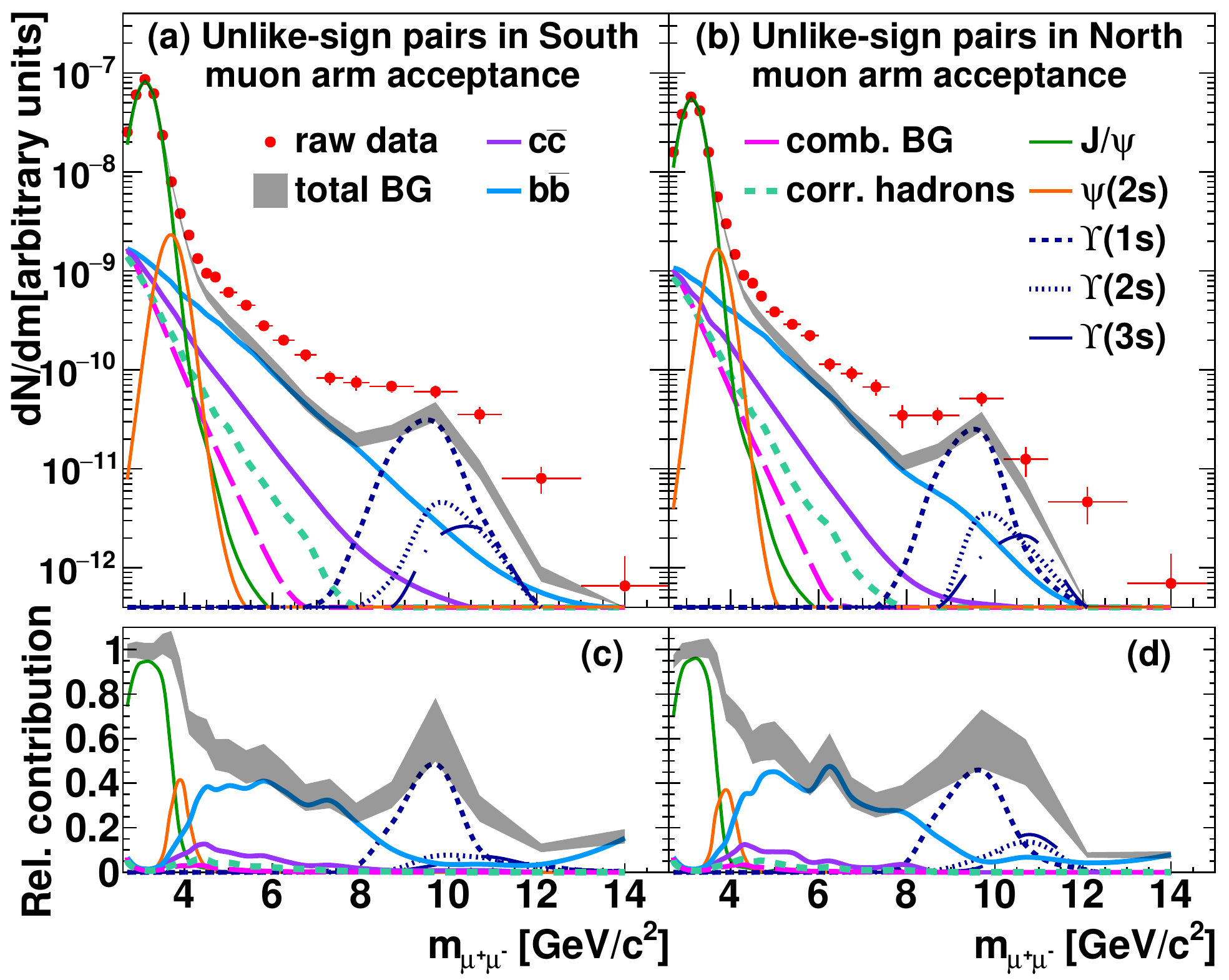}
\includegraphics[width=0.48\linewidth]{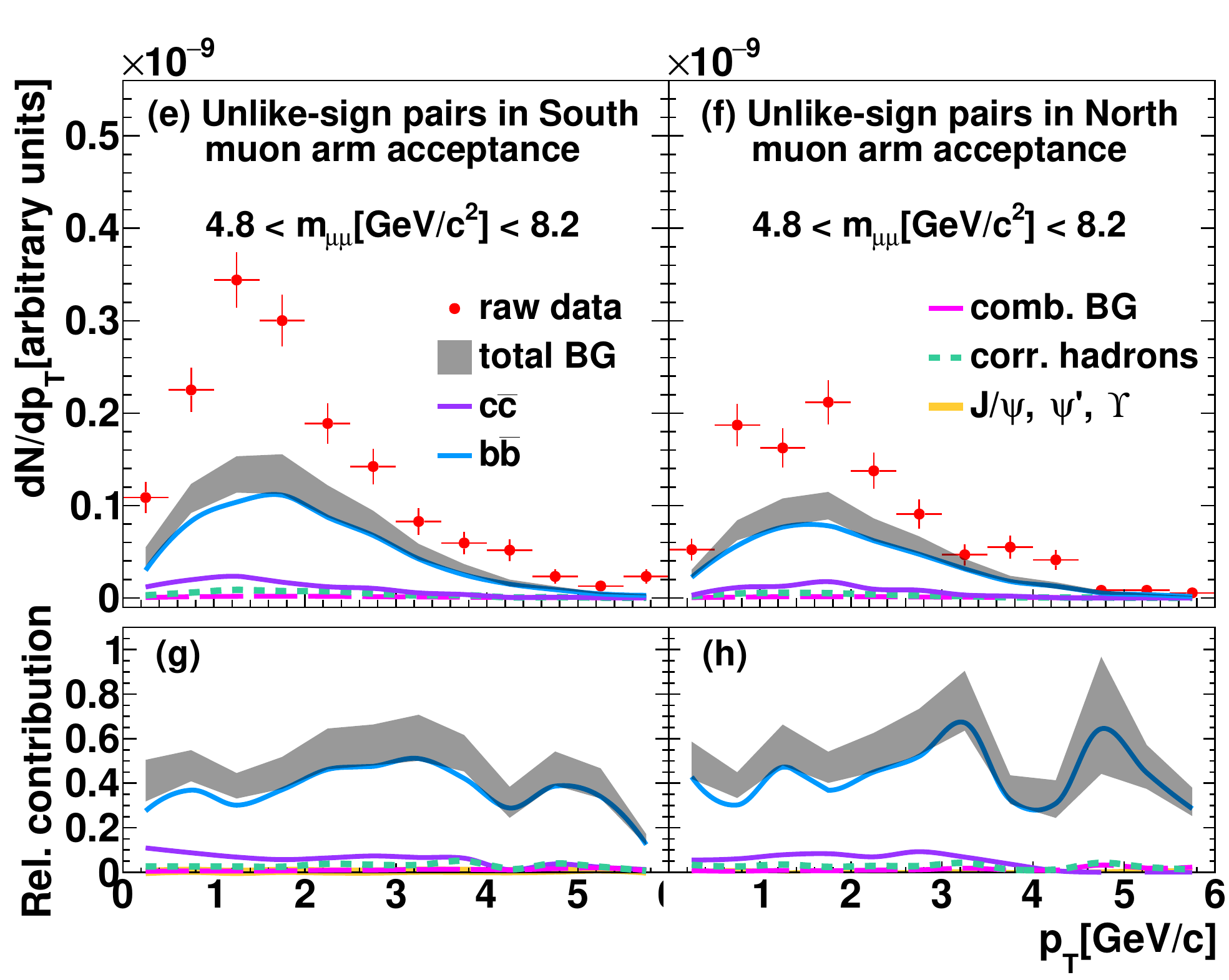}
\caption{ \label{Fig:dybackground} 
The unlike-sign \mumu pair data used to determine the Drell-Yan 
contribution as a function of (a,b) mass or (c,d) pair $p_{T}$ are 
shown. Contributions from all known sources other than the Drell-Yan 
process are also shown. Panels (c,d,g,h) give the ratio of different 
components to the total yield. Gray bands indicate the systematic 
uncertainty on the sum of all contributions.
}
\end{figure*}

The Drell-Yan yield is extracted in a mass region that excludes the 
$\Upsilon$ mass region. The primary sources of background pairs are from 
bottom and charm decays. The number of pairs from Drell-Yan 
($N^{+-}_{\rm DY}$) is obtained as:

\begin{equation}
\begin{split}
N^{+-}_{\rm DY} = N_{\rm incl}^{+-}-N_{b\bar{b}}^{+-}-N_{c\bar{c}}^{+-}-N_{J/\psi, \psi'}^{+-}\\
-N_{\Upsilon}^{+-}-N_{\rm cor}^{+-}-N_{\rm comb}^{+-},
\label{Eq:dybg}
\end{split}
\end{equation}

\noindent where $N_{\rm incl}^{+-}$ is the number of pairs passing all 
single and pair cuts in Tables~\ref{Tab:singlecuts} 
and~\ref{Tab:paircuts}, $N_{J/\psi, \psi'}^{+-}$ is the estimated number 
of pairs from $J/\psi$ and $\psi'$ decays, $N_{\Upsilon}^{+-}$ is the 
estimated number of pairs from the $\Upsilon$ family, $N_{\rm cor}^{+-}$ 
is the estimated number of pairs from correlated hadrons, and 
$N_{\rm comb}^{+-}$ is the estimated number of combinatorial pairs. The 
background contributions as a function of pair mass or $p_{T}$ are shown 
in Fig.~\ref{Fig:dybackground}.

\subsection{Acceptance and Efficiency Corrections}

To obtain a physical yield or cross section $\Gamma$, the raw yield 
$\Gamma_{raw}$ determined in the previous section, must be corrected for 
detector effects in multiple steps.

\begin{equation}
\begin{split}
\Gamma = \Gamma_{raw}
\cdot \frac{\sigma_{\rm BBC}}{N_{\rm BBC}\cdot\epsilon_{\rm bias}}
\cdot \frac{\alpha}{A \times \epsilon_{\rm rec}},
\end{split} \label{Eq:corraw}
\end{equation}

\noindent where $\Gamma$ and $\Gamma_{raw}$ can represent differential 
or integrated quantities. The raw yield is converted to yield per event 
by dividing by $N_{\rm BBC}$, the number of sampled MB events. The \pp 
cross section sampled by the BBC is $\sigma_{\rm BBC}=23.0\pm2.2$ mb at 
$\sqrt{s}=200$ GeV~\cite{Drees:2003zza}, it relates to the inelastic \pp 
cross section $\sigma_{pp}$:

\begin{equation}
\sigma_{pp} = \frac{\sigma_{\rm BBC}}{\epsilon_{\rm BBC}}, 
\end{equation}

\noindent where $\epsilon_{\rm BBC}=0.55\pm0.06$ is the fraction of 
inelastic \pp collisions recorded by the BBC. The BBC trigger bias for 
hard scattering events is 
${\epsilon_{\rm bias}}=0.79\pm0.02$~\cite{Adler:2006wg}.

The other factors in Eq.~\ref{Eq:corraw} are $\epsilon_{\rm rec}$, the 
pair reconstruction efficiency that accounts for efficiency losses due 
to track reconstruction, single track and pair cuts, the software trigger 
efficiency, and detector inefficiency; $A$, the detector acceptance; and 
$\alpha$, an additional normalization constant that accounts for effects 
not included in the Monte-Carlo simulation, which will be described 
in detail in Sec.~\ref{Sec:globalnorm}.

The acceptance $A$ has different meanings for the different measurements 
presented here. The azimuthal opening angle distributions for \mumu 
pairs from \cc and \bb are corrected up to the ideal muon arm 
acceptance, which requires that each muon has a momentum $p>3$ \gevc and 
falls in the pseudorapidity range $1.2<|\eta|<2.2$. For the \mumu pairs 
from Drell-Yan production the correction is for the muon pair to be in 
the rapidity range $1.2<|y^{\mu\mu}|<2.2$. To determine the \bb 
cross section we correct up to 4$\pi$, the full phase space as shown in 
Tab.~\ref{Tab:bbextrapolation}. In general, $A\times\epsilon_{\rm rec}$ 
is calculated using the default simulation framework. Input from the 
appropriate physics event generator is run through the simulation; the 
ratio of the reconstructed $\Gamma_{raw}^{MC}$ yield over the input 
yield $\Gamma^{MC}$ gives $A\times\epsilon_{\rm rec}$.

Finally, the factor $\alpha$ accounts for the combined effect of double 
interactions, $\alpha_{double}$; modifications of the reconstruction 
efficiency due to detector occupancy, $\alpha_{\rm occ}$; the change of 
the trigger livetime with luminosity, $\alpha_{live}$; and additional 
variations with luminosity, $\alpha_{lum}$; which are not included in 
the Monte-Carlo simulations. We determine $\alpha$ by comparing the 
measured $J/\psi$ cross section \cite{Adare:2011vq} with the result 
using Eq.~\ref{Eq:corraw} with $\alpha=1$. 
$\alpha=1.38$ for south and north muon arm, respectively. We obtain 
$\alpha=1.30\pm0.16$ and $\alpha=1.38\pm0.17$ for south and north muon 
arm, respectively. Our values are consistent with the product of the 
individual factors $\alpha_{double}\times \alpha_{\rm occ}\times 
\alpha_{live} \times \alpha_{lum}$ within the systematic uncertainties, 
where the individual factors are determined with data driven methods 
(see Sec.~\ref{Sec:globalnorm}).

\subsubsection{Azimuthal correlations and pair \pt of \mumu from \cc and \bb }

The fully corrected per event pair yield is given by Eq.~\ref{Eq:hfeff}.

\begin{equation}
\begin{split}
\frac{dN}{dX} = 
\frac{N_{\rm HF}}{\Delta X}
\cdot\frac{\epsilon_{\rm BBC}}{N_{\rm BBC}\cdot\epsilon_{\rm bias}}
\cdot\frac{\alpha}{\epsilon_{\rm rec}(X)},
\end{split} \label{Eq:hfeff}
\end{equation}

\noindent where $X$ is either $\Delta\phi$ or pair $p_{T}$, $\Delta X$ is 
the corresponding bin width, and $N_{\rm HF}$ refers to 
$N_{c\bar{c}}^{+-}$ or $N_{b\bar{b}}^{\pm\pm}$ given by 
Eq.~\ref{Eq:ccbg} and Eq.~\ref{Eq:bbbg}, respectively.  All other 
factors are the same as in Eq.~\ref{Eq:corraw}.

The pair reconstruction efficiency $\epsilon_{\rm rec}(X)$ is determined 
using input distributions from \pythia and \powheg and is computed by 
taking the ratio of reconstructed and generated yields with both 
generated tracks satisfying the condition of the ideal muon arm 
acceptance ($p>3$ \gevc and $1.2<|\eta|<2.2$). Here we correct the data 
up to the ideal muon arm acceptance. We do not correct up to \mumu pairs 
in $1.2<|y_{\mu\mu}|<2.2$ to avoid systematic effects from model 
dependent extrapolations. Systematic uncertainties for model dependent 
efficiency corrections are determined by comparing $\epsilon_{\rm 
rec}(X)$ using \pythia or \powheg as input distributions. This will be 
discussed in detail in Section~\ref{Sec:syserror}.

\subsubsection{Drell-Yan}

The differential cross section as a function of mass or $p_{T}$ is given 
by Eq.~\ref{Eq:dyeff} and Eq.~\ref{Eq:dypteff}.

\begin{equation}
\begin{split}
\frac{d^{2}\sigma}{dmdy} = 
\frac{N_{\rm DY}}{\Delta m\Delta y}
\cdot\frac{\sigma_{\rm BBC}}{N_{\rm BBC}\cdot \epsilon_{\rm bias}}
\cdot\frac{\alpha \cdot \beta(m,y)}{A\times \epsilon_{\rm rec}(m,y)}
,
\end{split} \label{Eq:dyeff}
\end{equation}

\begin{align} \label{Eq:dypteff}
\nonumber
\frac{1}{2\pi p_{T}}&\frac{d^{2}\sigma}{dydp_{T}} = \\
&\frac{N_{\rm DY}}{2\pi p_{T} \Delta y\Delta p_{T}}
\cdot\frac{\sigma_{\rm BBC}}{N_{\rm BBC}\cdot \epsilon_{\rm bias}}
\cdot\frac{\alpha \cdot \beta(y,p_{T})}{A\times \epsilon_{\rm rec}(y,p_{T})}
,
\end{align}

\noindent where $N_{\rm DY}$ is raw yield of pairs from Drell-Yan given 
by Eq.~\ref{Eq:dybg}. $\Delta m$, $\Delta p_{T}$, and $\Delta y$ are the 
bin widths in pair mass, pair $p_{T}$ and pair rapidity respectively. 
The factors $\beta(m,y)$ and $\beta(y,p_{T})$ correct the cross section 
averaged over the bin to the cross section at the bin center. These 
correction factors are estimated using \pythia simulations and lie 
between 0.97 and 1.03. All other factors are the same as in 
Eq.~\ref{Eq:corraw}.

The pair acceptance and efficiency $A\times \epsilon_{\rm rec}(m,y)$ and 
$A\times \epsilon_{\rm rec}(y,p_{T})$ are determined using input 
distributions generated using \pythia. It corrects the 
pair yield to one unit of rapidity at $1.2< |y^{\mu\mu}|<2.2$.

\subsubsection{Bottom cross section}

We also determine the \bb cross section from the measured \mumu pair 
yield from \bb. In the fitting procedure we determined the normalization 
$\kappa_{b\bar{b}}$, which was chosen such that it directly relates to 
the $\sigma_{b\bar{b}}$ cross section:

\begin{equation}
\label{Eq:bbnorm}
\sigma_{b\bar{b}}=
\frac{\alpha\cdot\sigma_{\rm BBC}}{N_{\rm BBC}\cdot \epsilon_{\rm bias}}\cdot\kappa_{b\bar{b}}.
\end{equation}

The acceptance and efficiency corrections, trigger efficiency, branching 
ratios, and oscillation parameters are all implicitly encapsulated in 
$\kappa_{b\bar{b}}$, because the templates for fitting already include 
all the aforementioned considerations.

We used two models \pythia and \powheg, to take into account a possible 
model dependence. The extrapolation from the limited phase space of our 
\mumu measurement to the entire kinematic region can be divided into 
four steps:

\begin{itemize}

\item Extrapolation from \lsmumu muon pairs with $m_{\mu\mu}>3$ \gevcc 
in the \textit{ideal} muon arm acceptance to all muon pairs 
($\mu^{\pm}\mu^{\pm}$ and $\mu^{+}\mu^{-}$) with $m_{\mu\mu}>3$ \gevcc 
in the \textit{ideal} muon arm acceptance.

\item Extrapolation to all muon pairs in the entire mass region 
($m_{\mumu}>0$ \gevcc) in the \textit{ideal} muon arm acceptance.

\item Extrapolation to all muon pairs with the pseudorapidity of each 
muon satisfying $1.2<|\eta_{\mu}|<2.2$.

\item Extrapolation to muon pairs in $4\pi$.

\end{itemize}

\begin{table}[h]
\caption{\label{Tab:bbextrapolation} 
Step by step reduction of phase space for \mumu pairs from \bb production; 
starting from all \mumu pairs produced to like-sign \mumu pairs 
with $m_{\mu\mu}>3$ \gevcc in the ideal muon arm acceptance. All numbers 
represent the number of \mumu pairs per generated \pythia or \powheg \bb 
event in the specified phase space. Each step is cumulative to the 
previous, i.e. each row includes one more restriction to the \mumu 
phase space. The factors in brackets quantify the
decrease of the number of pairs from the previous step.
}
\begin{ruledtabular} \begin{tabular}{cccc}
  &   \multicolumn{2}{c}{Event gen.}\\
condition & \pythia & \powheg\\
\hline
$4\pi$                  & $6.76\times10^{-2}~(15.4)$ & $6.73\times10^{-2}~(15.6)$\\
$1.2<|\eta_{\mu}|<2.2$  & $4.39\times10^{-3}~(10.7)$ & $4.32\times10^{-3}~(10.7)$\\
$p_{\mu}>3$ \gevc       & $4.11\times10^{-4}~(3.48)$ & $4.04\times10^{-4}~(3.39)$\\
$m_{\mu\mu}>3$ \gevcc   & $1.18\times10^{-4}~(3.19)$ & $1.19\times10^{-4}~(3.48)$\\
\lsmumu in PHENIX       & $3.71\times10^{-5}      $ & $3.42\times10^{-5}      $\\
\end{tabular} \end{ruledtabular} 
\end{table}

Table~\ref{Tab:bbextrapolation} quantifies each step. For clarity they 
are shown in reversed order. One can see that in each step, the 
difference between \pythia and \powheg is less than $8\%$, which is 
consistent with the observation from Ref.~\cite{Adare:2014iwg}, that the 
model dependence of the extrapolation is small because the \mumu(or 
$ee$) pair distributions from bottom are dominated by decay kinematics.

The differential cross section $d\sigma_{b\bar{b}}/dy_{b}|_{\langle 
y_{b} \rangle=\pm1.7}$ can be calculated as follows:

\begin{equation}
\frac{d\sigma_{b\bar{b}}}{dy_{b}}|_{\langle y_{b} \rangle=\pm1.7} = 
\frac{\alpha\cdot\sigma_{\rm BBC}}{N_{\rm BBC}\cdot \epsilon_{\rm bias}}
\cdot \frac{dN_{b}}{dy_{b}}|_{\langle y_{b}\rangle =\pm1.7} \cdot\kappa_{b\bar{b},\frac{N}{S}},
\end{equation}

\noindent where $dN_{b}/dy_{b}|_{\langle y_{b}\rangle =\pm1.7}$ is the 
rapidity density of $b$ quarks determined from the average of \pythia 
and \powheg, $\kappa_{b\bar{b},\frac{N}{S}}$ is the fitted normalization 
for bottom from the north (south) muon arm.

\section{Systematic uncertainties 
\label{Sec:syserror}}

We consider four types of sources of possible systematic uncertainties 
on the extraction of \mumu pairs from \cc, \bb, and Drell-Yan. These are 
uncertainties:
\begin{itemize}
\item on the shape of the template distributions,
\item on the normalization of template distributions,
\item on the acceptance and efficiency corrections,
\item and on the overall global normalization. 
\end{itemize}

The first three sources of systematic uncertainties are point-to-point 
correlated, but allow for a gradual overall change in the shape of the 
distributions. We refer to these uncertainties as type B. Global 
normalization uncertainties do not affect the shape of the distributions 
but only the absolute normalization; these are quoted separately as type 
C.

There are multiple contributors to each type of systematic error, for 
example the \cc and \bb templates are model dependent and can be 
determined with \pythia or \powheg. For each such case we repeat the 
full analysis with the various assumptions. The spread of the results 
around the default analysis is used to assign systematic uncertainties.

If we considered two assumptions, like in the example given, we quote 
the uncertainty as half the difference between the two assumptions. If 
there is a clearly preferred default case, we use the difference of 
results obtained with extreme assumptions to assign systematic 
uncertainties.

\begin{figure*}
\includegraphics[width=0.325\linewidth]{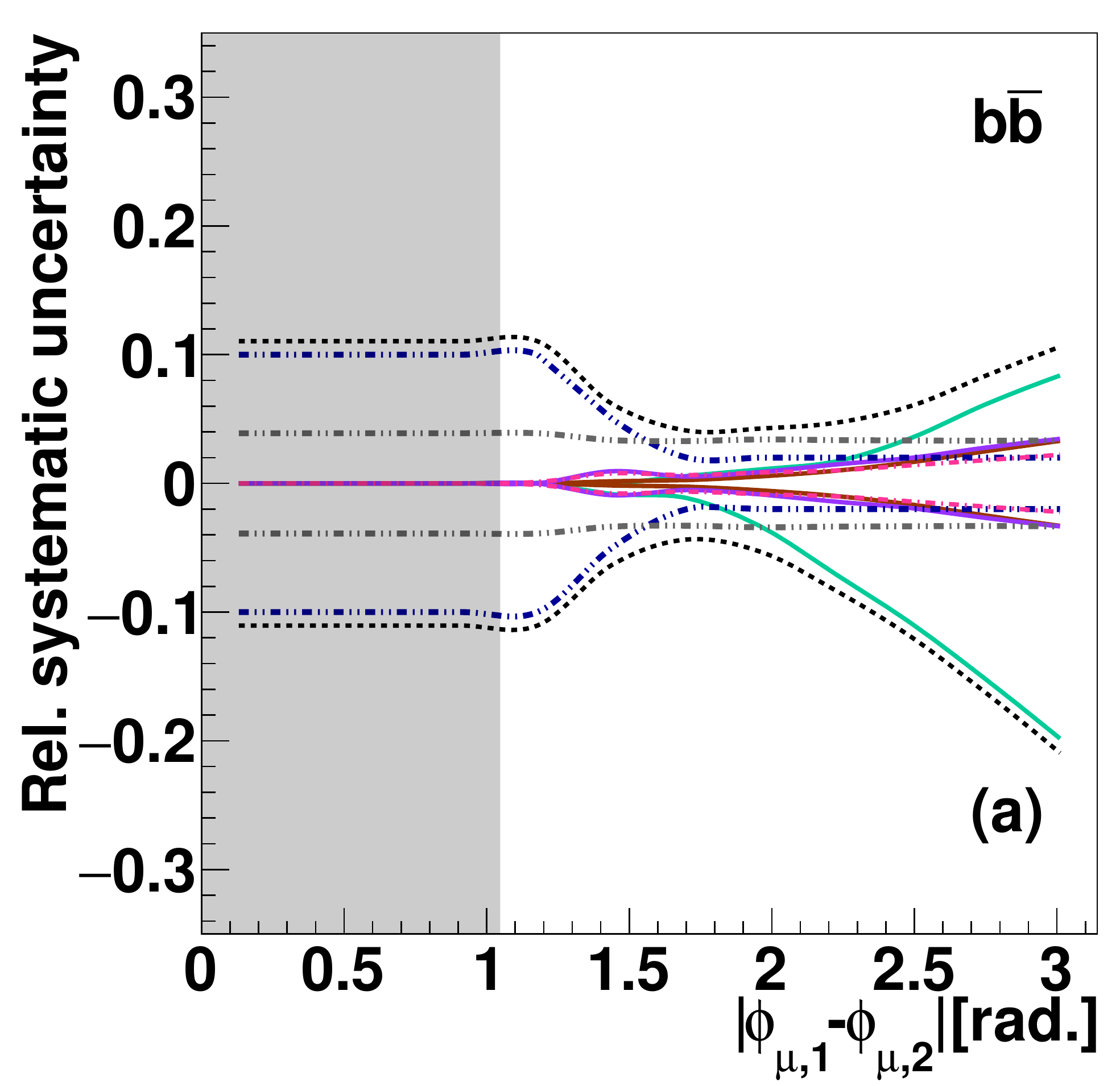}
\includegraphics[width=0.325\linewidth]{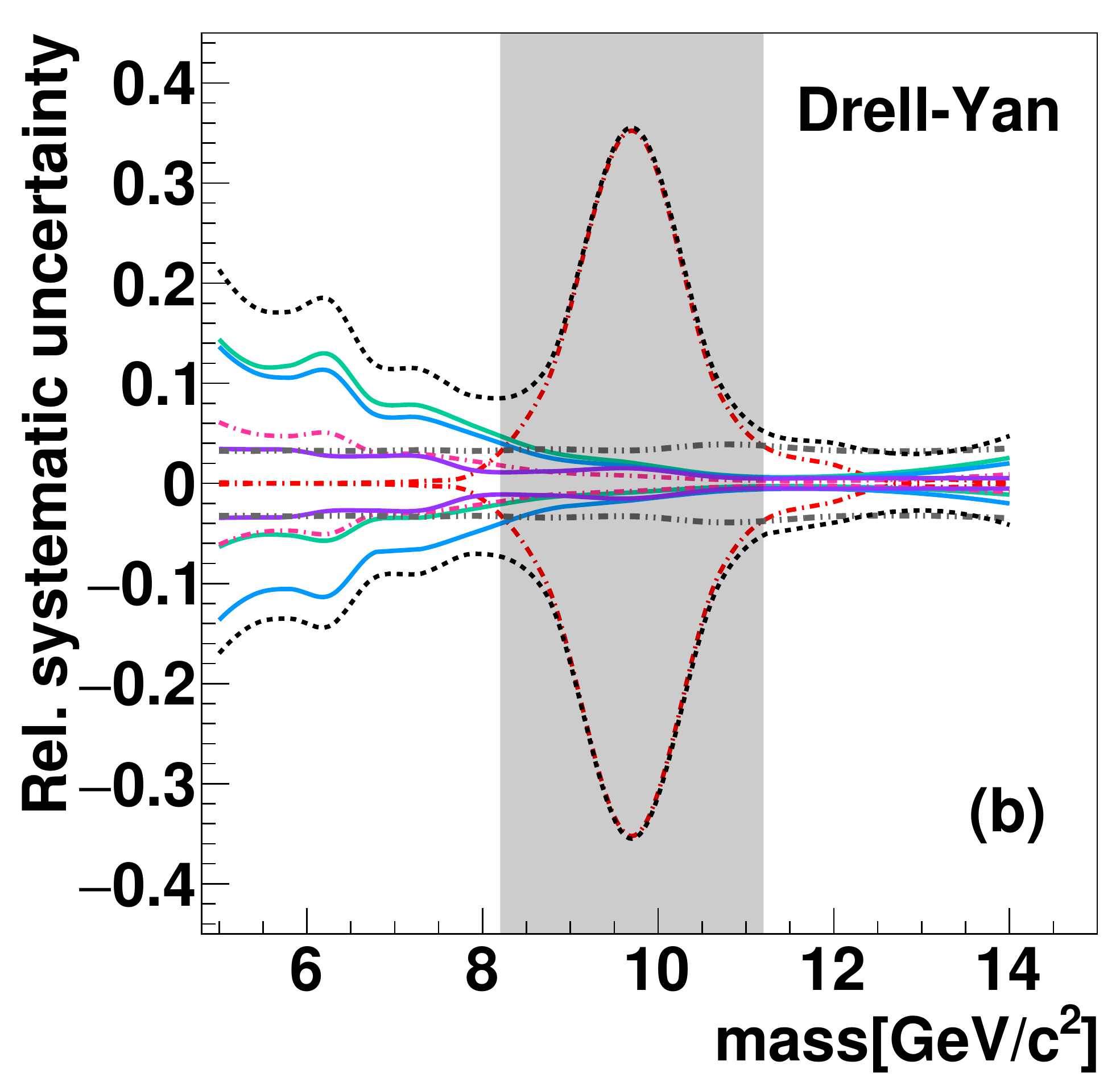}
\includegraphics[width=0.325\linewidth]{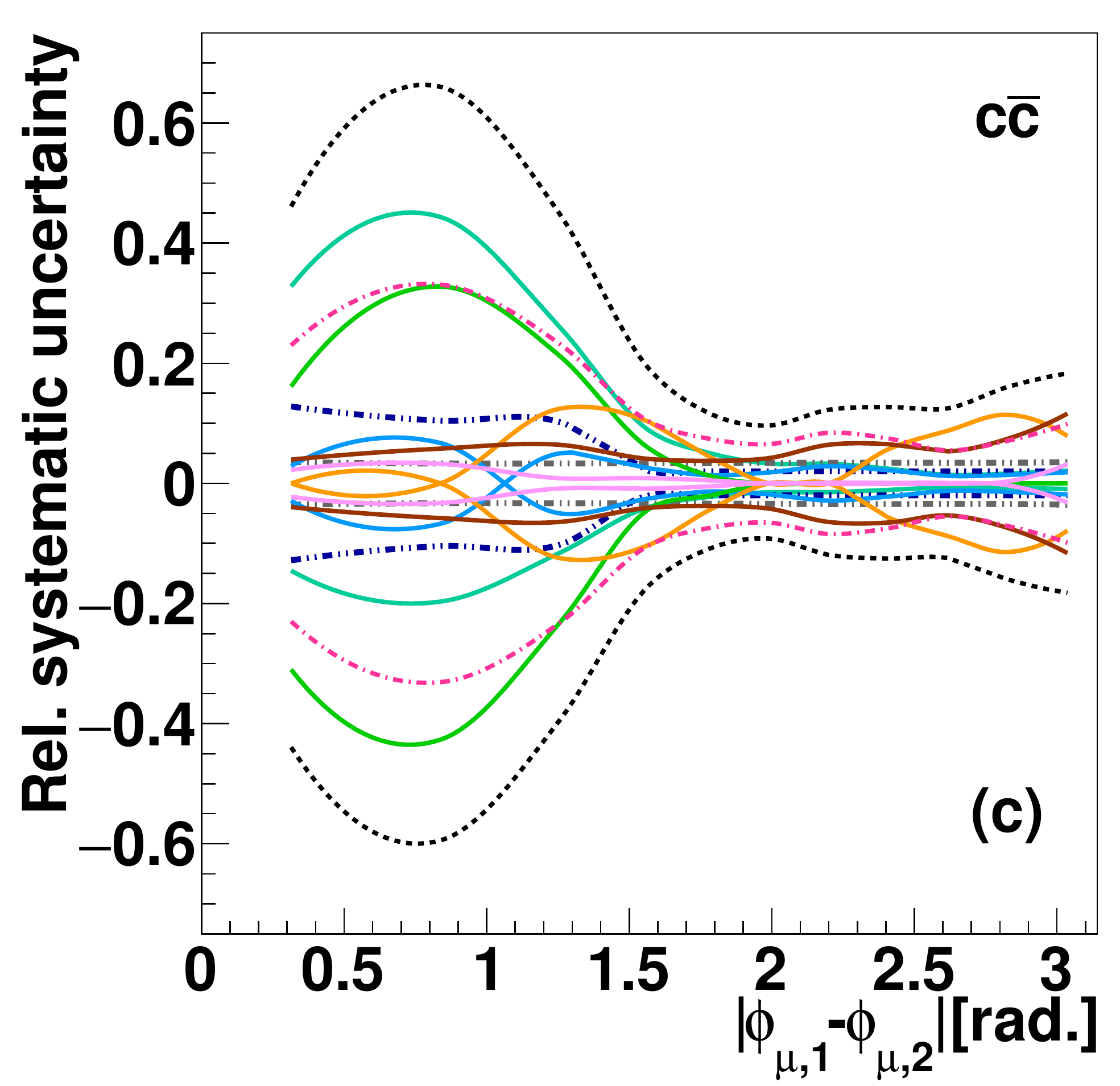}
\includegraphics[width=0.325\linewidth]{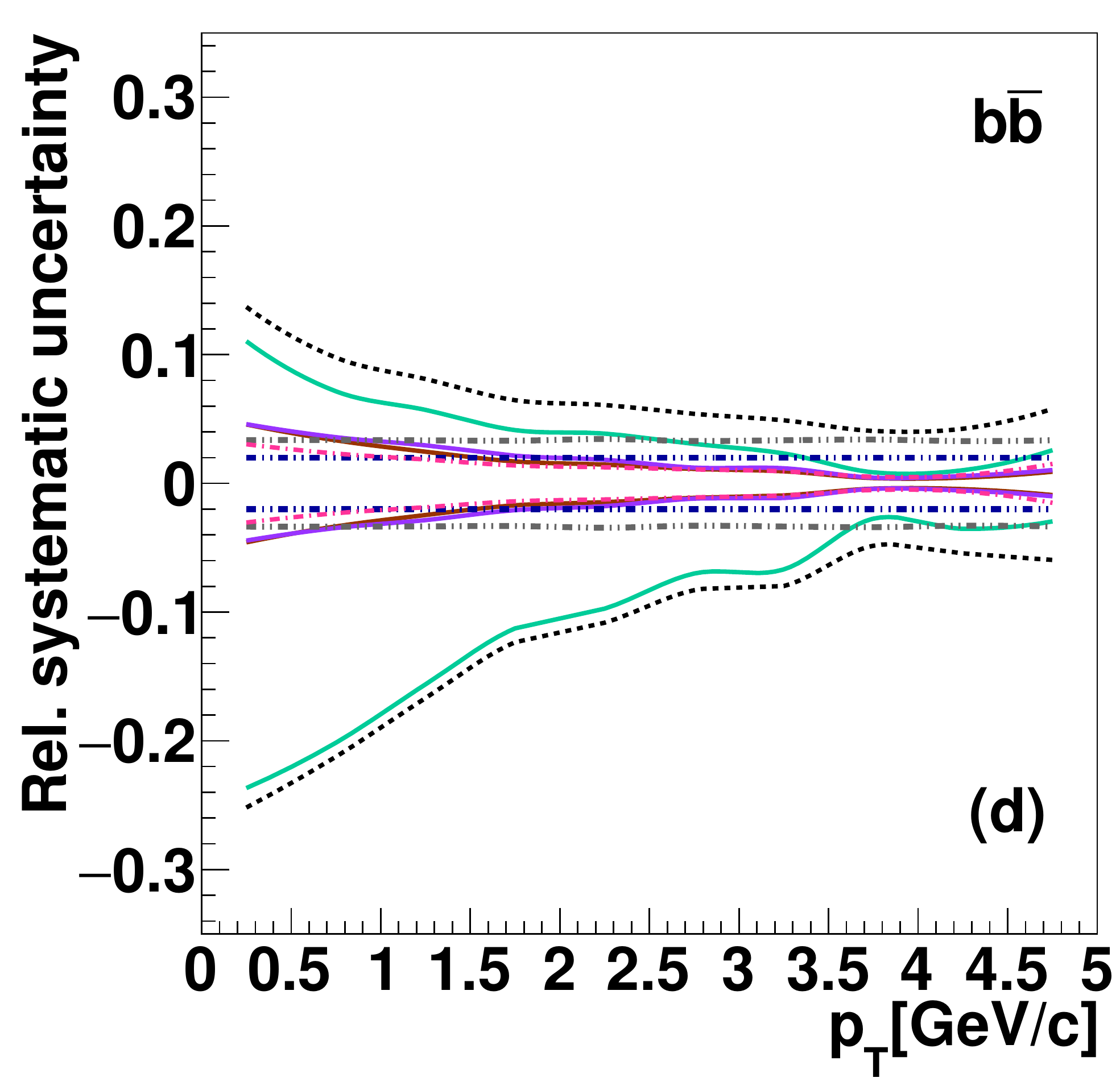}
\includegraphics[width=0.325\linewidth]{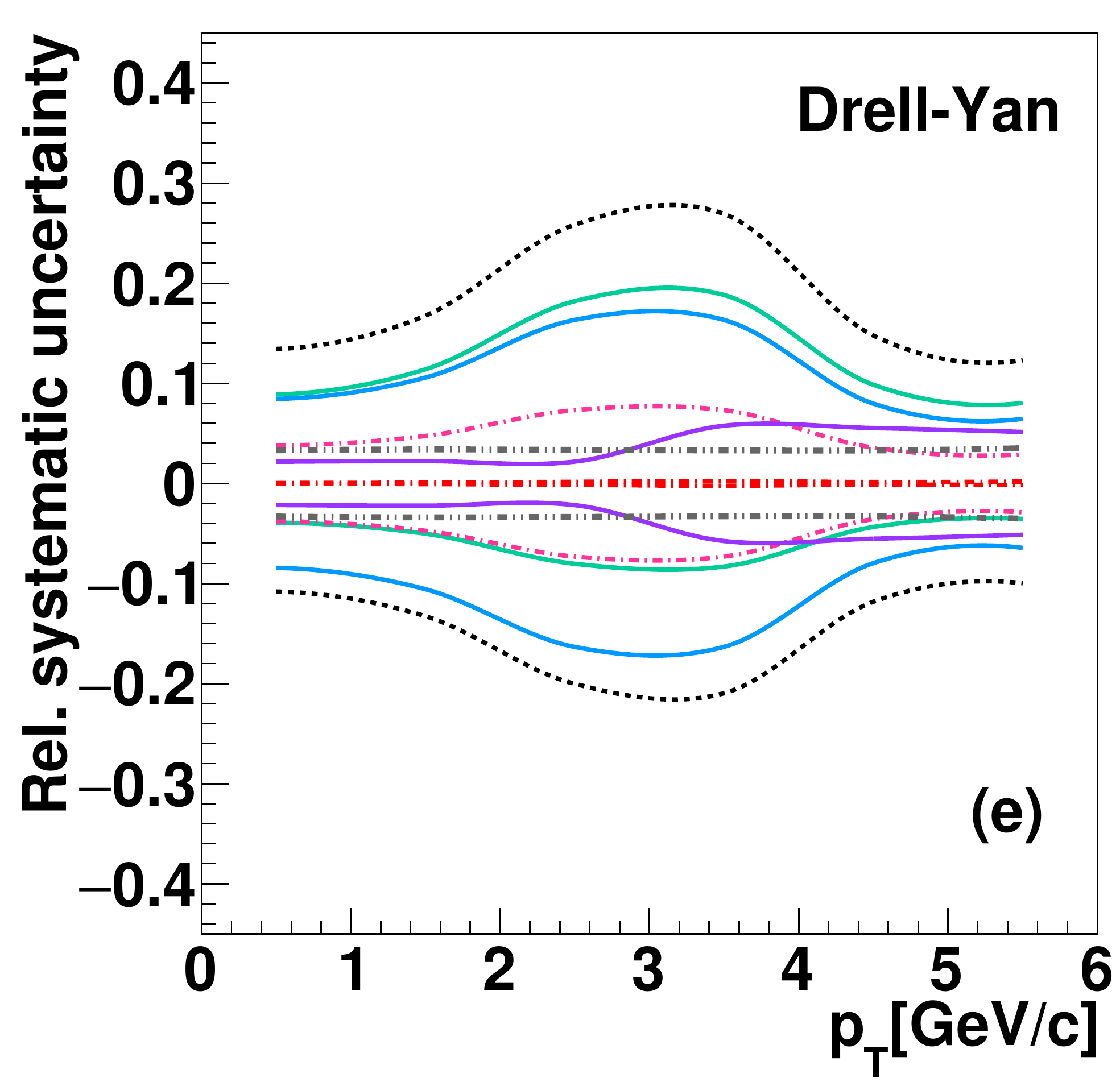}
\includegraphics[width=0.325\linewidth]{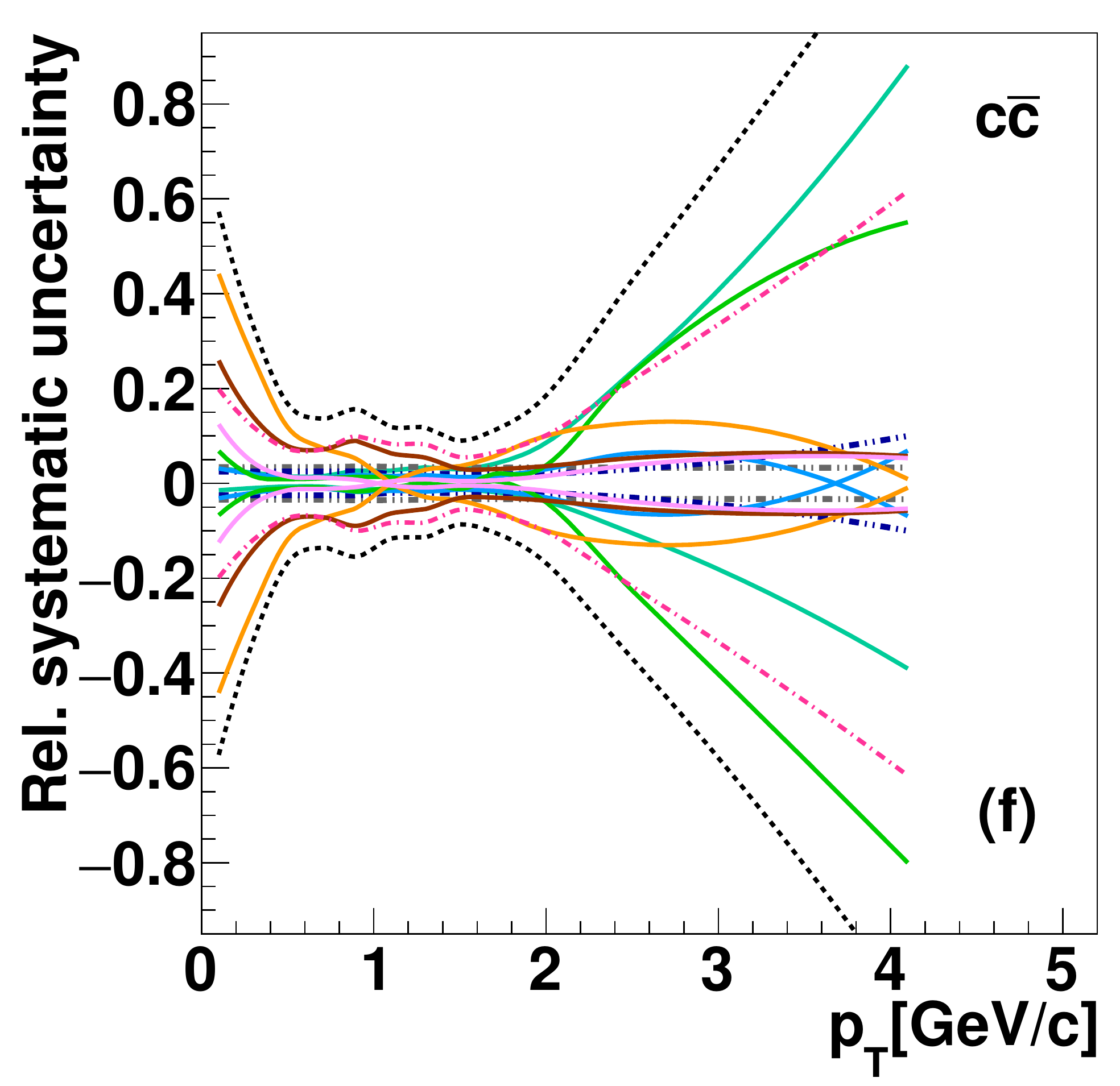}
\includegraphics[width=0.98\linewidth]{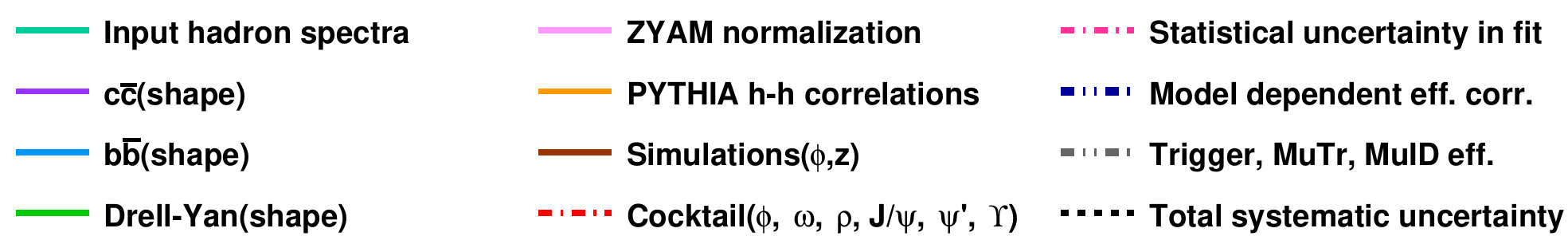}
\caption{ \label{Fig:sysbreakdown} Relative two-arm averaged systematic 
uncertainties for \cc and \bb measurements as a function of $\Delta\phi$ 
or pair $p_{T}$ and Drell-Yan measurement as a function of mass or 
$p_{T}$. The shaded regions are excluded from the respective 
measurements.}
\end{figure*}

We quantify all systematic uncertainties as standard deviations. The 
systematic uncertainties on the different measurements are summarized in 
Table.~\ref{Tab:syssummary}. For the differential distributions of \cc, 
\bb, and Drell-Yan, the systematic uncertainties vary with azimuthal 
opening angle, pair $p_{T}$ or mass as shown in 
Fig.~\ref{Fig:sysbreakdown}.

\begin{table*}[t]
\caption{\label{Tab:syssummary}
Summary of arm-averaged relative systematic uncertainties for the total 
bottom cross section $\sigma_{b\bar{b}}$, the differential Drell-Yan 
cross section $d^{2}\sigma_{DY\rightarrow\mu\mu}/dmdy$, and the \bb(\cc) 
differential yields 
$dN_{b\bar{b}(c\bar{c})\rightarrow\mu\mu}/d\Delta\phi$. The systematic 
uncertainty type is indicated in the second column and is applicable 
only to the differential measurements. The uncertainties for the 
differential measurements vary with azimuthal opening angle, pair $p_T$, 
or mass. Asymmetric uncertainties are quoted in bracketed values. For 
the \cc measurement, the regions $\Delta\phi<\pi/2$, $p_{T}<0.5$ \gevc 
and $p_{T}>2.0$ \gevc are excluded because the yield approaches zero and 
relative systematic uncertainties diverge. With these regions excluded, 
the difference between the systematic uncertainties of all measurements 
for the south and north muon arms differs by no greater than $2\%$ for 
all systematic uncertainties sources.
}
\begin{ruledtabular} \begin{tabular}{ccccccccc}
& type & $\sigma_{b\bar{b}}$   &$\frac{dN_{b\bar{b}\rightarrow\mu\mu}}{d\Delta\phi}$   
&$\frac{dN_{b\bar{b}\rightarrow\mu\mu}}{dp_{T}}$& $\frac{d^{2}\sigma_{DY\rightarrow\mu\mu}}{dmdy}$   
& $\frac{1}{2\pi p_{T}}\frac{d^{2}\sigma_{DY\rightarrow\mu\mu}}{dyp_{T}}$   & $\frac{dN_{c\bar{c}\rightarrow\mu\mu}}{d\Delta\phi}$ & $\frac{dN_{c\bar{c}\rightarrow\mu\mu}}{dp_{T}}$\\
\hline
Input hadron spectra & B & $+$4.7\%         & $+$($<$6\%)          & $+$($<$12\%)  & $+$($<$14\%)         &  $+$($<$20\%)  &  $+$($<$9\%)   & $+$($<$9\%) \\
                     &   & $-$11.0\%        & $-$($<$19\%)         & $-$($<$25\%)  & $-$($<$7\%)          &  $-$($<$9\%)   &  $-$($<$4\%)  & $-$($<$4\%) \\ 
Hadron simulation    & B & 2\%            & $<$1\%             & $<$1\%     & $<$1\%             &  $<$1\%      & $<$1\%     & $<$1\% \\
$c\bar{c}$ (shape)           & B & 2\%            &  $<$4\%            & $<$5\%     & $<$4\%            &  $<$6\%    & -          & - \\
$b\bar{b}$ (shape)           & B & -                & -                  & -          & $<$14\%            &  $<$17\%     & $<$3\%     & $<$3\%\\
Drell-Yan (shape)     & B & $<$1\%          & $<1$\%             &  $<$1\%    & -                  &  -           & $<$6\%     & $<$5\% \\
ZYAM normalization   & B & $<$1\%           & $<$1\%             &  $<$1\%    & $<$1\% 			   &   $<$1\%     & $<$2\%     & $<$3\%\\
\pythia $h$-$h$ correlations 
                     & B & -                & -                  & -          & -                  &  -           & $<$14\%    & $<$13\%\\
Simulations($\phi$,\z) 
                     & B & $<$1\%           & $<4$\%             & $<5$\%     & $<$1\%             &  $<$1\%      & $<8$\%     &  $<$8\%\\
\\
Fitting range        & B & $2$\%            & $<$1\%             & $<$1\%     & $<1$\%             &  $<1$\%      & $<$1\%     & $<$1\% \\
$\phi, \omega, \rho, J/\psi, \psi',\Upsilon$ norm.
                     & B & -                & -                  & -          & $<$2\%             &  $<1$\%      & $<$1\%     &  $<$1\% \\
Statistical uncertainty in fit
                     & B & -                & $<$4\%             & $<$4\%     &  $<$6\%            &  $<$8\%      & $<$10\%    & $<$10\%  \\
\\
\bb model dep. extrapolation 
                     & - & \multirow{2}{*}{6.5\%}              & - & - &-               & - & -  & - \\
Model dep. eff. corrections
                     & B &                  &  $<$10\%           & $<$3\%& -                  & - & $<$5\%     & $<$4\%\\

Trigger efficiency   &B&1.5\%&1.5\%&1.5\%&1.5\%&1.5\%&1.5\%&1.5\%\\
MuTr efficiency            &B&4\%&4\%&4\%&4\%&4\%&4\%&4\%\\
MuID efficiency            &B&2\%&2\%&2\%&2\%&2\%&2\%&2\%\\
\\
Sum of type B      	 & \multirow{2}{*}{-} & $+$9.3\%       
& $+$(4\%--11\%)         
& $+$(6\%--14\%) 
& $+$(4\%--21\%)        
& $+$(13\%--28\%) 
& $+$(10\%--28\%) 
& $+$(10\%--20\%) \\
systematic uncertainties     &   
& $-$13.2\%       
& $-$(4\%--22\%)         
& $-$(6\%--26\%) 
& $-$(4\%--17\%)        
& $-$(11\%--22\%)
& $-$(10\%--20\%)  
& $-$(8\%--16\%) \\
\\
Global normalization     & C  & 12\%  & 12\%  & 12\% 
&  12\%        & 12\% & 12\%  & 12\% \\
\end{tabular} \end{ruledtabular}
\end{table*}

\subsection{Shape of simulated distributions}
\label{Sec:sysshape}

The \cc, \bb, Drell-Yan, and hadronic pair background components are 
correlated through the fitting procedure, thus an uncertainty on the 
shape for any one template distribution will affect the fit results of 
all four components simultaneously. For example, if one increases the 
hardness of the input pion \pt spectrum, the number of high mass 
like-sign hadron-hadron pairs will increase, which will lead to a 
smaller \lsmumu pair yield from \bb. Because \bb is the main competing 
source to the Drell-Yan process in the high \ulmumu pair mass region, 
this will in turn lead to a larger Drell-Yan yield. Drell-Yan and bottom 
both contributes to the intermediate mass region where \cc is extracted, 
and hence will also modify the \cc yield.

In the following we will discuss the uncertainties on the shape of 
individual contributions and how these uncertainties propagate to the 
measurement of all components.

\subsubsection{Input hadron spectra}
\label{Sec:syshadronspectra}
The input pion and kaon \pt spectra are tuned to match PHENIX and BRAHMS 
data at $\langle y\rangle=0$ and $\langle y\rangle=2.95$, respectively.  
This is achieved by applying weighting factors ($w_h(y)$) to the \pt 
spectra from \pythia, which are determined by a linear interpolation 
between the two ratios of \pythia to the data at $\langle y\rangle=0$ 
and $\langle y\rangle=2.95$ (see Fig.~\ref{Fig:wgtfactors}). To estimate 
the systematic uncertainties on the input hadron \pt spectra, we vary 
the weighting function. We use either $w_{h}(\langle y\rangle=0)$ for 
all light hadrons, which gives a harder \pt spectra than the default 
case, or $w_{h}(\langle y\rangle=2.95)$, which gives a softer \pt 
spectra. The shape of the hadron-hadron pair mass distribution changes 
significantly only for masses above 3 \gevcc.

We take the difference of the cross sections obtained using these two 
sets of \pt spectra and the default \pt spectra as a systematic 
uncertainty on the input hadron spectra. For \sigmabb, this is 
determined to be ${+4.7\%}$ and ${-11.0\%}$. The uncertainties are also 
propagated to the \bb and \cc azimuthal opening angle distributions and 
the Drell-Yan yields. In all cases this is a dominant contributor to the 
systematic uncertainties (see Table~\ref{Tab:syssummary}).

We have also considered using the bands shown in 
Fig.~\ref{Fig:wgtfactors} as limits for the weighting factors, which 
lead to smaller uncertainties and we choose to quote the more 
conservative estimate. Uncertainties related to the choice of parton 
distribution function (PDF) are estimated by evaluating the differences 
obtained with simulations using the CTEQ5, CTEQ6, 
MRST2001(NLO)~\cite{Martin:2001es} and GRV98(LO)~\cite{Gluck:1998xa} 
parton distribution functions. The differences are negligible compared 
to the uncertainty due to shapes of the light hadron \pt spectra.

\subsubsection{ Hadron simulation}

The default PHENIX {\sc geant4} simulation utilizes the standard HEP 
physics list QGSP-BERT. For hadronic interactions of pions, kaons and 
nuclei above 12 GeV, the quark gluon string model (QGS) is applied for 
the primary string formation and fragmentation. At lower energies, the 
Bertini cascade model (BERT) is used, which generates the final state 
from an intranuclear cascade.

To estimate possible uncertainty due to the description of the hadronic 
interactions in the absorbers, we have used two other physics lists: The 
(i) FTFP-BERT list, which replaces QGS with the Fritiof model (FTF) for 
high energies. The FTF uses an alternative string formation model 
followed by the Lund fragmentation model. And (ii) QGSP-BIC where the 
low energy approach is replaced by the binary cascade model (BIC), which 
was optimized to describe proton and neutron interactions, but is less 
accurate for pions.

Using these different physics lists leads to a 2\% difference of 
\sigmabb, and a negligible difference to the charm and Drell-Yan 
normalizations.

\subsubsection{Charm and bottom simulation}

There are potential model dependencies of the \mumu and muon-hadron 
templates for \cc and \bb. To estimate these we compare the \mumu and 
muon-hadron templates obtained using \pythia and \powheg. Systematic 
uncertainties on charm and bottom are assumed to be uncorrelated and are 
added in quadrature.

Due to the large mass of the bottom quark, decay kinematics govern the 
shape of the distributions, hence the difference between \pythia and 
\powheg is small (see Fig.~\ref{Fig:anahfmodel}). The largest effect of 
this uncertainty is exhibited at mass $\sim5$ \gevcc for the Drell-Yan 
measurement where the contribution of \bb is around 40\% of the total 
yield.

For charm, the model dependence is larger than that of bottom, 
particularly for $m<1$ \gevcc. In the high mass region \powheg tends to 
predict higher yields for both \mumu and muon-hadron templates, which is 
likely due to a harder single muon \pt spectrum. However, this has a 
small effect on the extraction of bottom and Drell-Yan yields in the 
high mass region where the contribution of charm is less than 10\%.

\subsubsection{ Drell-Yan}
\label{Sec:dyktsys}

\begin{figure}[h]
\includegraphics[width=1.0\linewidth]{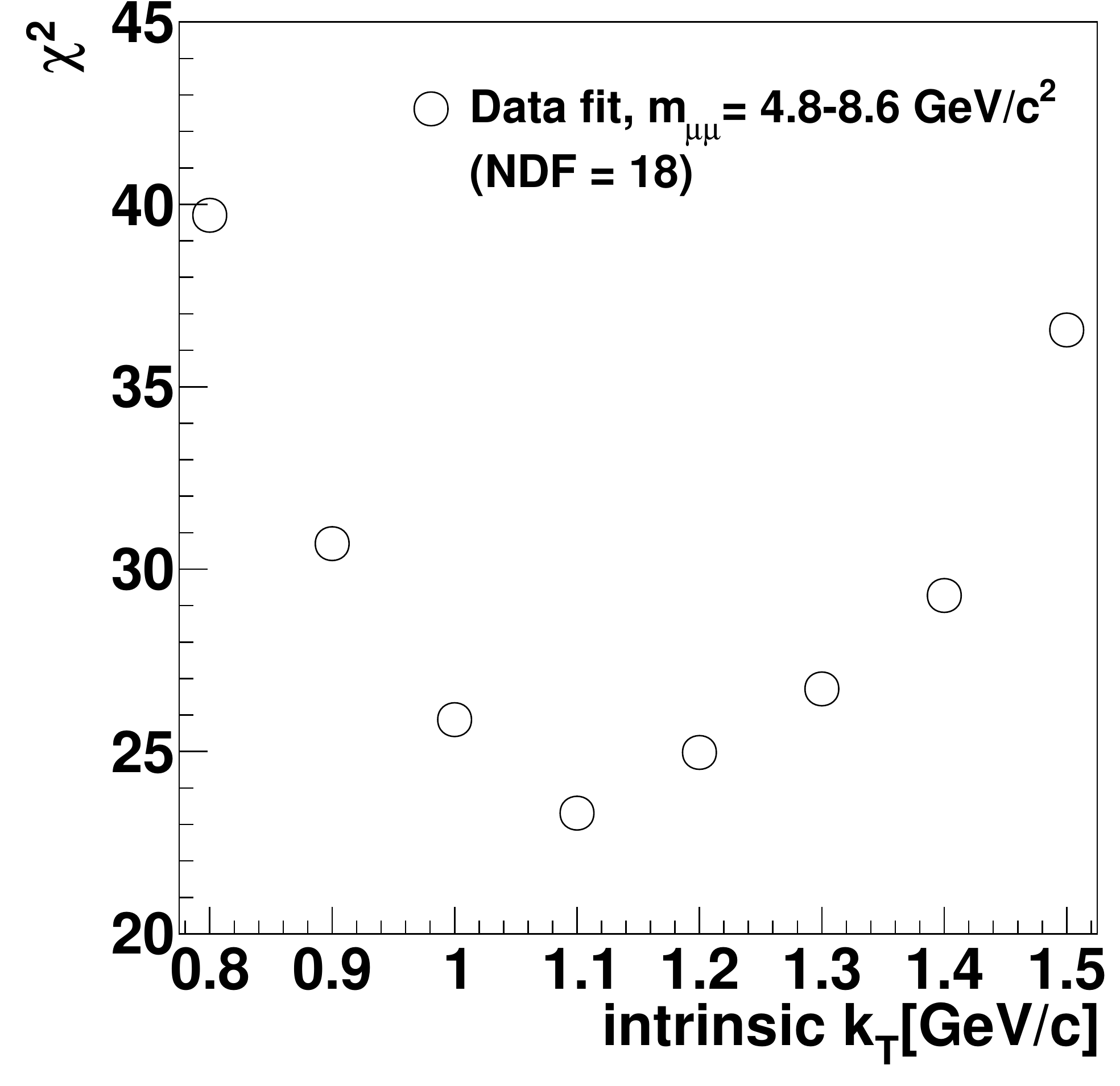}
\caption{\label{Fig:dykt} 
The $\chi^{2}$ for the pair $p_{T}$ spectrum of Drell-Yan pairs in the 
mass region 4.8--8.6 \gevcc compared to \pythia simulations with 
different intrinsic $k_{T}$. The $\chi^{2}$ is minimized at a $k_{T}$ of 
1.1 \gevc.
}
\end{figure}

The intrinsic $k_T=1.1$ \gevc used in the \pythia simulations is 
determined by minimizing $\chi^2$ of the \pt distribution of Drell-Yan 
pairs in the Drell-Yan mass region, between data and simulations. 
Background components (mostly from \cc and \bb) are normalized using 
cross sections obtained from the procedure and subtracted as a function 
of $p_{T}$.  We find that an intrinsic $k_{T}$ of 1.1 \gevc best 
describes the \pt distribution of Drell-Yan pairs in the high mass 
region (see Fig.~\ref{Fig:dykt}).

We vary the $k_T$ by $\pm0.1$ \gevc where the $\chi^2$ changes by 
$\sim1$ to estimate uncertainties in the Drell-Yan distributions. The 
uncertainty mainly affects the \cc yield at $\Delta\phi<\pi/2$ and 
$p_{T}>2$ \gevc and is negligible elsewhere.

\subsubsection{ ZYAM normalization}
\label{sec:syszyam}

To estimate the effect of varying the relative contributions between 
correlated and uncorrelated pairs, we have varied the mass region which 
we use for the $\Delta\phi_{prim}$ distribution. Instead of the default 
normalization region $\mathcal{M}$ below 3 \gevcc, we picked 3 separate 
regions: 0.7--1.3 \gevcc, 1.3--1.6 \gevcc, 1.6--2.2 \gevcc. This results 
in a variation of the ratio of correlated to uncorrelated pairs by 
$\pm$10\%. The relative effect on the sum of correlated and uncorrelated 
pairs is less than 2\% over the entire mass region, and has a negligible 
effect on the determination of \bb, \cc, and Drell-Yan cross sections.

\subsubsection{ Hadron-hadron correlations from \pythia}

For the measurement of \cc yields as a function of $\Delta\phi$ or pair 
$p_{T}$, correlated hadron pairs are a major background source. To 
estimate the uncertainty in the description of Tune A \pythia, we 
compare distributions of like-sign pairs between data and simulation in 
the same mass region (1.5--2.5 \gevcc) where other contributions, 
including \bb are negligible. We observe that the width of the 
back-to-back peak at $\Delta\phi=\pi$ is slightly wider in data compared 
to \pythia simulation. This is seen in the $p_{T}$ distributions as 
well, because $p_{T}$ is strongly correlated with $\Delta\phi$. The 
discrepancy is strictly less than 12\% and varies with $\Delta\phi$ or 
$p_{T}$. One data driven approach would be to apply an additional weight 
to the unlike-sign hadronic pair background as a function of 
$\Delta\phi$ or $p_{T}$, where the weight is computed by taking the 
ratio between data and simulations using the like-sign pairs as a 
function of $\Delta\phi$ or $p_{T}$ in the same mass region. This is 
motivated by the fact that the like-sign pairs are dominated by hadronic 
contributions in the mass region of interest.

Here we take the average between the Tune A setup and 
this data driven modification to be our central value, and assign a 
systematic uncertainty on the \cc yields as the difference between these 
two approaches. The resultant systematic uncertainty is strongly 
$\Delta\phi$ and $p_{T}$ dependent, ranging from $0\%$ to $14\%$.

\subsubsection{ Azimuthal angle($\phi$) description in simulations}

We compare the $\phi$ distributions of single tracks in data, 
simulations with the default framework, and the FastMC. We find 
reasonable agreement between data and the default simulation and 
conclude that the uncertainty from the default simulation framework is 
negligible. However, for simulations using the FastMC, we approximated 
the relative $\phi$ dependent efficiency by a weighting strategy in 
$\phi$ bins of finite width, which gives rise to a small smearing in the 
$\phi$ (and hence $\Delta\phi$ and to a lesser extent $p_{T}$) 
distributions (see Fig.~\ref{Fig:fastmccheck}). We assign $5\%$ 
uncertainty to the $\Delta\phi$ distributions generated using the 
FastMC, which is estimated by comparing $\Delta\phi$ distributions of 
mixed pairs between FastMC and real data. This in turn gives rise to an 
average of $5\%$ and $3\%$ to the \cc and \bb differential yields 
respectively.

\subsubsection { $z$-vertex description of simulations}
\label{Sec:sysz}

\begin{figure*}
\includegraphics[width=0.99\linewidth]{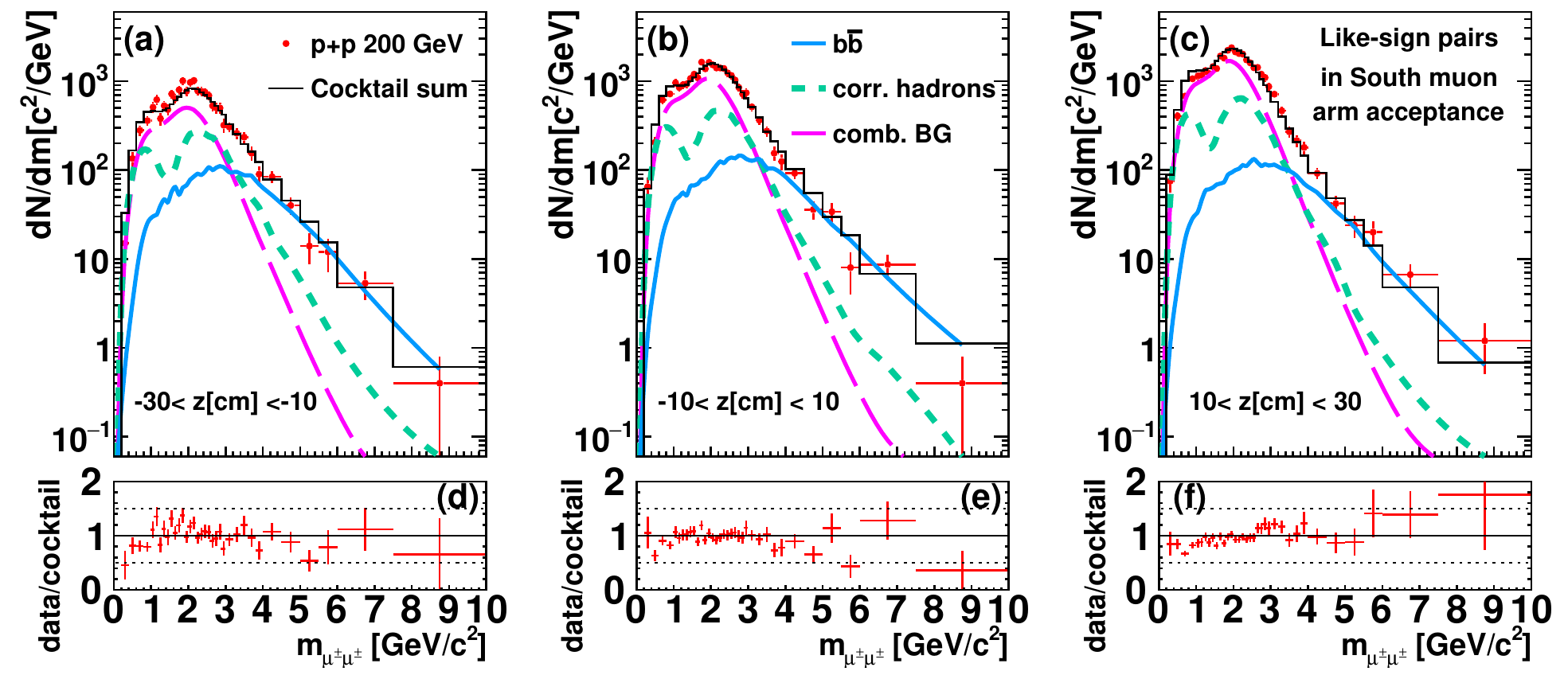}
\includegraphics[width=0.99\linewidth]{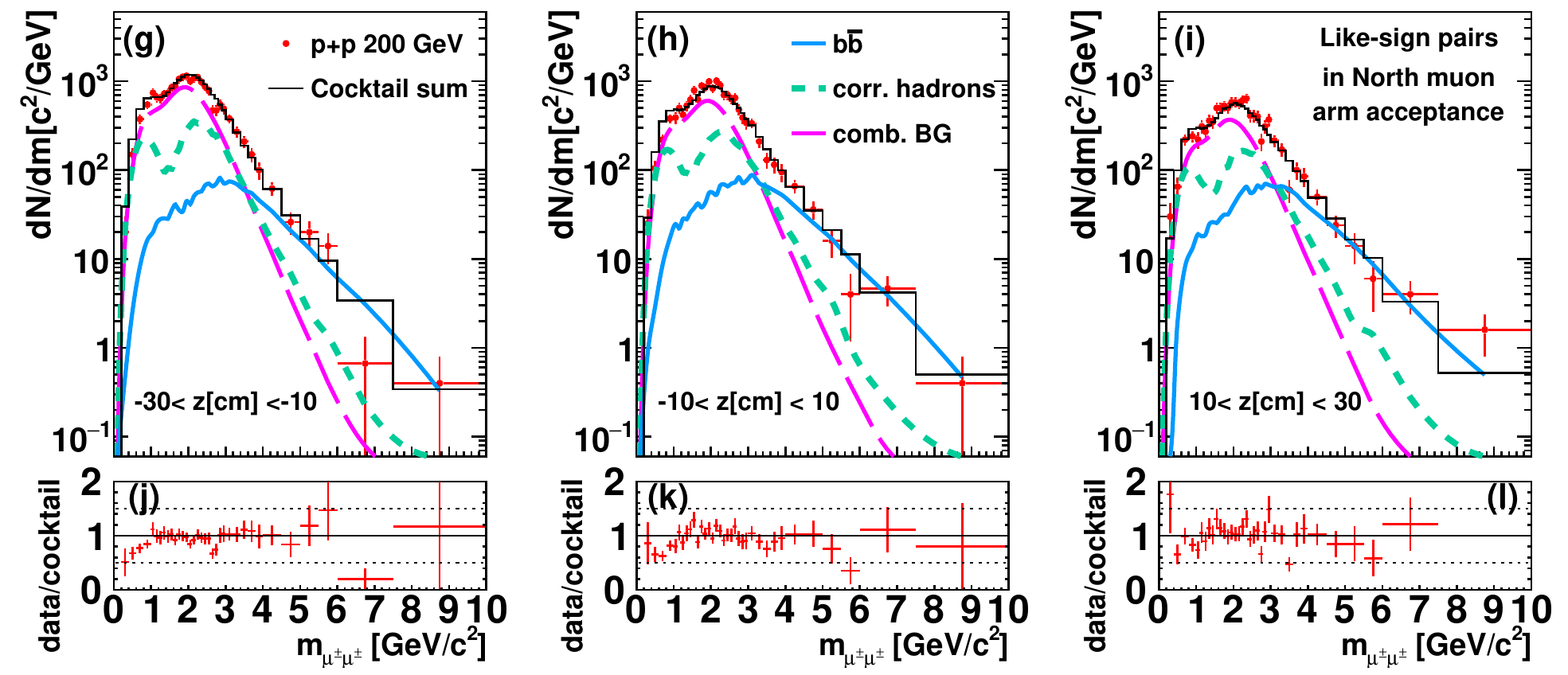}
\caption{ \label{Fig:lsmassz} Inclusive like-sign \mumu pair yield from 
\pp collisions at $\sqrt{s}=200$ GeV as a function of mass in three \z 
vertex bins for the south and north muon arms. The data are compared to 
the contributions from \bb decays, and the correlated \& combinatorial 
contribution from hadronic pairs.}
\end{figure*}

We have generated hadronic pairs in discrete $z_{\rm vtx}$ regions that 
cover 1/4 of the full collision $z_{\rm vtx}$ region using the FastMC. 
Figure~\ref{Fig:lsmassz} shows a comparison of data and simulations in 
different $z_{\rm vtx}$ regions after the initial normalization 
(Sec.~\ref{Sec:normhadrons}) and iterative fitting procedure 
(Sec.~\ref{Sec:fit}). We see good agreement between the simulations and 
data in all $z_{\rm vtx}$ regions; there is no indication that the 
approximations in the $z_{\rm vtx}$ description of correlated hadrons is 
biasing the fit of the like-sign pairs.

To estimate the systematic uncertainty on this approximation, recall 
that the yield of decay muons varies linearly with $z_{\rm vtx}$, 
whereas the yield of prompt muons is constant \cite{Aidala:2017pum}. 
Thus, the main effect of the $z_{\rm vtx}$ approximation is the 
uncertainty on the prompt muon to decay muon ratio. In the FastMC the 
ratio is determined in three vertex bins of 5~cm width at $z_{\rm vtx}=$ 
-20, 0, and 20~cm, instead of the full 20~cm $z_{\rm vtx}$ slices. We 
assign a systematic uncertainty by varying the prompt muon to decay muon 
ratio separately for each $z_{\rm vtx}$ region. Because prompt muons are 
dominated by charm decays, we estimate this effect by varying the charm 
cross section by $\pm15\%$ for one particular \z slice separately. The 
effect on the fitted \bb cross section is $\sim1\%$ and is negligible 
compared to other sources of systematic uncertainties.

\subsection{Normalization of simulated distributions}

In addition to uncertainties due to the shape of distributions, 
uncertainties on the normalization of one component can affect the yield 
of other components. We list sources of such uncertainties in this 
section.

\subsubsection{Fitting}

To estimate uncertainties in the fit range, we vary the lower bound of 
the fit range of like-sign pairs from $m=1.0$ \gevcc to $m=2.0$ \gevcc. 
The variation in \sigmabb is around $2\%$ and is assigned as the 
systematic uncertainty on the fit range. The unlike-sign fit range is 
also varied to diagnose possible effects due to non Gaussian tails of 
the mass distribution of \ulmumu pairs from resonance decays. The 
variation of $\kappa_{c\bar{c}}$ is less than 5\% with different fit 
ranges in the unlike-sign, and this $\kappa_{c\bar{c}}$ variation 
propagates into $<1$\% variation in \sigmabb.

We estimate possible uncertainties due to the stability of the fit by 
varying the binning of distributions. The variations are negligible 
compared to the statistical uncertainty. We therefore do not assign 
systematic uncertainties on fit stability.

\subsubsection{Normalization of cocktail components}

The vector mesons $\phi,~\omega,~\rho,~J/\psi,~\psi'$, and $\Upsilon$ 
are background components to determine $N_{c\bar{c}}^{+-}$ and $N_{\rm 
DY}^{+-}$ in Eq.~\ref{Eq:ccbg} and~\ref{Eq:dybg}, respectively. Their 
normalizations are fixed using previous measurements. The normalization 
of each component has associated statistical and systematic 
uncertainties from those measurements. We add these uncertainties in 
quadrature and vary normalizations of these background components to 
estimate propagated uncertainties in $N_{c\bar{c}}^{+-}$ and 
$N_{\rm DY}^{+-}$. Because the template fit excludes all mass regions 
dominated by resonance decays, the uncertainty from the normalizations 
of the resonances only have a minor effect of less than 2\% on the fit 
results, which is negligible compared to other sources of uncertainties.

\subsubsection{Statistical uncertainty in fit result}

Charm, bottom, and hadronic pairs are background components for 
$N_{\rm DY}^{+-}$. The statistical uncertainties on fitted values of 
$\kappa_{c\bar{c}}$, $\kappa_{b\bar{b}}$, and $\kappa_{h}$ become a 
source of systematic uncertainty for $N_{\rm DY}^{+-}$. Similarly, 
systematic uncertainties for $N^{+-}_{c\bar{c}}$ arise from statistical 
uncertainties on $\kappa_{h}$, $\kappa_{\rm DY}$, and $\kappa_{b\bar{b}}$, 
and $N_{b\bar{b}}^{\pm\pm}$ from $\kappa_{h}$ and $\kappa_{c\bar{c}}$. 
The statistical uncertainties for $\kappa_{b\bar{b}}$ and $\kappa_{\rm DY}$ 
is $\sim8\%$, and for $\kappa_{h}$ is $\sim2\%$ for each arm. The 
associated systematic uncertainty depends heavily on the signal to 
background ratio and varies from measurement to measurement.

\subsection{Extrapolation, acceptance and efficiency}

This section details systematic uncertainties related to acceptance and 
efficiency.

\subsubsection{Model dependence on \bb}

We use the high mass like-sign pairs to constrain \sigmabb, hence a 
determination of $d\sigma_{b\bar{b}}/dy$ involves an extrapolation to 
zero mass at forward rapidity, whereas the determination of \sigmabb 
involves a further extrapolation to the full rapidity region. This is 
dependent on correlations between \mumu pairs from bottom as well as the 
oscillation parameters and branching ratios. To quantify the uncertainty 
in the extrapolation, we take the average of the fitted cross section 
\sigmabb using \pythia and \powheg and assign the difference 
($\pm6.5\%$) as the systematic uncertainty. We note that the difference 
between the default values of the time-integrated probability for a 
neutral $B^{0}_d$ ($B^{0}_s$) to oscillate $\chi_{d}$ ($\chi_{s}$) of 
\pythia and the values from the PDG, $\chi_{d}=0.1860\pm0.0011$ 
($\chi_{s}=0.499304\pm0.000005$) \cite{Patrignani:2016xqp} is less than 
$2\%$ and hence much less than the assigned uncertainty.

\subsubsection{ Model dependence on efficiency correction}

The charm and bottom azimuthal opening angle distributions are corrected
to represent \mumu pairs the ideal muon arm acceptance. To assess the 
sensitivity to different input distributions we compare the efficiency 
as a function of $\Delta\phi$ calculated using \pythia and 
\powheg. No model dependence of the efficiency corrections is observed 
for \mumu pairs with $\Delta\phi>1.5$ from \cc and \bb. For 
$\Delta\phi<1.5$, we assign an additional uncertainty based on the 
difference of the efficiency corrections calculated by \pythia and 
\powheg.

The charm and bottom pair $p_{T}$ spectra are also corrected to represent
the muon arm acceptance. No model dependence of the efficiency corrections 
is observed for \mumu pairs in the measured 
$p_{T}$ range. We assign an uncertainty based on the statistical uncertainty 
of the calculated efficiency corrections.

For Drell-Yan, we estimate the model dependence of the acceptance and 
efficiency corrections by varying the intrinsic $k_{T}$ settings of 
\pythia within the systematic limits as described in 
Sec.~\ref{Sec:dyktsys}. No model dependence of the acceptance and 
efficiency corrections is observed. We assign an uncertainty based on 
the statistical uncertainty of the calculated efficiency corrections.

\subsubsection{ Trigger efficiency}

The possible discrepancy between the software trigger emulator and the 
hardware trigger is quantified by comparing the real data trigger 
decision with the offline software trigger. We find that they differ by 
within $1.0\%$ and $1.5\%$ for the south and north arm, respectively. 
We use these values as estimates of the associated systematic 
uncertainty.

\subsubsection{ Reconstruction efficiency}

The muon track reconstruction and muon identification used in this 
analysis is the standard PHENIX muon reconstruction chain. The 
systematic uncertainties have been previously studied. We assign MuTr 
($4\%$) and MuID ($2\%$) as systematic uncertainties on reconstruction 
efficiency based on the work published in \cite{Aidala:2017pum}.

\subsection{Global normalization uncertainties}
\label{Sec:globalnorm}

The absolute normalization of the \mumu pair spectra is set by the 
measured $J/\psi$ yield \cite{Adare:2011vq}, which is measured with an 
accuracy of 12\%. This is the systematic uncertainty on the scale for 
all results presented in this paper.

The normalization is expressed in Eq.~\ref{Eq:corraw} by the factor 
$\alpha$, which accounts for the combined effect of the change of the 
trigger livetime with luminosity $\alpha_{live}$, modifications of the 
reconstruction efficiency due to detector occupancy $\alpha_{\rm occ}$, 
additional variations of the efficiencies with luminosity 
$\alpha_{lum}$, and the effect of double interactions $\alpha_{double}$.

As a cross-check, these individual factors were determined separately. 
The trigger livetime was monitored during data taking and the correction 
was found to be 1.35 (1.30) for the south(north) arm, respectively. The 
occupancy effect was studied by embedding simulated \mumu pairs in \pp 
events and results in $\alpha_{\rm occ}=1.06 \ (1.04) $. In addition, 
there is a drop of the detector efficiency with increasing beam 
intensity that was found to give $\alpha_{lum}=1.04 \ (1.07)$.

Finally, the approximately 20\% double interactions in the sample 
increase the pair yield by about 11\%, resulting in 
$\alpha_{double}=0.90$. The yield increase is smaller than the number of 
double interactions mostly for two reasons. Diffractive events 
contribute to events with double interactions but do not contribute 
significantly to the pair yield. Events with double interactions contain 
collisions more than 40--50~cm away from the nominal collision point; 
pairs from these events have significantly reduced reconstruction 
efficiency. The combination of both effects approximately cancel the 
efficiency losses due to detector occupancy and high interaction rates.

The product of individual corrections to the normalization is 
$\alpha_{double}\times \alpha_{\rm occ}\times \alpha_{live} \times 
\alpha_{lum}$ = 1.33 (1.34) for the south (north) arm. These values are 
consistent within uncertainties with $1.30\pm0.16$ ($1.38\pm0.17$), the 
values based on the $J/\psi$ measurement.

\section{Results}
\label{Sec:results}

\subsection{Azimuthal opening angle and pair $p_{T}$ distributions for 
\mumu pairs from \cc and \bb}

\begin{figure*}
\includegraphics[width=0.48\linewidth]{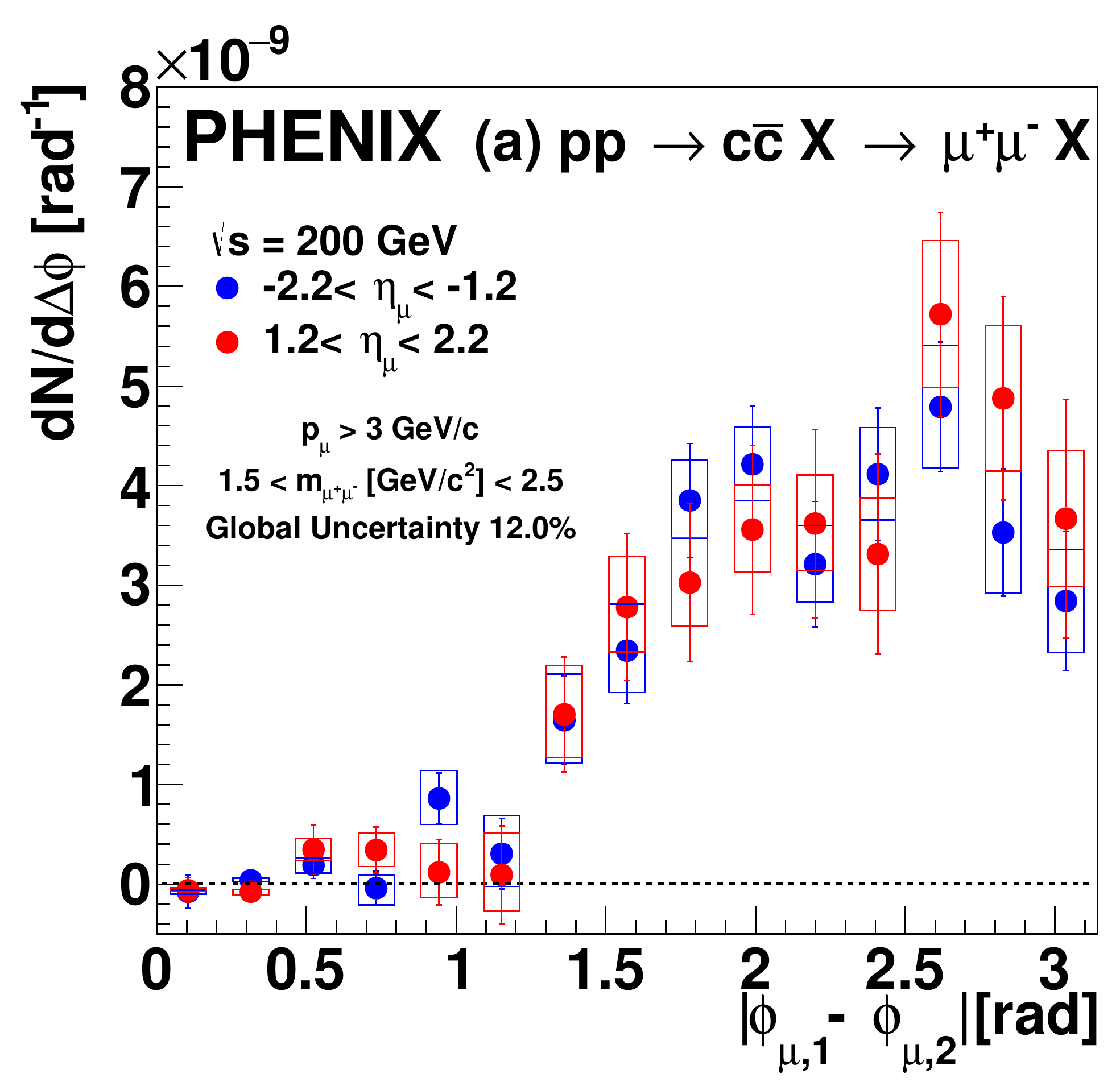}
\includegraphics[width=0.48\linewidth]{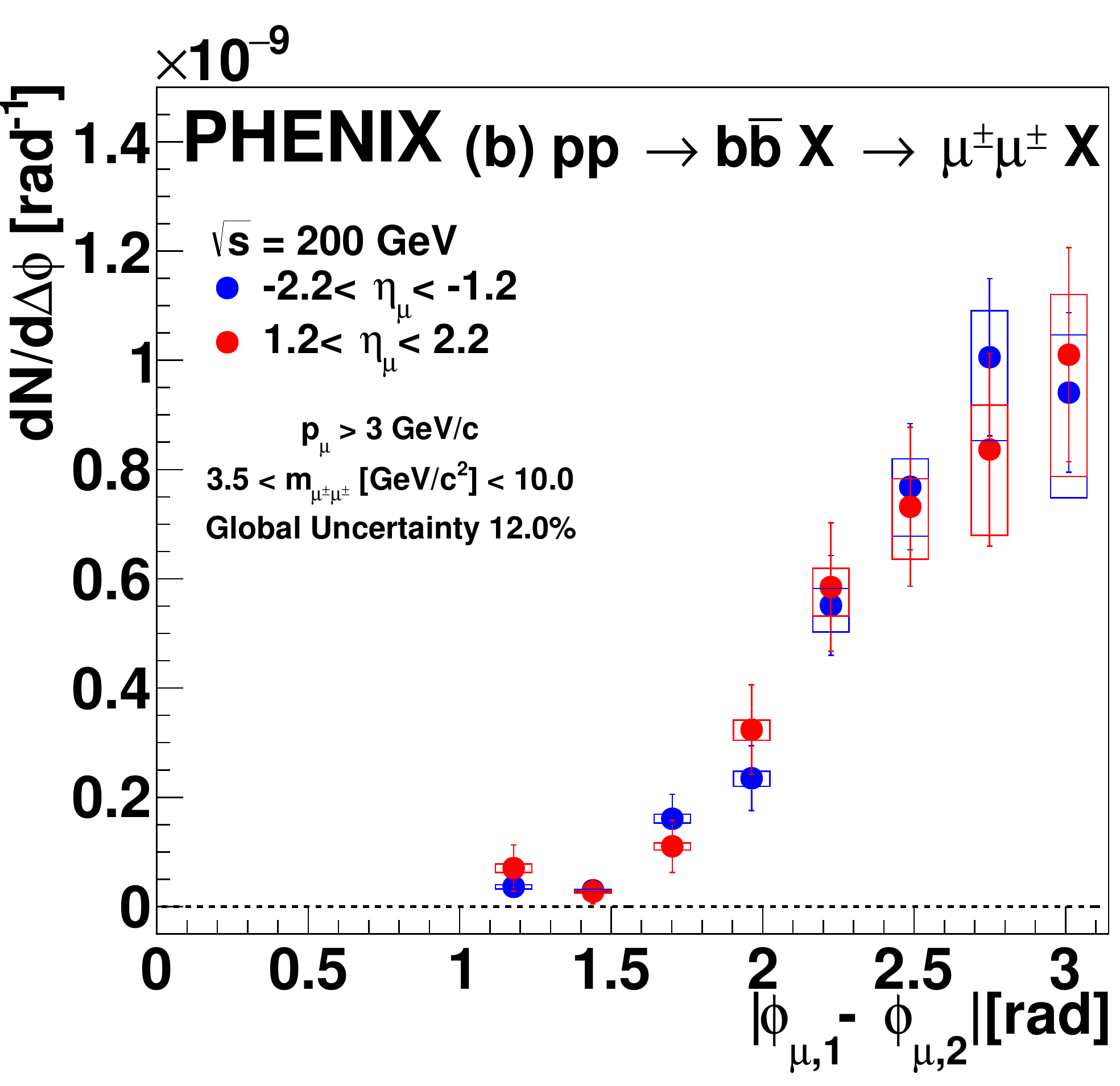}
\caption{ \label{Fig:hfdphiarm01} The corrected \mumu yield as a 
function of \dphi from (a) charm and (b) bottom decays. The error bars 
correspond to statistical uncertainties, and the boxes correspond to the 
type B systematic uncertainties. The $12.0\%$ type C systematic 
uncertainty is not shown. Results are given separately for the south and 
north muon arms.}
\end{figure*}

\begin{figure*}
\includegraphics[width=0.48\linewidth]{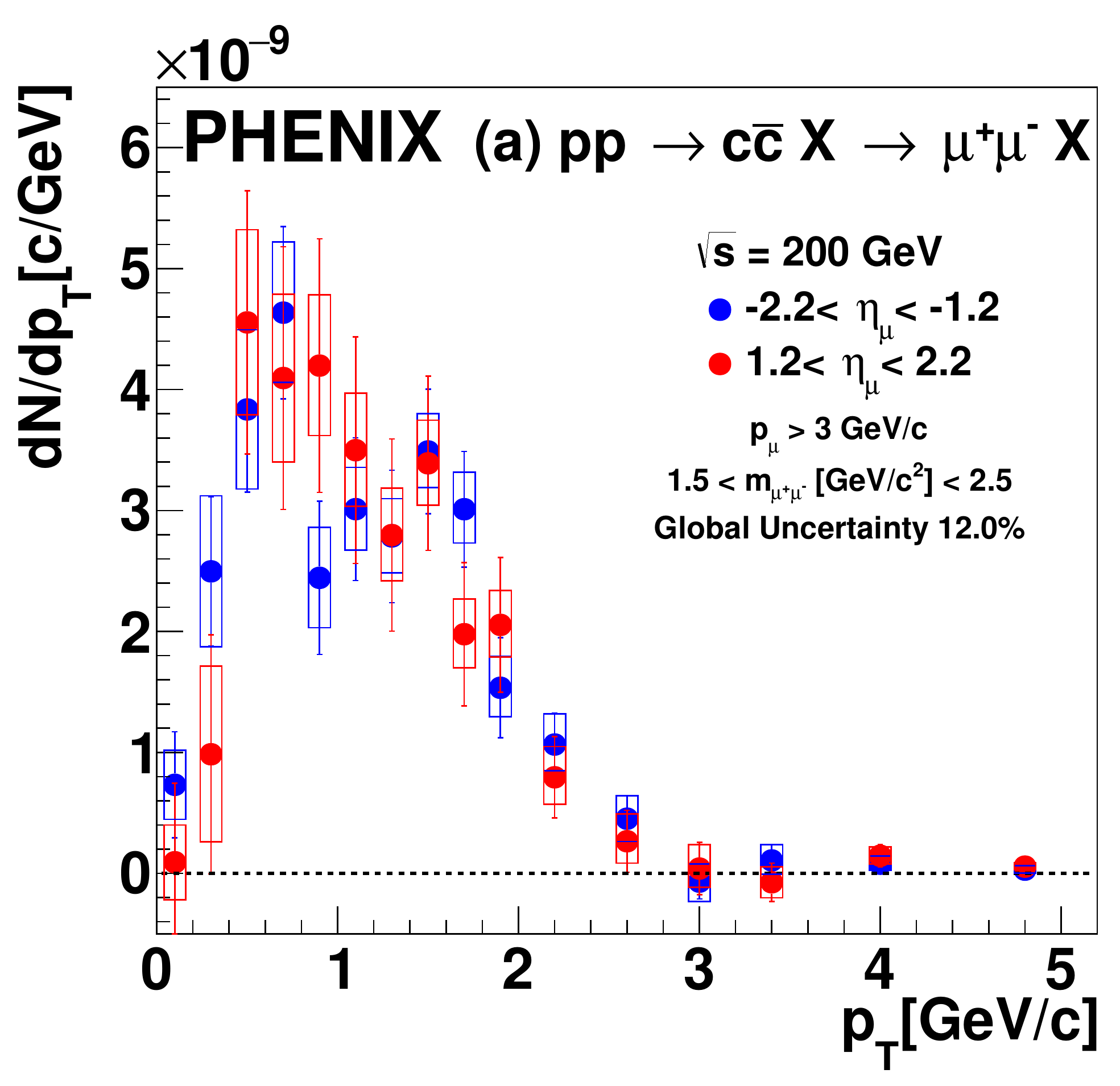}
\includegraphics[width=0.48\linewidth]{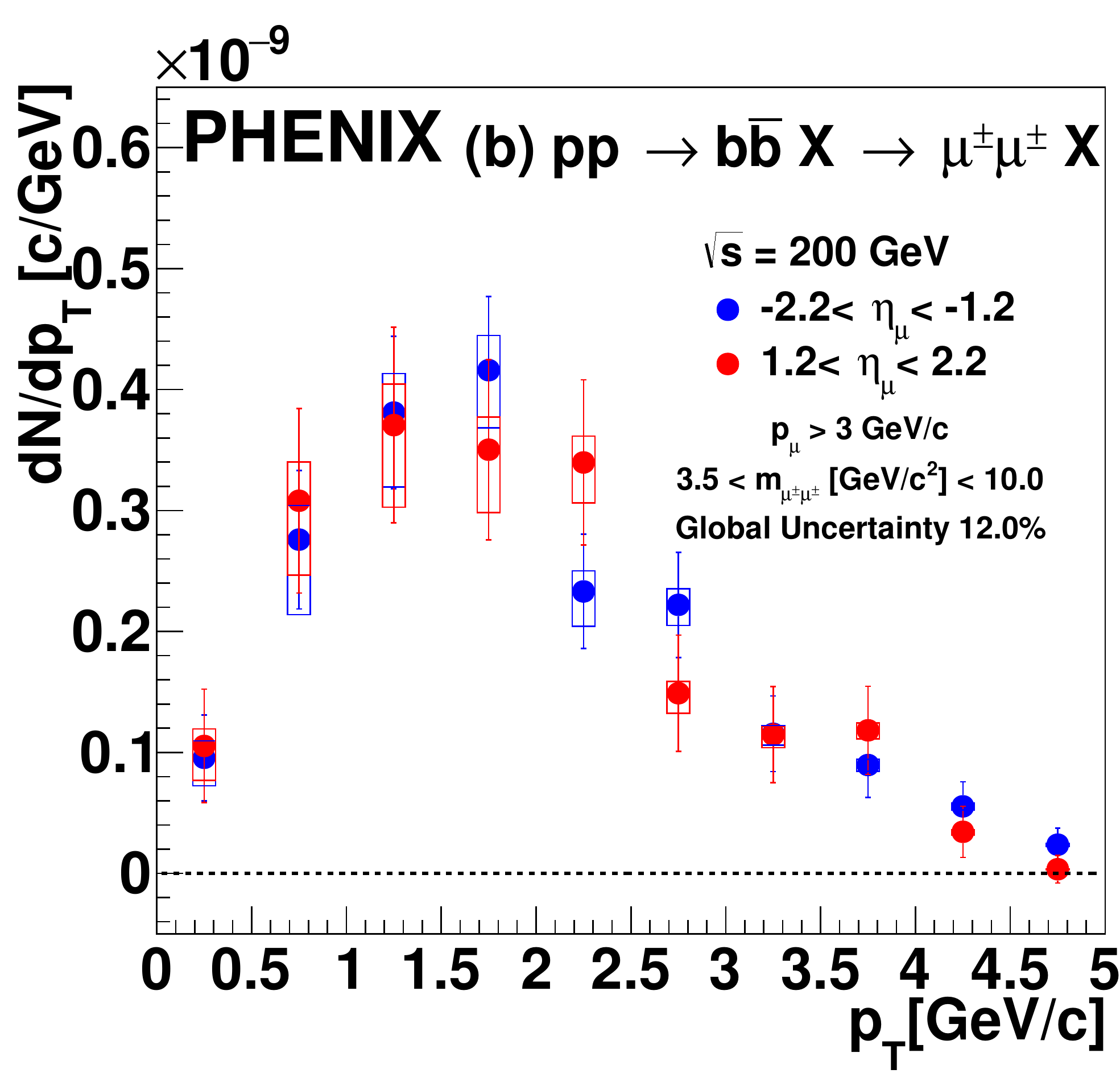}
\caption{ \label{Fig:hfptarm01} The corrected \mumu yield as a function 
of pair $p_{T}$ from (a) charm and (b) bottom decays. The error bars 
correspond to statistical uncertainties, and the boxes correspond to the 
type B systematic uncertainties. The additional $12.0\%$ type C 
systematic uncertainty is not shown. Results are presented separately 
for the south and north muon arms.}
\end{figure*}

The fully corrected \mumu pair yield from \cc and \bb decays are shown 
in Figs.~\ref{Fig:hfdphiarm01} and~\ref{Fig:hfptarm01} as a function of 
\dphi and pair $p_{T}$. The muons are in the nominal acceptance of $p>3$ 
\gevc and \mbox{$1.2<|\eta|<2.2$}. The pairs are in selected mass ranges 
of $1.5<m_{\mu^{+}\mu^{-}}<2.5$ \gevcc and 
$3.5<m_{\mu^{\pm}\mu^{\pm}}<10.0$ \gevcc for \cc and \bb, respectively. 
The yields for the two pseudorapidity regions are consistent with each 
other. Due to the mass selection, the $\Delta\phi$ and $p_{T}$ 
distributions are highly correlated with each other.

\begin{figure*}
\includegraphics[width=0.48\linewidth]{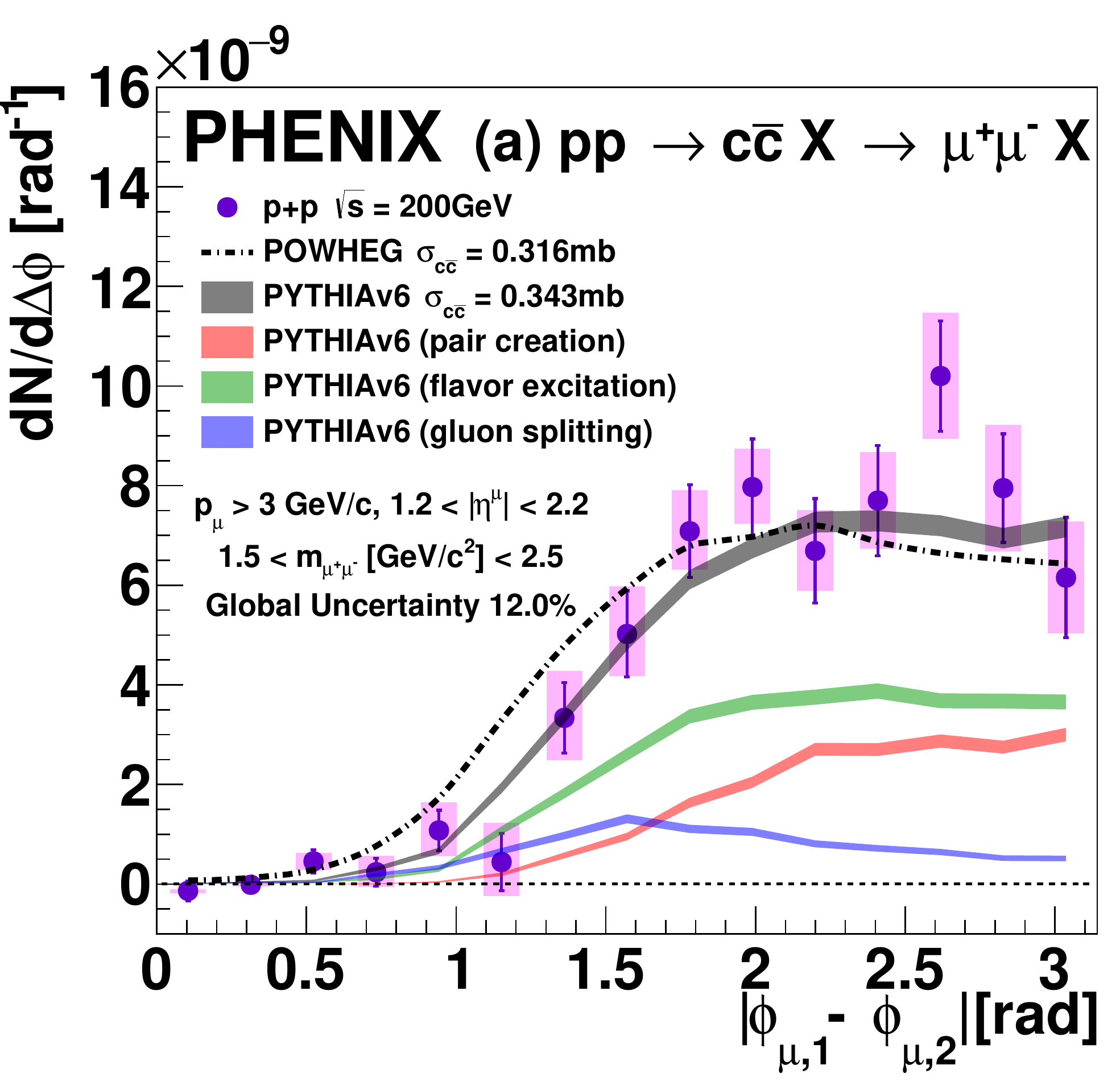}
\includegraphics[width=0.48\linewidth]{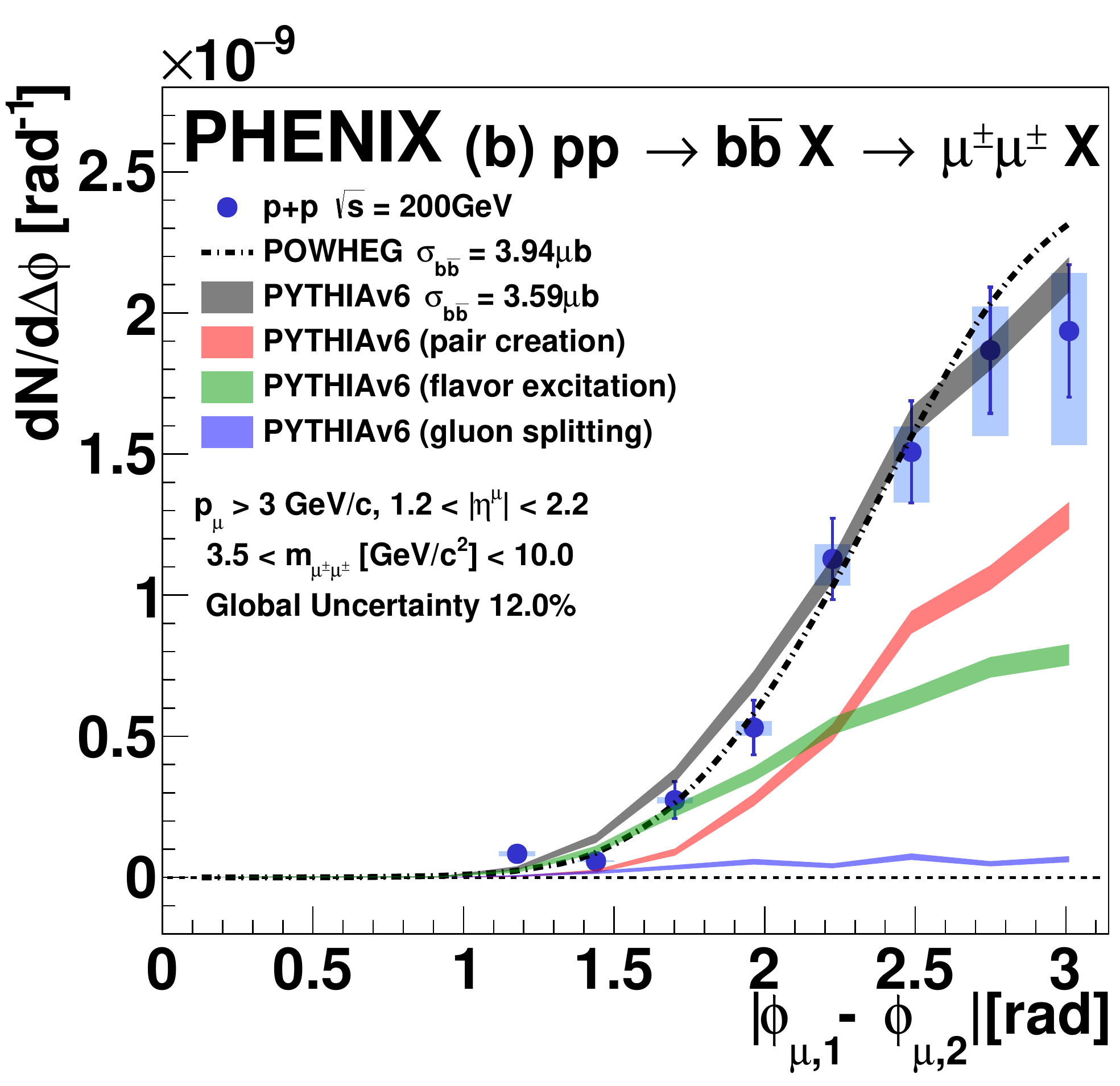}
\caption{\label{Fig:hfdphiarm2} The corrected \mumu yield as a function 
of azimuthal opening angle from (a) charm and (b) bottom decays. The 
data are compared to the distributions calculated with \powheg and 
\pythia. The model calculations are normalized to the data (see text for 
details). For \pythia the \mumu pair yield is broken down into 
contributions from pair creation, flavor excitation, and gluon 
splitting. 
}
\end{figure*}

\begin{figure*}
\includegraphics[width=0.48\linewidth]{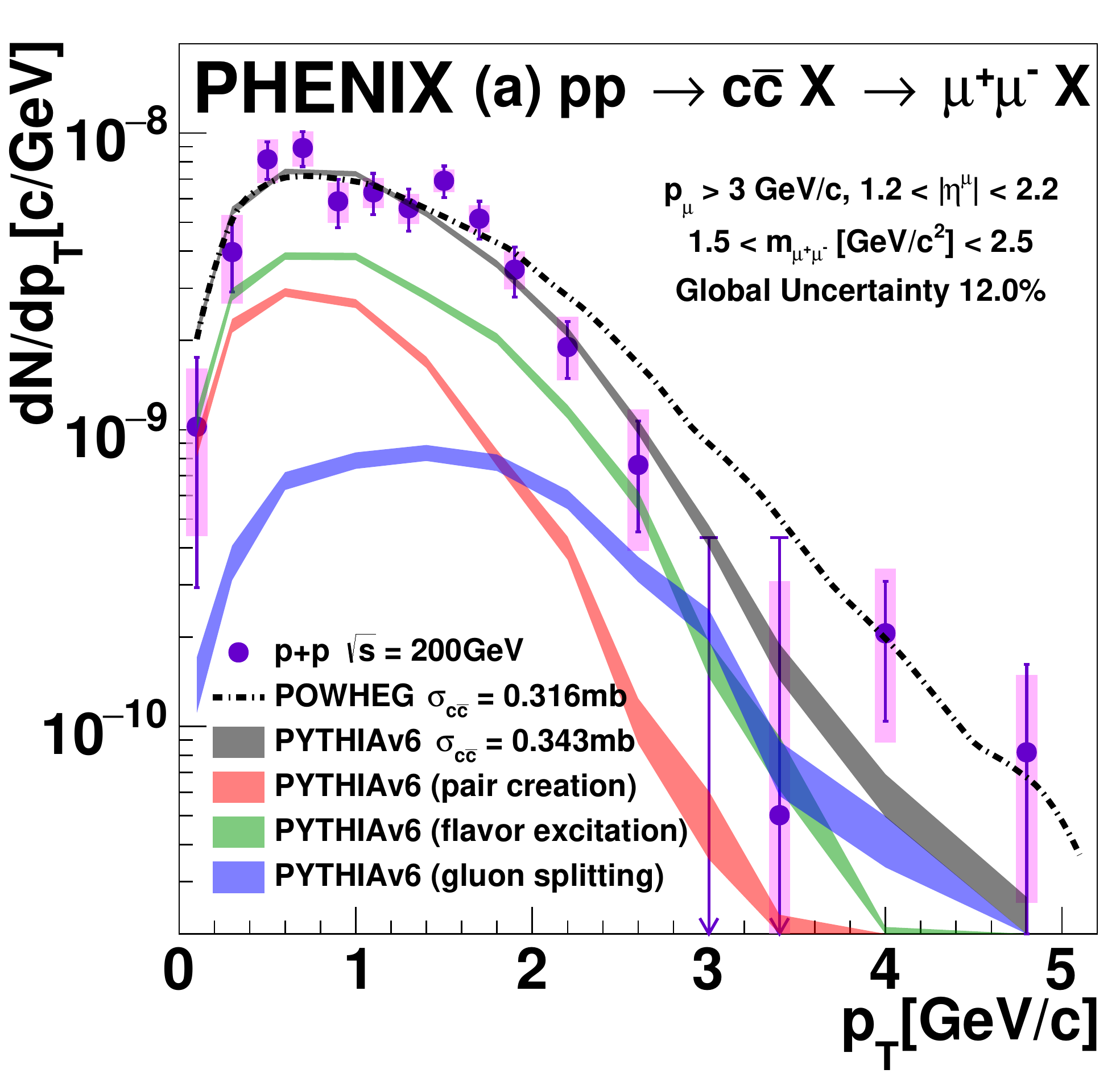}
\includegraphics[width=0.48\linewidth]{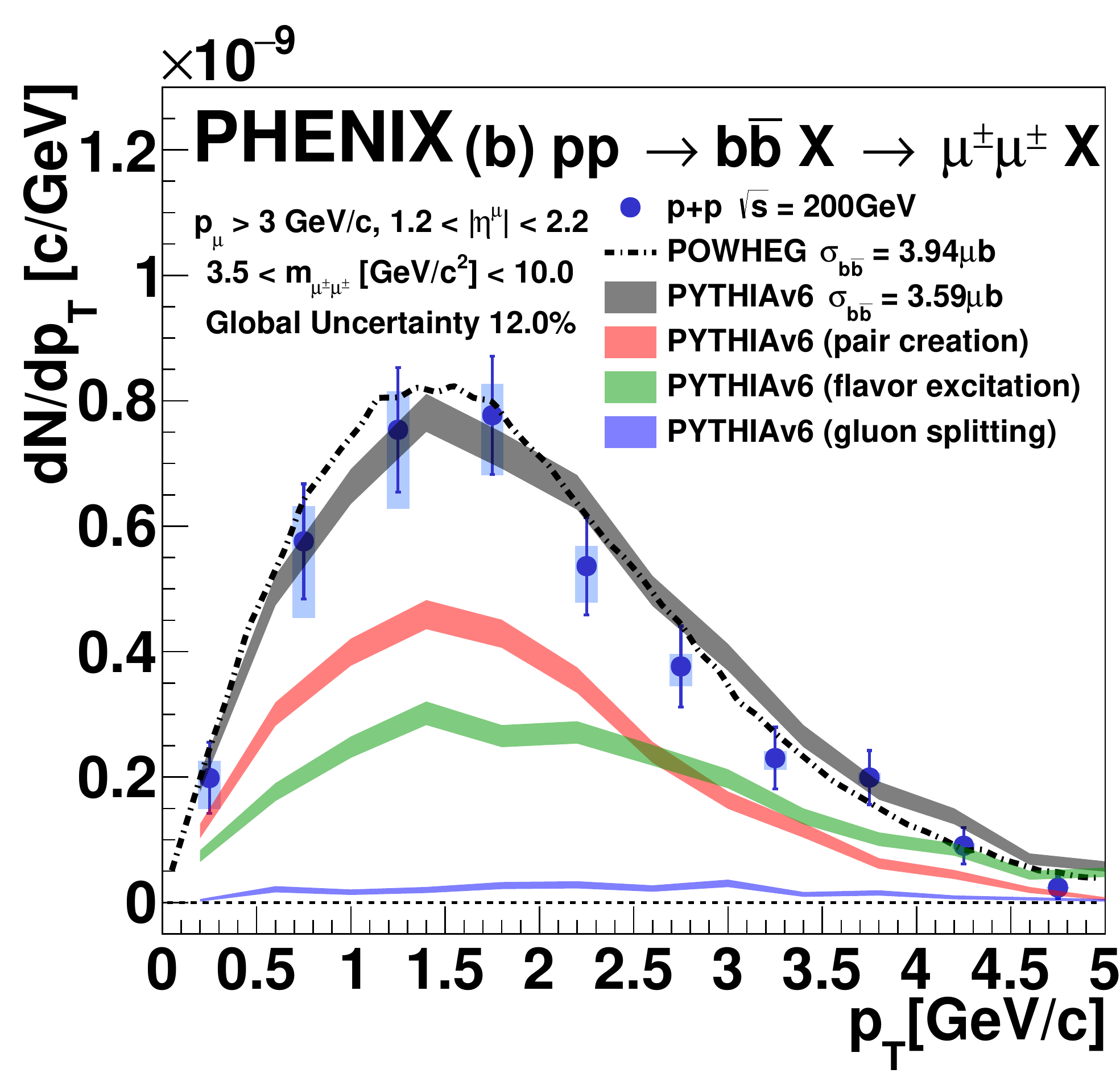}
\caption{\label{Fig:hfptarm2} The corrected \mumu yield as a function of 
pair $p_{T}$ from (a) charm and (b) bottom decays. Presentation of the 
comparison to \powheg and \pythia is the same as 
Fig.~\ref{Fig:hfdphiarm2}. The upper limits on panel (a) indicate 95\% 
confidence level (For a data point with value $d$ and statistical 
uncertainty $\sigma$, the upper limit $u$ is determined by the following 
relation: $\int_{0}^{u}f/\int_{0}^{\infty}f=0.95$, where $f$ is a 
Gaussian distribution with mean $d$ and width $\sigma$.)
}
\end{figure*}

The spectra for the two pseudorapidity regions are combined using the 
method documented in Appendix~\ref{app:average} and compared to model 
calculations based on \pythia and \powheg. The comparison is shown in 
Figs.~\ref{Fig:hfdphiarm2} and~\ref{Fig:hfptarm2}. Pairs generated by 
the models are filtered with the same kinematic cuts that are applied in 
the data analysis.  The model curves are normalized using the fitting 
procedure outlined in Sec.~\ref{Sec:fit}.

For \cc the model calculations are normalized in the kinematic region 
$1.4< m <2.5$ \gevcc and $p_{T}<2$ \gevc to the data. Consequently, as 
seen in Fig.~\ref{Fig:hfptarm2}, the $p_{T}$ spectrum is adequately 
described by both \pythia and \powheg for $p_{T}<2$ \gevc. However, for 
$p_{T}>2$ \gevc, the yield predicted by \powheg is systematically higher 
than the data, while the yield from \pythia is more consistent with the 
data.

The larger yield predicted by \powheg also manifests itself in the 
$\Delta\phi$ projection at $\Delta\phi<1.5$. For \cc, the azimuthal 
correlation determined with \powheg is significantly wider compared to 
the one from \pythia. Again the data favor \pythia in the probed 
kinematic region. This is particularly apparent at $\Delta\phi<\pi/2$.

Because both \pythia and \powheg use the \pythia fragmentation scheme 
and very similar parton distribution functions, the differences between 
the model calculations must result from the underlying correlation 
between the $c$ and $\bar{c}$ quarks that originate from the pQCD 
differential-cross-section calculation. Our data are more consistent 
with \pythia than with \powheg. We note that this preference is not 
limited to data taken in the kinematic region accessible in this 
analysis; it also holds true for the mid-forward kinematic region probed 
by the PHENIX electron-muon measurement \cite{Adare:2013xlp} and mid-mid 
kinematic region probed by the PHENIX dielectron measurement 
\cite{Adare:2017caq}.

For \bb, \pythia shows a slightly wider peak in $\Delta\phi$ than 
\powheg. However, within uncertainties the data are well described by 
both generators in $\Delta\phi$ and $p_{T}$. The smaller model 
dependence can be traced back to the larger $b$ quark mass, which is 
much larger than the muon mass \cite{Adare:2017caq}. For the bulk of $B$ 
meson decays, the momentum of the muon is nearly uncorrelated to the 
momentum of the decay muon. Therefore, the opening angle between two 
muons from \bb is randomized.  In other words, the distributions of 
\mumu pairs from \bb are mostly determined by the decay kinematics and 
are less sensitive to the correlation between the $b$ and $\bar{b}$ 
quark.

For the \pythia calculation we can distinguish heavy flavor production 
from different processes, specifically pair creation, flavor excitation, 
and gluon splitting. To separate these we access the ancestry 
information using the \pythia event record. Despite the fact that the 
measured azimuthal opening angle and pair $p_{T}$ distributions are 
constrained due to the limited acceptance and the mass selection, there 
are clear differences between the shapes generated by different 
processes. The leading order pair creation features a strong 
back-to-back peak, whereas next-to-leading-order processes exhibit much 
broader distributions. For \bb, \pythia predicts negligible contribution 
from gluon splitting, whereas for \cc, there is significant contribution 
from gluon splitting, particularly for $\Delta\phi<1$ and $p_{T}>3$ 
\gevc. For both \cc and \bb, the default ratios and shapes of the three 
different processes from \pythia describe the data well.

Although for \powheg a similar separation is not possible, it seems as 
if contributions from higher order processes with characteristics 
similar to gluon splitting are more frequent in \powheg than in \pythia, 
leading to a broader azimuthal opening angle distribution and a harder 
\pt spectrum for pairs from \cc. More constraints on the \cc 
correlations, which seem to drive the observed model differences, could 
be obtained from a quantitative and systematic study of heavy flavor 
correlations for \pp collisions at $\sqrt{s}=200$ GeV obtained from 
different kinematic regions. A simultaneous analysis of the $ee$ 
\cite{Adare:2017caq}, $e\mu$ \cite{Adare:2013xlp} and $\mu\mu$ data can 
provide stronger discriminating power to different theoretical models. 
Such an analysis is presented in \cite{Aidala:2018dqb}.

\section{Bottom cross section}

To determine heavy flavor production cross sections, the \mumu pair data 
need to be extrapolated from the small kinematic region covered by the 
experiment to the full phase space. This extrapolation has to rely on 
model calculations. For the case of charm, there are significant 
discrepancies between the differential distributions calculated by 
different models, hence an extrapolation to full phase space is model 
dependent \cite{Adare:2017caq}.  However, this is less of an issue for 
bottom production. The distributions of \mumu pairs from \bb are 
dominated by decay kinematics and model dependent systematic 
uncertainties on the extrapolation are much less dominant. In the 
following we determine the average of the bottom cross sections obtained 
from \pythia and \powheg using the fitting procedure, and assign 
systematic uncertainties according to the difference between models.

\begin{figure}[htb]
\includegraphics[width=1.0\linewidth]{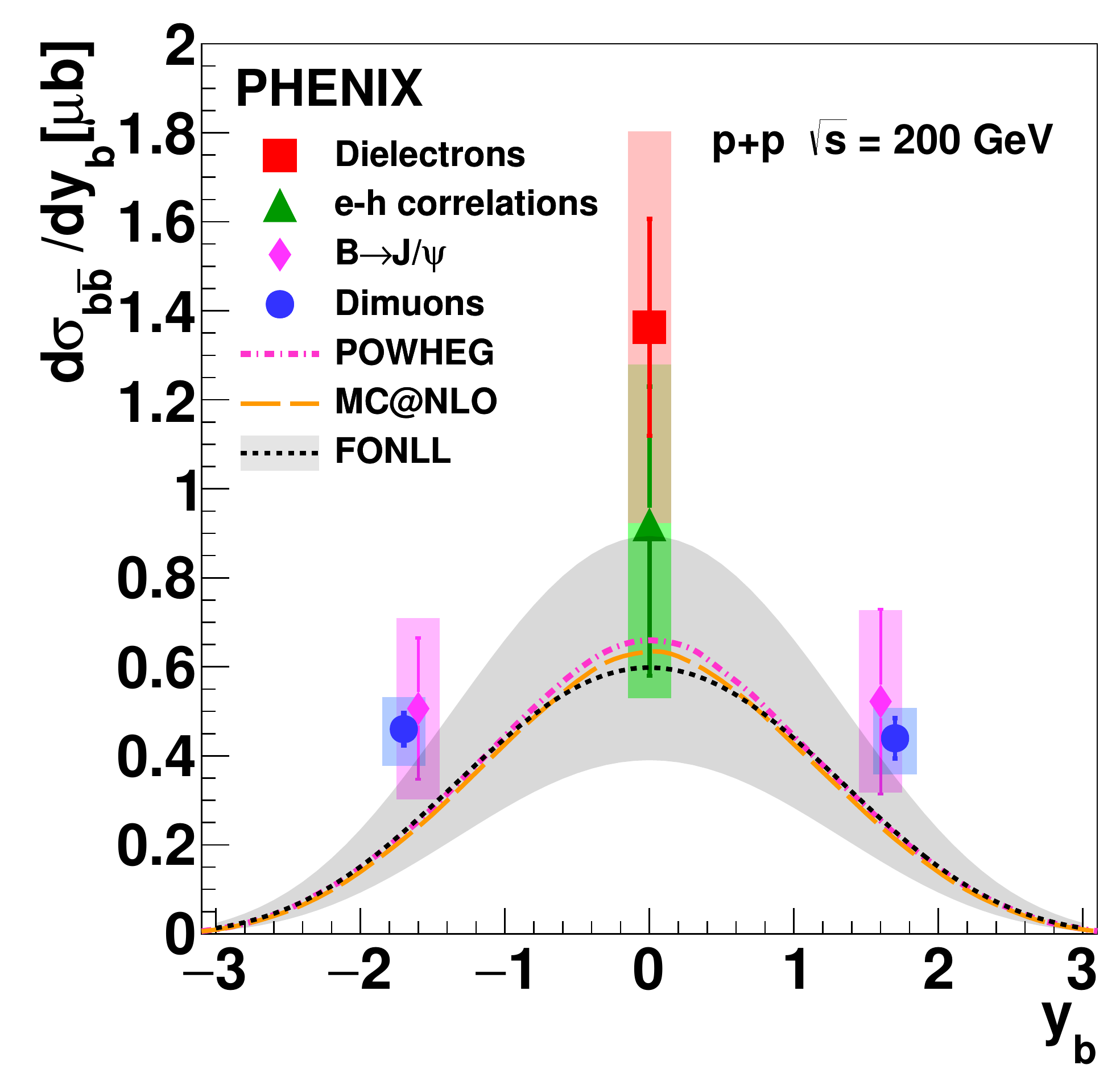}
\caption{ \label{Fig:bbdndy} Rapidity density 
$d\sigma_{b\bar{b}}/dy_{b}$ in \pp collisions at $\sqrt{s}=200$ GeV 
measured in PHENIX via various channels compared to theoretical 
calculations. Here $y_{b}$ is the rapidity of a $b$ quark.}
\end{figure}

The extracted cross sections using \pythia and \powheg are listed in 
Table.~\ref{Tab:bottomxsec}. The first two columns display the 
cross sections obtained by fitting data from the south and north muon 
arm at backward and forward rapidity, respectively. These values are 
then converted rapidity $d\sigma_{b\bar{b}}/dy$ at $y=-1.7$ and $y=+1.7$,
corresponding to the average rapidity of the south and north muon arms.

\begin{table}[tbh]
\caption{\label{Tab:bottomxsec}$\sigma_{b\bar{b}}$ from fit using 
different models. Only statistical uncertainties are shown.}
\begin{ruledtabular} \begin{tabular}{cccc}
& south & north & combined\\
\hline
\pythia  \sigmabb[$\mu$b] & $3.71 \pm 0.29$ & $3.42 \pm 0.35$ & $3.59 \pm 0.22$\\
\powheg \sigmabb[$\mu$b] & $3.94 \pm 0.31$ & $3.94 \pm 0.40$ & $3.94 \pm 0.25$\\
    \\
average \sigmabb[$\mu$b] & $3.82 \pm 0.30$ & $3.65 \pm 0.38$ & $3.75 \pm 0.24 $ \\   
\end{tabular} \end{ruledtabular}
\end{table}

\begin{figure}[htb]
\includegraphics[width=1.0\linewidth]{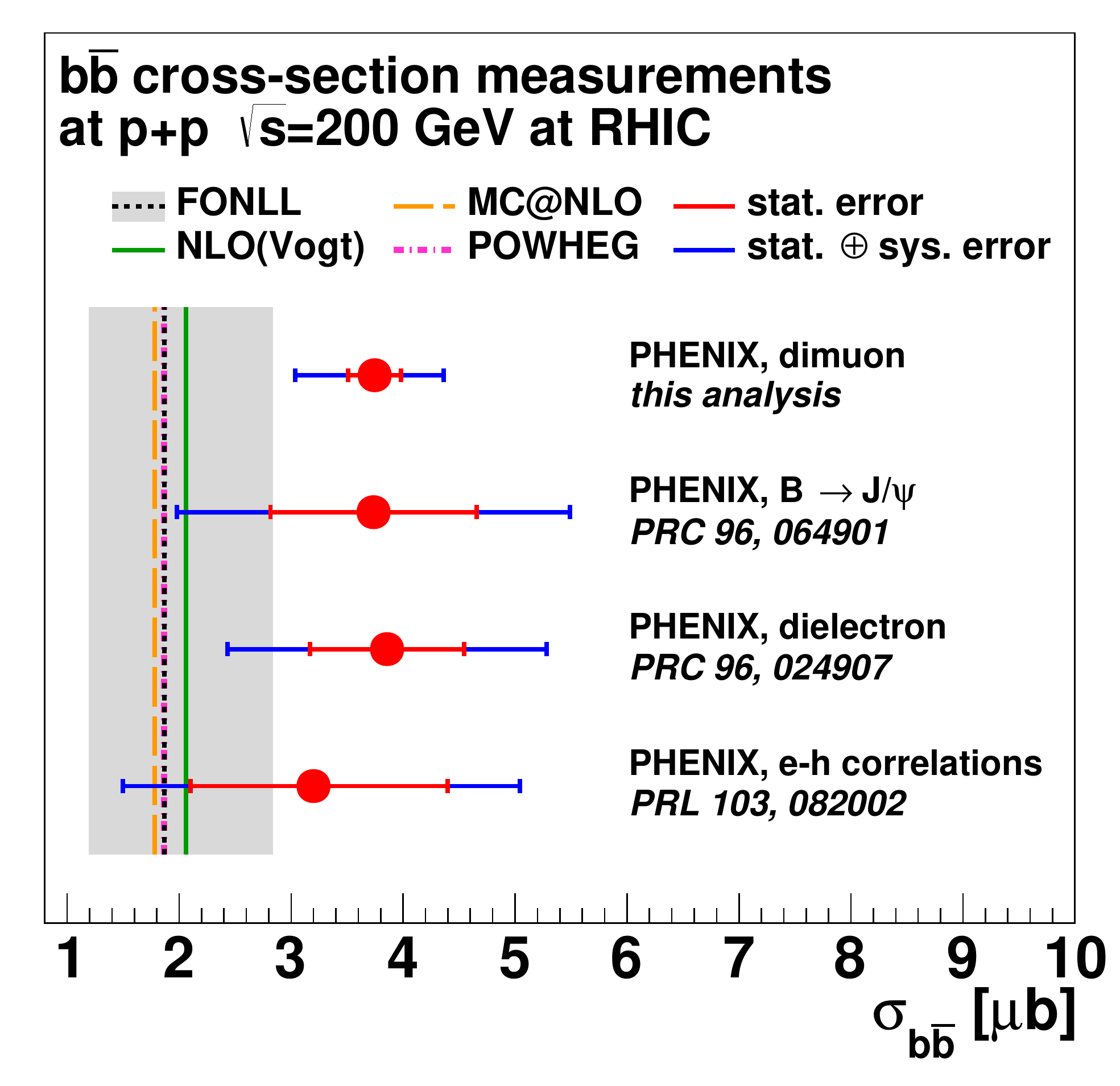}
\caption{ \label{Fig:localbbxsec} Bottom cross section 
$\sigma_{b\bar{b}}$ in \pp collisions at $\sqrt{s}=200$ GeV measured at 
RHIC via various channels compared to NLL and NLO calculations. 
The gray band represents the systematic uncertainty in the FONLL calculation.}
\end{figure}

\begin{figure}[htb]
\includegraphics[width=0.98\linewidth]{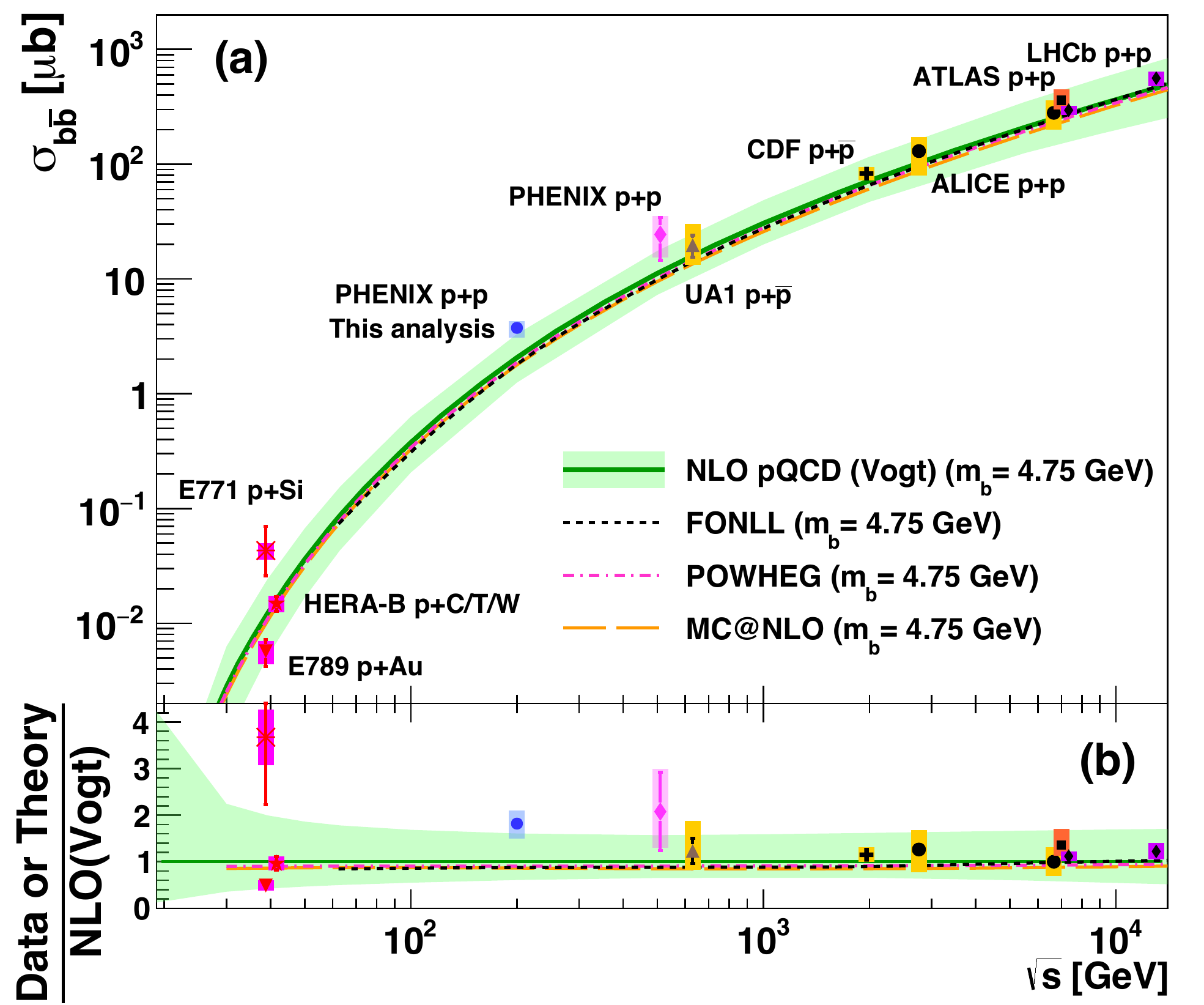}
\caption{ \label{Fig:globalbbxsec} 
Bottom cross section $\sigma_{b\bar{b}}$ as a function of $\sqrt{s}$. 
Uncertainties due to rapidity extrapolation are not included in the LHCb 
measurements. Measured cross sections are compared to NLL and NLO 
calculations.}
\end{figure}

The results are shown in Fig.~\ref{Fig:bbdndy} and compared to other 
PHENIX bottom-cross-section measurements via various channels 
($B\rightarrow J/\psi$~\cite{Aidala:2017yte}, 
dielectrons~\cite{Adare:2017caq}, $e$-$h$ 
correlations~\cite{Adare:2009ic}), and differential cross sections 
computed using fixed-order-plus-next-to-leading-log 
(FONLL)~\cite{Cacciari:1998it}, \mcnlo~\cite{Frixione:2003ei} and 
\powheg~\cite{Frixione:2007nw}. In all three calculations, we adopted 
the ``standard" value of $m_{b}=4.75$ \gevcc~\cite{Vogt:2007aw}. This 
choice of the bottom quark mass is mainly motivated by the mass of 
$\Upsilon(1S)$. It has been shown in previous studies that the NLO pQCD 
calculations with this standard value of $m_{b}$ can reproduce the 
$p$$+A$ and $\pi$+$p$ bottom cross sections at low energies fairly well 
to within large experimental and theoretical 
uncertainties~\cite{Vogt:2002ve}. The large theoretical uncertainties 
arise from the renormalization and factorization scale, bottom quark 
mass and PDF choices. We observe that the model dependence on the 
differential bottom cross section as a function of rapidity is small 
($<10$\%); it is mainly due to the uncertainties in the PDFs. The shaded 
band correspond to theoretical uncertainties estimated using a FONLL 
calculation, which includes uncertainties on the renormalization and 
factorization scales, bottom quark mass (varied between $4.5$ and $5.0$ 
\gevcc), and PDFs, added in quadrature. The measurements at 
$\sqrt{s}=$200 GeV tend to prefer the upper limit of this uncertainty 
band.

The measurements using the two muon arms can be combined to give a more 
precise measurement of the total bottom cross section, 
$\sigma_{b\bar{b}}[\mu b] = 3.75 \pm 0.24 (stat) \pm ^{0.35}_{0.50} 
(syst) \pm 0.45 (global)$, which is the most precise measurement of the 
bottom cross section at $\sqrt{s}=200$ GeV to date. In 
Fig.~\ref{Fig:localbbxsec}, our measurement is compared to all other 
RHIC measurements.

As can be seen from Figs.~\ref{Fig:bbdndy} and~\ref{Fig:localbbxsec}, 
all RHIC bottom-cross-section measurements are remarkably consistent 
with each other. We compare to the total cross sections from various 
next-to-leading order (NLO) or next-to-leading logarithmic (NLL) 
calculations, including the NLO calculation from 
Ref.~\cite{Vogt:2007aw}, again using the value $m_{b}=4.75$ \gevcc for 
the bottom quark mass. The total bottom cross section is around a factor 
of two higher than all theoretical calculations with $m_{b}=4.75$ 
\gevcc.

These measurements can be compared to the global trend of the \bb cross 
section as a function of $\sqrt{s}$ \cite{Alexopoulos:1997zx, 
Jansen:1994bz, Abt:2005qs, Acosta:2004yw, Abelev:2012sca, 
Albajar:1990zu, Aaij:2016avz, Aidala:2017iad, Abelev:2014hla, 
Aad:2012jga}, as shown in Fig.~\ref{Fig:globalbbxsec}. Interestingly, 
the variation of different theoretical calculations is less than 8\% 
despite spanning 5 orders of magnitude in cross section and 3 orders of 
magnitude in beam energy. At beam energies larger than 2 TeV, the data 
points from the Tevatron and LHC are in good agreement with the central 
values of the theoretical calculations, in contrast to measurements at 
$\sqrt{s}=200$ GeV at RHIC. Following the unconstrained averaging 
procedure adopted by the PDG ~\cite{Patrignani:2016xqp}, the weighted 
average of the $\sigma_{b\bar{b}}$ measurements at RHIC is $3.8\pm0.5\ 
\mu b$, and is $>3 \sigma$ higher than the theoretical central values 
(see Fig.~\ref{Fig:localbbxsec}). This may suggest that while the 
current central/default settings of these theoretical calculations may 
reasonably describe bottom cross sections at high beam energies, they 
fail to describe the cross section at $\sqrt{s}=200$ GeV.

An input bottom quark mass $m_{b}=4.12\pm 0.11$ \gevcc is required for 
\powheg to reproduce the bottom cross section measured at $\sqrt{s}=200$ 
GeV. This mass is significantly lower than the pole mass of the bottom 
quark, 4.78 \gevcc~\cite{Patrignani:2016xqp}, hence it is unlikely that 
this discrepancy can be explained solely by the uncertainty in the 
bottom quark mass.

This measurement indicates that an effect which is more visible at lower 
beam energies may still be missing in current theoretical calculations. 
Future measurements at beam energies between $\sim10$ GeV and $\sim1000$ 
GeV with higher precision should help shed light on this issue.

\subsection{Drell-Yan differential cross section}

\begin{figure}
\includegraphics[width=1.0\linewidth]{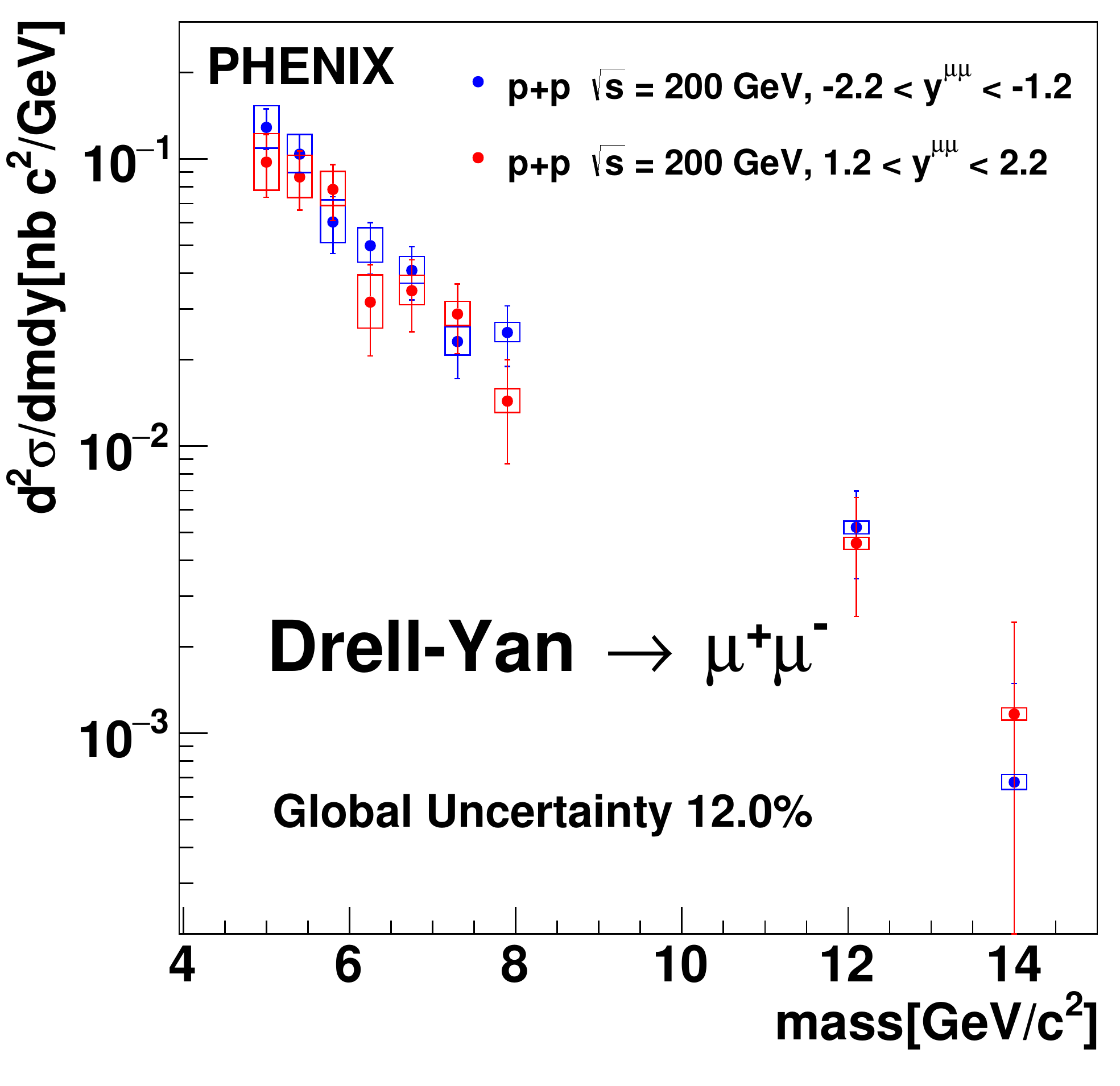}
\caption{\label{Fig:dyyieldarm01} The corrected \mumu yield from 
Drell-Yan in pair rapidity region $1.2<|y^{\mumu}|<2.2$ as a function or 
pair mass. Results are shown separately for the south and north muon 
arms.}
\end{figure}

\begin{figure}
\includegraphics[width=1.0\linewidth]{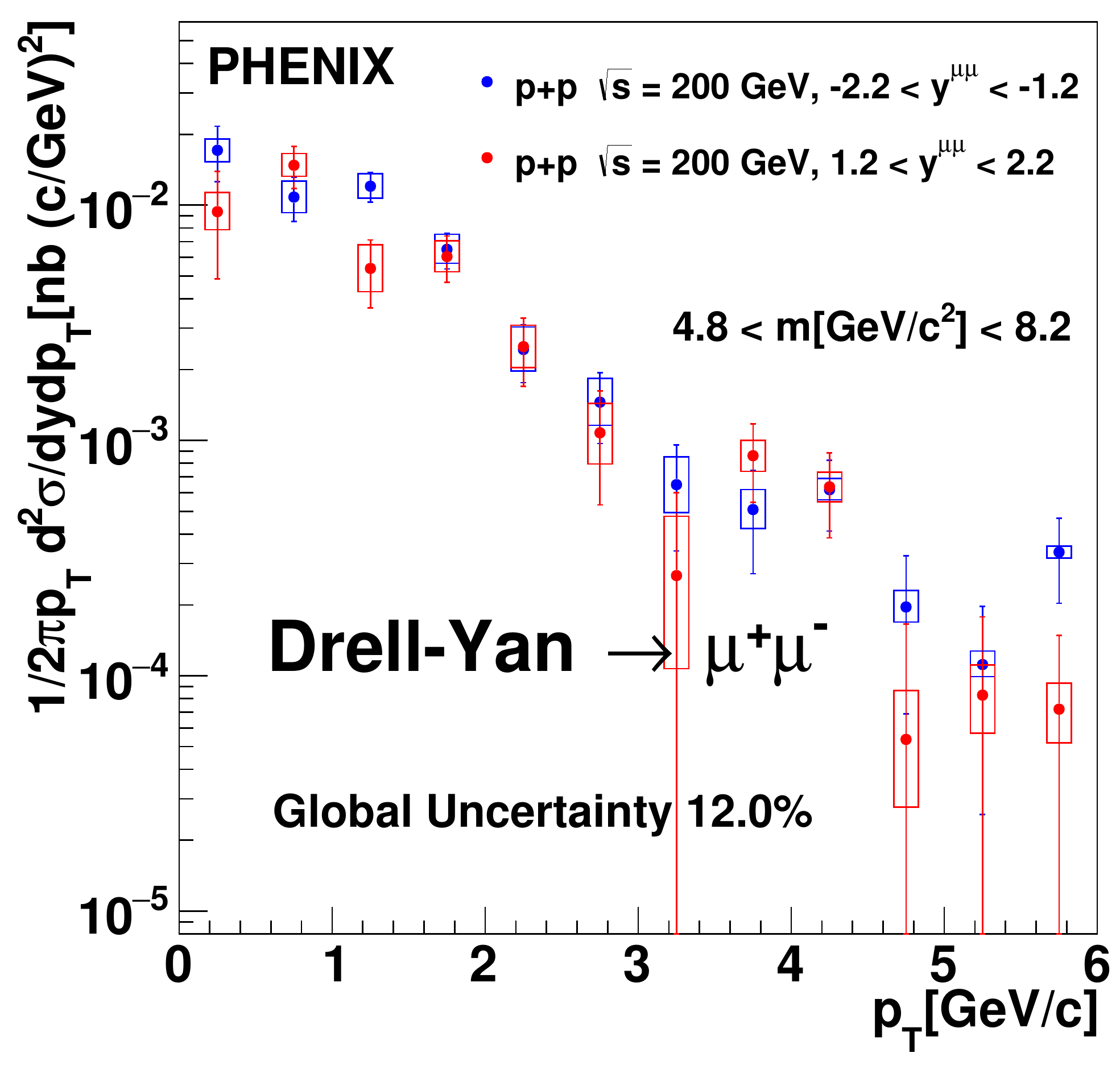}
\caption{\label{Fig:dyptyieldarm01} The corrected \mumu yield from 
Drell-Yan in pair rapidity region $1.2<|y^{\mumu}|<2.2$ and mass region 
$4.8<m <8.2$ \gevcc as a function of pair $p_{T}$. Results are shown 
separately for the south and north muon arms.}
\end{figure}

The fully corrected \mumu pair cross section from the Drell-Yan process 
in the pair rapidity region \mbox{$1.2<|y^{\mumu}|<2.2$}, as a function 
of mass, and a function of $p_{T}$ for pairs in the mass region $4.8<m$ 
[\gevcc]$<8.2$ are shown in Figs.~\ref{Fig:dyyieldarm01} 
and~\ref{Fig:dyptyieldarm01}, respectively. The kinematic region covered 
by the measurement corresponds to a Bjorken-$x$ value of 
$\approx5\times10^{-3}$. The measured differential Drell-Yan cross 
section at forward and backward rapidities are consistent with each 
other.

\begin{figure}
\includegraphics[width=1.0\linewidth]{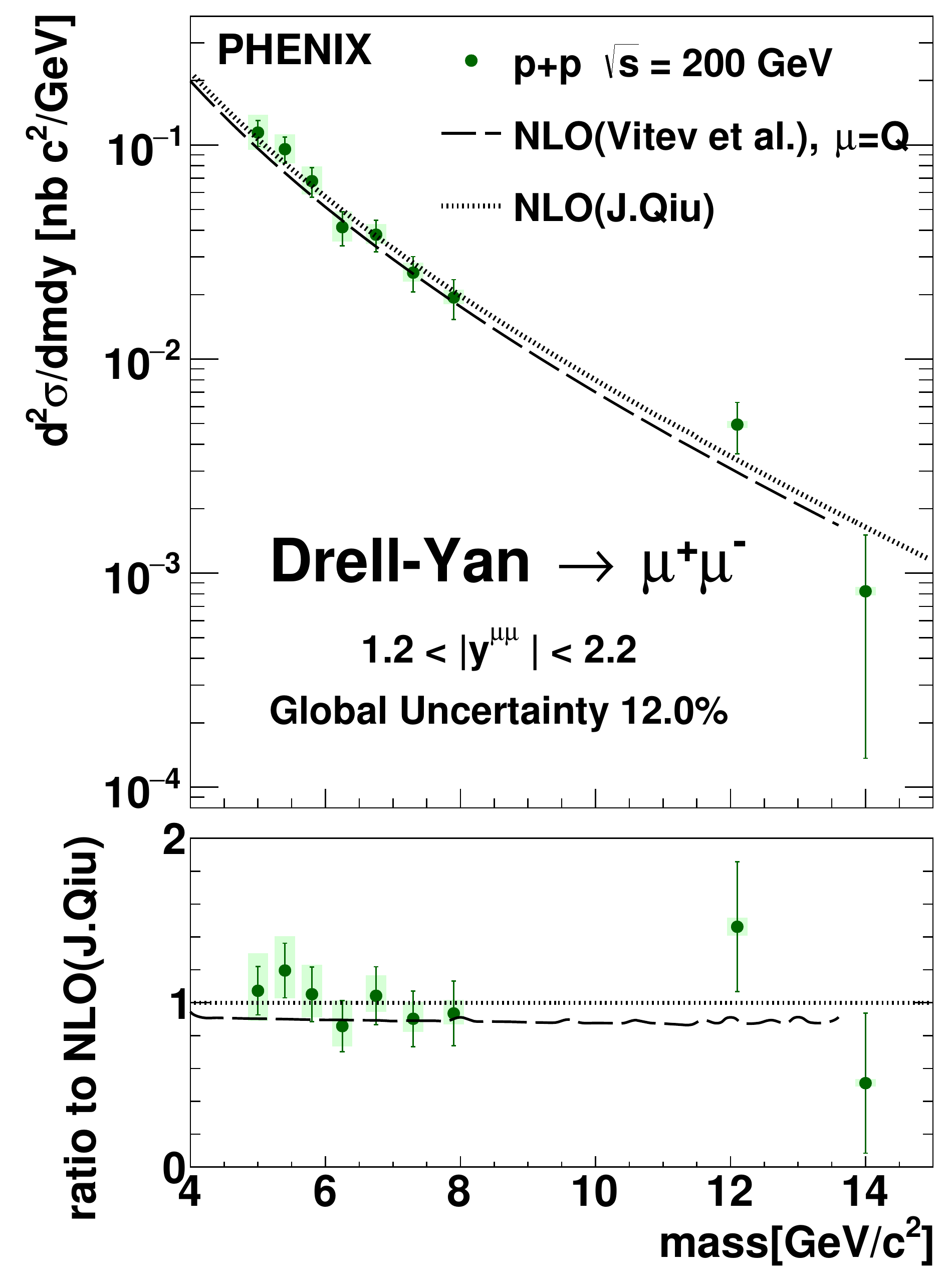}
\caption{\label{Fig:dyyieldarm2} Panel (a) shows the corrected \mumu 
yield from Drell-Yan in pair rapidity region $1.2<|y^{\mumu}|<2.2$. Data 
are compared to NLO calculations. Panel (b) gives the ratio of the data 
to one of the NLO calculations. }
\end{figure}

\begin{figure}
\includegraphics[width=1.0\linewidth]{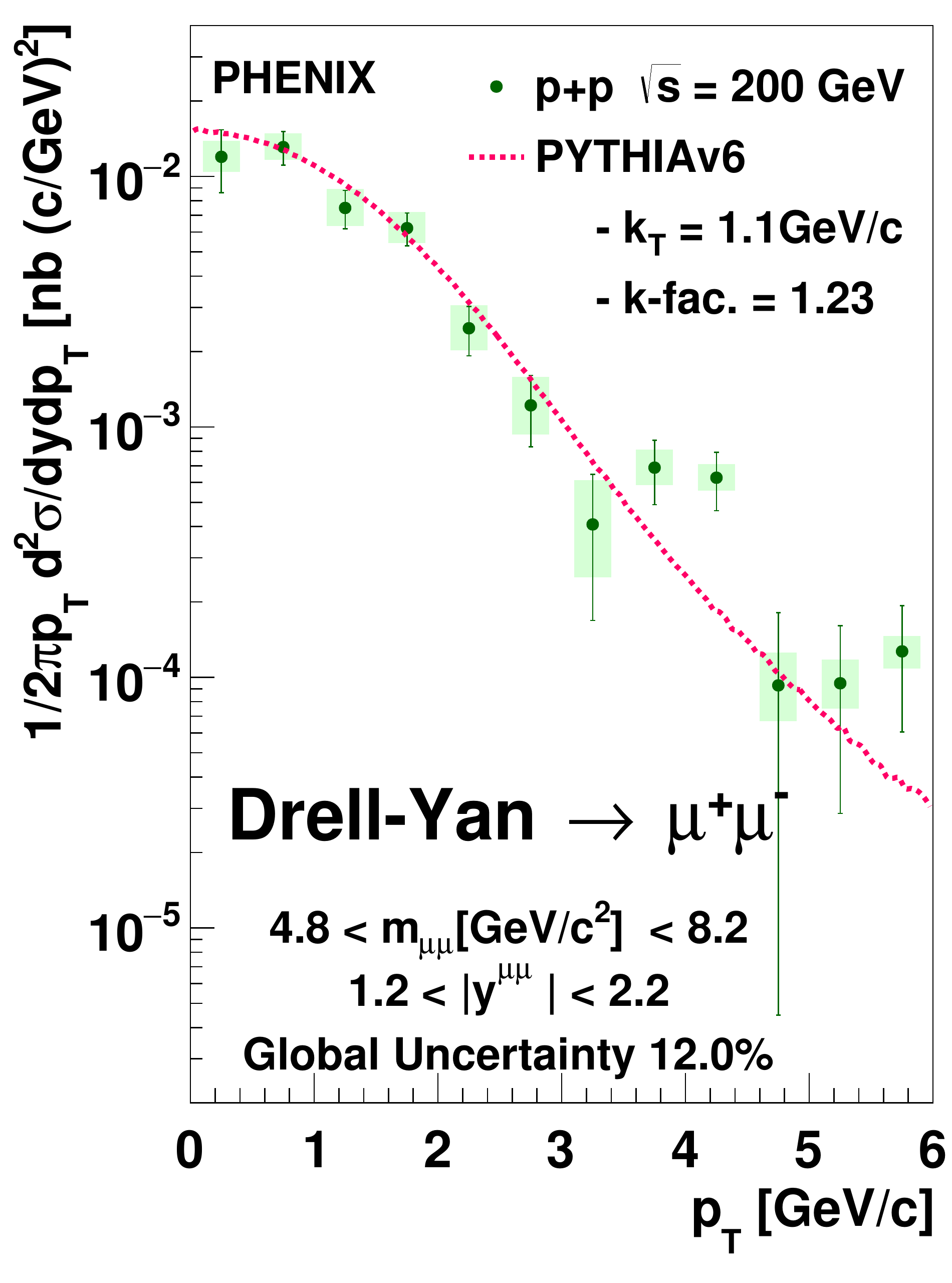}
\caption{\label{Fig:dyptyieldarm2} The corrected \mumu yield from 
Drell-Yan in pair rapidity region $1.2<|y^{\mumu}|<2.2$ and mass region 
$4.8<m<8.2$ \gevcc as a function of pair $p_{T}$. Data are compared 
\pythia calculations under settings used for this analysis.}
\end{figure}

We combine the measurements from the two rapidity regions. The mass 
spectrum is then compared with NLO calculations from Vitev 
\cite{Neufeld:2010dz} and Qiu J. et al \cite{Fai:2004tg} in 
Fig.~\ref{Fig:dyyieldarm2}. Both calculations adopt the factorization 
approach where higher orders are evaluated order-by-order in 
perturbation theory. Within experimental uncertainties, the data are 
well reproduced by NLO calculations. The $p_{T}$ spectrum of Drell-Yan 
muon pairs in the mass region 4.8--8.2 \gevcc is shown in 
Fig.~\ref{Fig:dyptyieldarm2} and compared to \pythia, where the 
intrinsic $k_{T}$ is tuned from the procedure described in 
\ref{Sec:dyktsys}, and normalized from the fitting procedure as 
documented in the above text. We find that an intrinsic $k_{T}$ of 1.1 
\gevc and a k-factor of 1.23 best describe the data. To date this is the 
first Drell-Yan measurement at RHIC energies. As Drell-Yan is a common 
background to various physics processes involving dileptons, the 
presented data may give a constraint for the background estimation of 
such measurements. The Drell-Yan cross section as a function of 
invariant mass and $p_{T}$ can also provide constraints on the 
unpolarized transverse-momentum-dependent parton distribution functions 
(TMD PDFs), which is of critical importance to understanding the 
internal structure of the proton. This measurement gives input to a 
previously unexplored phase space and serves as a solid baseline for 
future measurements.

\section{Summary}
\label{Sec:summary}

We present \mumu pair measurements from open heavy flavor decays and the 
Drell-Yan mechanism in \pp collisions at $\sqrt{s}=200$ GeV.

Invariant yields of \mumu pairs from \cc and \bb are measured as a 
function of $\Delta\phi$ and $p_{T}$ and compared to different models, 
\pythia and \powheg. Within experimental uncertainties, the azimuthal 
opening angle and pair $p_{T}$ distributions from \bb are well described 
by these models. For \cc, the data favor the \pythia description, while 
the \powheg calculations predict a systematically higher yield than 
\pythia at smaller opening angles in the probed kinematic region.

We find that the high mass like-sign pairs are dominated by decays from 
open bottom, which provides a strong constraint to the bottom cross 
section. The measured total bottom cross section is consistent with RHIC 
measurements at the same energy, and is around a factor of two higher 
than the central value of NLL and NLO calculations with an input bottom 
quark mass of $m_{b}=4.75$ \gevcc.

The Drell-Yan cross section as a function of mass in 4.8--15.0 \gevcc 
is presented and compared to NLO calculations from Vitev and Qiu. Within 
uncertainties we find good agreement between NLO calculations and data. 
The Drell-Yan $p_{T}$ cross section in the mass region 4.8--8.2 \gevcc 
is also presented, along with the \pythia tune that best describes the 
data.

\section*{Acknowledgements}


We thank the staff of the Collider-Accelerator and Physics
Departments at Brookhaven National Laboratory and the staff of
the other PHENIX participating institutions for their vital
contributions.  We acknowledge support from the 
Office of Nuclear Physics in the
Office of Science of the Department of Energy,
the National Science Foundation, 
Abilene Christian University Research Council, 
Research Foundation of SUNY, and
Dean of the College of Arts and Sciences, Vanderbilt University 
(U.S.A),
Ministry of Education, Culture, Sports, Science, and Technology
and the Japan Society for the Promotion of Science (Japan),
Conselho Nacional de Desenvolvimento Cient\'{\i}fico e
Tecnol{\'o}gico and Funda\c c{\~a}o de Amparo {\`a} Pesquisa do
Estado de S{\~a}o Paulo (Brazil),
Natural Science Foundation of China (People's Republic of China),
Croatian Science Foundation and
Ministry of Science and Education (Croatia),
Ministry of Education, Youth and Sports (Czech Republic),
Centre National de la Recherche Scientifique, Commissariat
{\`a} l'{\'E}nergie Atomique, and Institut National de Physique
Nucl{\'e}aire et de Physique des Particules (France),
Bundesministerium f\"ur Bildung und Forschung, Deutscher
Akademischer Austausch Dienst, and Alexander von Humboldt Stiftung (Germany),
J. Bolyai Research Scholarship, EFOP, the New National Excellence
Program ({\'U}NKP), NKFIH, and OTKA (Hungary),
Department of Atomic Energy and Department of Science and Technology (India), 
Israel Science Foundation (Israel), 
Basic Science Research Program through NRF of the Ministry of Education (Korea),
Physics Department, Lahore University of Management Sciences (Pakistan),
Ministry of Education and Science, Russian Academy of Sciences,
Federal Agency of Atomic Energy (Russia),
VR and Wallenberg Foundation (Sweden), 
the U.S. Civilian Research and Development Foundation for the
Independent States of the Former Soviet Union, 
the Hungarian American Enterprise Scholarship Fund,
the US-Hungarian Fulbright Foundation,
and the US-Israel Binational Science Foundation.

\appendix

\section{DETAILED DESCRIPTION OF SIMULATION FRAMEWORKS}
\label{app:simframework}

\begin{figure}[h]
\includegraphics[width=0.98\linewidth]{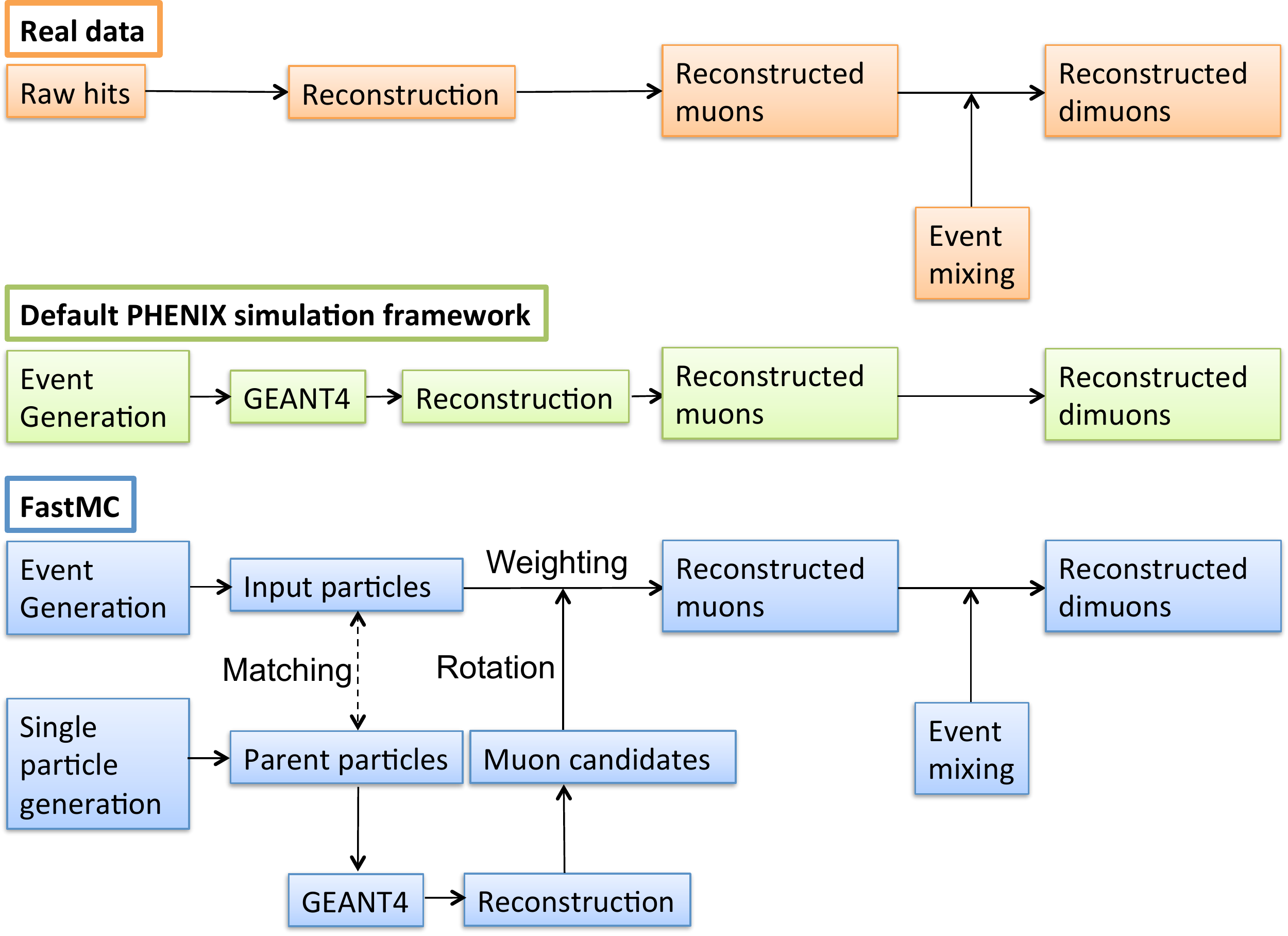}
\caption{\label{Fig:simframework} Flow chart of the analysis chain of the
dimuon reconstruction for real data, default PHENIX simulation framework, 
and FastMC framework. }
\end{figure}

Details of the two simulation chains used in this analysis, namely the 
default PHENIX simulation framework and the fastMC, are discussed in the 
following. The flowchart shown in Fig.~\ref{Fig:simframework} summarizes 
a comparison between the data reconstruction framework and the two 
simulation chains.

\begin{figure}[h]
\includegraphics[width=0.99\linewidth]{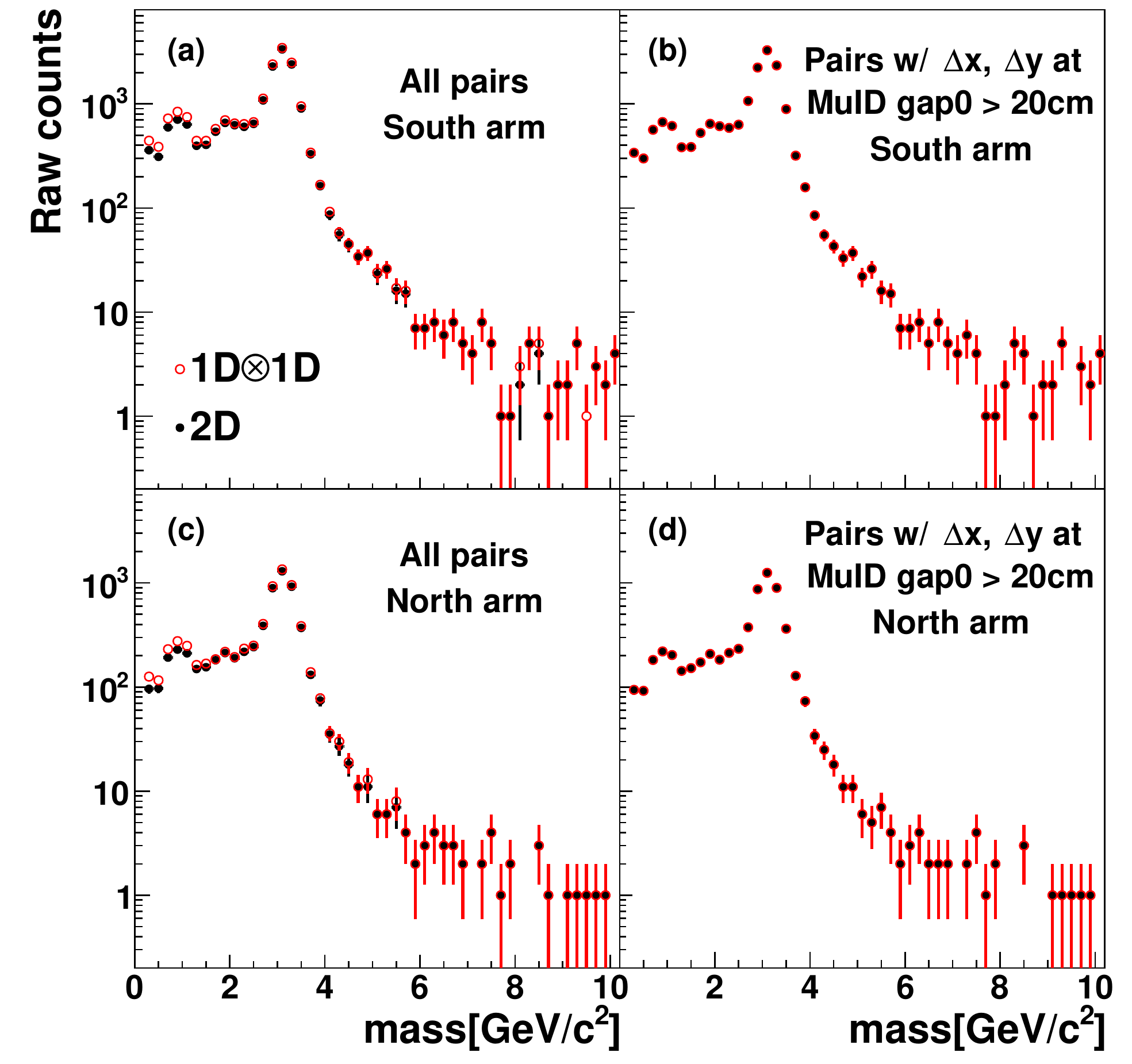}
\caption{\label{Fig:dataanatrig} The mass spectra of MuIDLL1-1D 
triggered data for the (a,b) south arm and (c,d) the north arm are shown 
separately. Open circles are pairs in which both associated tracks 
satisfy the MuIDLL1-1D condition, while closed circles are pairs in 
which the associated tracks satisfy the MuIDLL1-2D condition. Panels 
(a,c) show all pairs, while the panels (b,d) show only pairs with a 
spatial separation exceeding 20~cm at MuID gap~0. }
\end{figure}

\begin{figure}[h]
\includegraphics[width=0.48\linewidth]{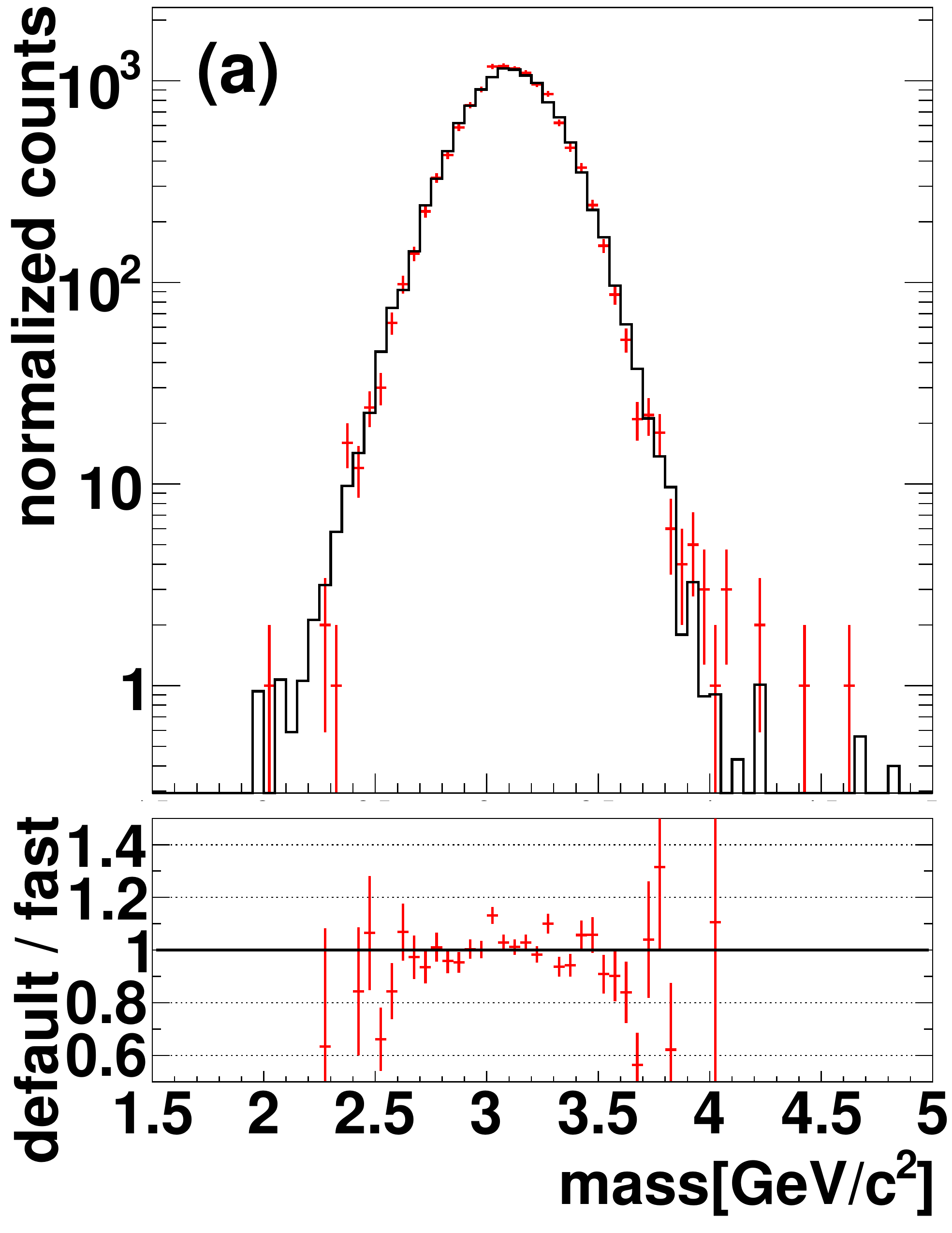}
\includegraphics[width=0.48\linewidth]{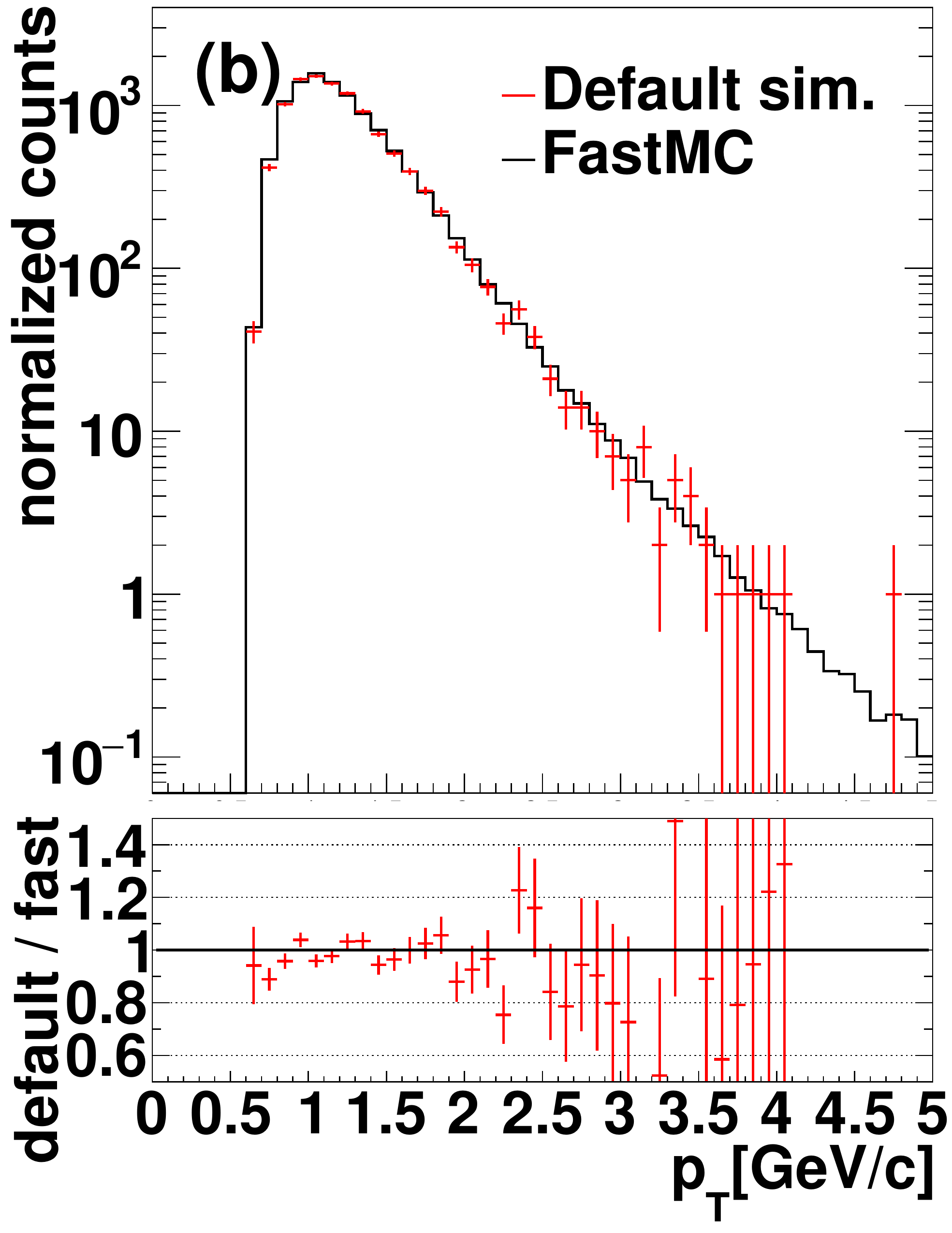}
\includegraphics[width=0.48\linewidth]{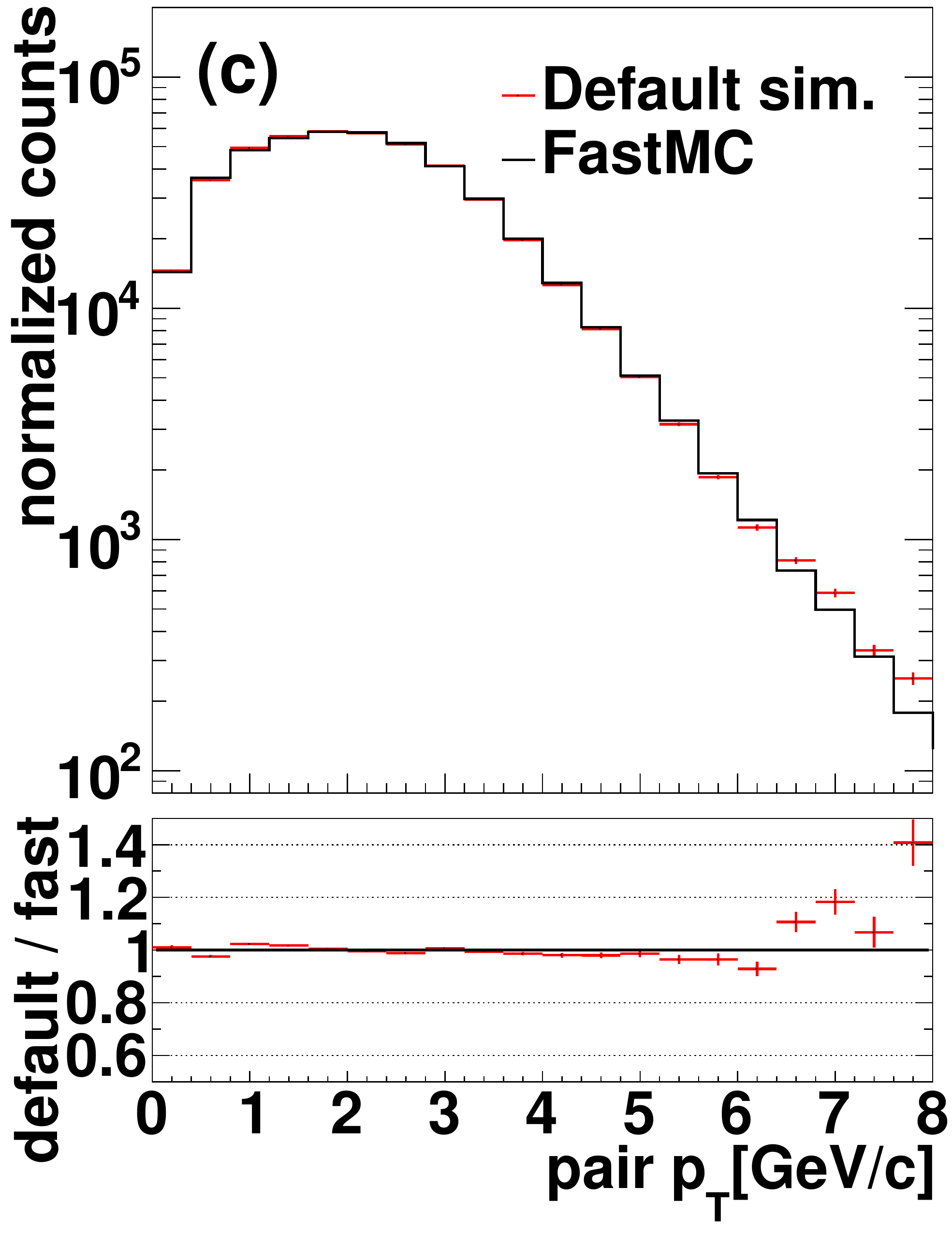}
\includegraphics[width=0.48\linewidth]{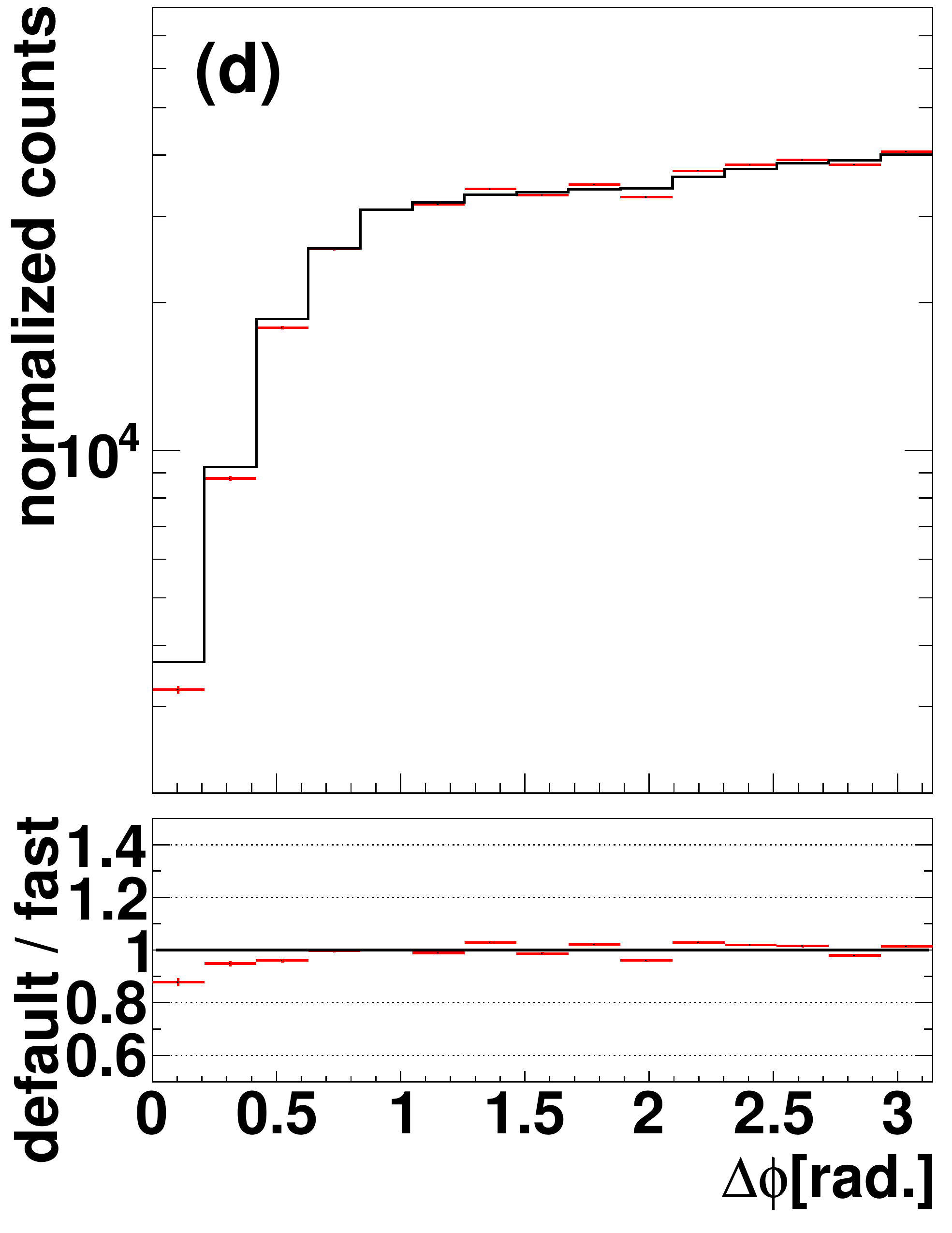}
\caption{\label{Fig:fastmccheck} Comparison of distributions from FastMC 
and default PHENIX simulation framework. (a) mass spectrum of $J/\psi$ 
muon pairs; (b) single $p_{T}$ spectrum of muons from $\pi^{\pm}$ and 
$K^{\pm}$ with realistic input $p_{T}$ spectra; (c) pair $p_{T}$ 
spectrum of muon pairs from \bb; (d) $\Delta\phi$ of muon pairs from 
\bb.}
\end{figure}

\subsection{Default PHENIX simulation framework}

The default PHENIX simulation is based on a detailed {\sc 
geant4}~\cite{Agostinelli:2002hh} implementation of the muon arms. This 
framework takes into account the detector's geometrical acceptance and 
all inefficiencies from dead channels. To account for variations of 
detector performance during the data taking period, the data are split 
into run groups with similar performance. For each group a map of dead 
channels is created for the MuTr. The simulation randomly selects these 
maps according to the sampled luminosity for each run group.

Muon pairs from physical sources are simulated with a \z-vertex 
distribution taken from MB \pp data. Once the pairs are processed 
through the detector simulation, they are reconstructed using the same 
procedure and filtered with the same cuts as used for real data. Thus, 
all detector effects including acceptance, dead areas, track 
reconstruction, and analysis cuts are taken properly into account.

Because the analyzed data are triggered with the MuIDLL1-2D trigger, the 
effects of the trigger also need to be accounted for. To achieve this, 
we apply an offline software trigger to all simulated tracks, which is 
an exact replication of the online hardware MuIDLL1-1D trigger. We 
require that both tracks of a pair fulfill the MUIDLL1-1D trigger 
condition. Here we make use of the fact that after enforcing a spatial 
separation of 20~cm between two MuID tracks, the MuIDLL1-2D pair trigger 
is reduced to a logical AND of the MuIDLL1-1D single track triggers. The 
separation cut necessary to achieve this factorization was determined 
from experimental data. In Fig.~\ref{Fig:dataanatrig}(a,c)  
a $\sim$20--30\% difference between the mass distribution from data 
triggered with the MuIDLL1-2D and the data requiring each track fulfills 
the MuIDLL1-1D is visible at low masses.  Once the separation cut is 
applied the difference disappears, as seen in panels (b) and (d).

\subsection{FastMC \label{Sec:fastmc}}

In spite of the large hadron rejection power ($\sim1/1000$) of the muon 
arms, a significant fraction of the reconstructed muons are from decays 
of light-flavor mesons ($\pi^{\pm}$, $K^{\pm}$, and $K^{0}$). Using the 
default Monte-Carlo to simulate these pairs is unpractical, because for 
every 1,000,000 generated pairs of particles in the detector acceptance, 
only one muon pair would be reconstructed from the simulation. In the 
FastMC approach we separate the generation of particles that result in 
reconstructed \mumu pairs from the simulation of the detailed detector 
response to an individual particle. The FastMC proceeds in four steps: 
(i) generation of a repository of possible detector responses to an 
individual particle using the default simulation framework, (ii) 
creation of events with multiple muons from the sources discussed in 
Sec.~\ref{Sec:expectedsources}, here the repository created in step (i) 
is used to determine the detector response, (iii) weighting each 
reconstructed muon with the appropriate probability for being 
reconstructed and not rejected by the analysis cuts, and (iv) finally 
forming muon pairs and calculating their mass, \pt and azimuthal opening 
angle.

\subsubsection{Detector response to individual particles}

For each particle species ($\pi^{\pm}$, $K^{\pm}$, $K^{0}$, and 
$\mu^{\pm}$) $\sim10^{9}$ particles were simulated. All particles are 
propagated through the full {\sc geant4} simulation and reconstruction 
chain. Light hadrons ($\pi^{\pm}$, $K^{\pm}$, and $K^{0}$) may give rise 
to a reconstructed muon either via (i) decaying to a muon in flight 
(decay muons), or (ii) penetrating all absorber layers (punch-through 
hadrons).  The contribution from protons is negligible ($<1\%$) compared 
to kaons and pions and hence neglected in this study.

These \textit{parent particles} are generated with flat distribution in 
momentum $p$ and polar angle $\theta$, and uniform distribution in 
$\phi$. Simulations are performed in three uniform \z-vertex regions, 
$(-22.5,-17.5~{\rm cm})$, $(-2.5,2.5~{\rm cm})$, and $(17.5, 22.5~{\rm 
cm})$, to account for variances in detector response along $z_{\rm 
vtx}$. Improvements by expanding to full collision $z_{\rm vtx}$ 
coverage in simulations is expected to be minimal(see 
Sec.~\ref{Sec:sysz}). All reconstructed variables are stored along with 
the generated vertex and parent momentum information. These \textit{muon 
candidates} are grouped into pools according to parent particle species 
and parent $p$ and $\theta$, where $p$ and $\theta$ ranges from 2 to 32 
\gevc and 0 to 0.8 radians respectively, which covers the kinematic 
region relevant for this analysis. One single pool covers the kinematic 
region $\Delta p \times \Delta \theta = 0.1 \times 0.02$ [\gevc rad]. 
The minimum number of muon candidates in one pool is $\sim10$. These 
pools are used as repository for the possible detector response to 
parent particles in the subsequent steps of the FastMC.

\subsubsection{Events with reconstructed muons}

To create an event with reconstructed muons, we first generate events of 
particles as discussed in Sec.~\ref{Sec:corrhadrons}. For each event the 
list of particles is filtered so that only $\pi^{\pm}$, $K^{\pm}$, 
$K^{0}$, and $\mu^{\pm}$ in the vicinity of the muon arm acceptance are 
kept, and the momentum information of these particles is stored. We will 
refer to these particles as \textit{input particles}.

A given input particle is matched to a pool of muon candidates in the 
repository for that particle species, and the input particle's $p$ and 
$\theta$. We randomly choose one muon candidate from the pool and use 
the reconstructed variables from that muon candidate for the input 
particle. The repository pools were generated from parent particles with 
a uniform $\phi$ distribution. While the input particles are matched to 
the muon candidate in parent $p$ and $\theta$, they are not matched in 
$\phi$. We therefore rotate all reconstructed variables in the azimuthal 
plane from the $\phi$ of the parent of the muon candidate to the $\phi$ 
of input particle.

At this point we have created a \textit{reconstructed muon} with all the 
characteristics that could have resulted from propagating the input 
particle through the default simulation framework. In particular, 
because the matching of input particles to muon candidates is completely 
random, the relative contributions and momentum distribution of decay 
muons and punch-through hadrons are properly accounted for. This 
procedure is repeated for all input particles in an event.

\subsubsection{Weighting each reconstructed muon with its probability}

So far each input particle leads to a reconstructed muon. This does not 
take into account the hadron rejection of the muon arms and the 
reconstruction efficiencies. Rejection and efficiency are encapsulated 
in weighting factors that are applied to each reconstructed muon. We 
factorize the weight into two components weight$_{\rm reco}$ and 
weight$_{\phi}$, which are discussed in the following. The final weight 
is calculated as:

\begin{align}
{\rm weight} = {\rm weight}_{\rm reco}\times {\rm weight}_{\phi}.
\end{align}

\subsubsection*{Weighting in $p$ and $\theta$}

The survival probability of a decay muon is highly dependent on the 
momentum of the muon, as well as the amount of material it traverses in 
the absorber, which in turn is dependent on the input particle's 
momentum $p$ and the polar angle $\theta$. We associate a weighting 
factor weight$_{\rm reco}(p,\theta,z)$ to each muon candidate. This factor 
is the probability that an input particle with momentum $p$ and polar 
angle $\theta$, produced at vertex $z$, results in the reconstructed 
muon candidate, averaged over $\phi$. The weight is computed by dividing 
the number of reconstructed muons in each pool by the number of parent 
particles generated to create the corresponding pool.

\subsubsection*{ Weighting in $\phi$}

In addition to weight$_{\rm reco}(p,\theta,z)$, we also need to weight in 
$\phi$ direction, weight$_{\phi}$, to account for the $\phi$ dependent 
relative survival probability and reconstruction efficiency. These 
mainly depend on the geometry of the MuTr, thus the weighting factors 
are determined by a combination of variables $(\phi^{\rm MuTr}, 
p_{T}^{\rm MuTr}, p_{z}^{\rm MuTr})$, which are the azimuthal position, 
transverse momentum, and longitudinal momentum evaluated at MuTr Station 
1. To determine weight$_{\phi}$, we generate single muons with a 
realistic momentum distribution and propagate these muons through the 
default simulation framework. Because the overall survival probability 
is factored into weight$_{\rm reco}$, weight$_{\phi}$ is normalized by 
requiring the average value of weight$_\phi$ to be one, i.e.

\begin{align}
\nonumber
& {\rm weight}_{\phi}(\phi^{\rm MuTr}, p_{T}^{\rm MuTr}, p_{z}^{\rm MuTr})\\
&=\frac{N_{\rm reco}(\phi^{\rm MuTr}, p_{T}^{\rm MuTr}, p_{z}^{\rm MuTr})\int^{+\pi}_{-\pi}d\phi^{\rm MuTr}}{\int^{+\pi}_{-\pi}d\phi^{\rm MuTr}N_{\rm reco}(\phi^{\rm MuTr}, p_{T}^{\rm MuTr}, p_{z}^{\rm MuTr})}.
\end{align}

\subsubsection{Constructing muon pairs}

In each event all reconstructed muons are combined to pairs. The pair 
variables are constructed from the reconstructed muon information 
following the exact same procedure as in real data. The weighting factor 
for a muon pair is the product of the weighting factors of the two 
reconstructed muons:

\begin{align}
{\rm weight}_{12} = {\rm weight}_{1}\times {\rm weight}_{2}.
\end{align}

This assumes that the pair reconstruction efficiency is a product of 
single track reconstruction efficiencies, which is true for tracks that 
are spatially separated in the MuTr and MuID. The latter is assured by 
the pair cuts we apply.

To estimate the accuracy of the FastMC, which is used to simulate 
muon-hadron and hadron-hadron pairs, we propagate \mumu pairs and single 
hadrons through the default simulation framework and FastMC and compared 
the resulting distributions. We find that the mass resolution, 
$\Delta\phi$, single and pair \pt distributions are well reproduced by 
the FastMC (see Fig.~\ref{Fig:fastmccheck}). Small discrepancies are 
observed in the azimuthal opening angle distribution $\Delta\phi$ between 
the two muons for small $\Delta\phi$. This is likely due to the $\phi$ 
weighting procedure. The related systematic uncertainties will be 
discussed in Sec.~\ref{Sec:syserror}.

\section{WEIGHTED AVERAGE OF SOUTH AND NORTH MUON ARM RESULTS}
\label{app:average}

We calculate a weighted average of the results from the south and 
north muon arms to obtain final results. The same method of weighting 
is used for all combined quantities, including the bottom cross 
section, angular and momentum distributions for \cc and \bb muon 
pairs, and the Drell-Yan \pt and mass distributions. Each quantity 
$\Gamma$, which can represent a yield in a given bin or a cross 
section, is calculated as a weighted average of the measured values 
$\Gamma_{S}$ and $\Gamma_{N}$ using the south and north arm respectively, according to: 

\begin{equation}
\Gamma = \frac{w_{S} \Gamma_{S} + w_{N} \Gamma_{N}}{w_{{\rm tot}}}. 
\end{equation}

\noindent The weights for the south and north arms, $w_{j}$ ($j=S,N$), are calculated from 
the inverse of the quadrature sum of statistical and uncorrelated systematic uncertainties, i.e. 
those systematic errors that are not common for the south and north measurements. Denoting $\sigma^2_{j,{\rm syst}}$, 
$\sigma^2_{j,{\rm uncorr}}$ and $\sigma^2_{j,{\rm corr}}$ as the total, uncorrelated, and 
correlated systematic uncertainties such that $\sigma^2_{j,{\rm syst}}=\sigma^2_{j,{\rm uncorr}}+\sigma^2_{j,{\rm corr}}$, 
the weights $w_{j}$ are given by the relation: 

\begin{equation}
w_{j} = \frac{1}{\sigma^2_{j,{\rm stat}} + \sigma^2_{j,{\rm uncorr}}}. 
\end{equation}

\noindent The weight $w_{{\rm tot}}$ is the sum of the weights for south and 
north arm, i.e. $w_{{\rm tot}}=w_S+w_N$. For the statistical uncertainties on 
the weighted average we quote:

\begin{equation}
\sigma^2_{{\rm stat}} = \frac{w_{S}^{2}\sigma^2_{S,{\rm stat}} + w_{N}^{2}\sigma^2_{N,{\rm stat}}}{w_{{\rm tot}}^{2}}.  
\end{equation}

\noindent The systematic uncertainties that are fully correlated between 
south and north arms are treated separately from uncorrelated systematic 
uncertainties as specified below:

\begin{align}
\nonumber \sigma^2_{{\rm syst}} = &\frac{w_{S}^{2}\sigma^2_{S,{\rm uncorr}} + w_{N}^{2}\sigma^2_{N,{\rm uncorr}}}{w_{{\rm tot}}^{2}} \\
&+ \frac{(w_{S}\sigma_{S,{\rm corr}}+w_{N}\sigma_{N,{\rm corr}})^{2}}{w_{{\rm tot}}^{2}}, 
\end{align}

\noindent These systematic uncertainties are calculated separately for 
the upper and lower boundaries.

\section{SIMULATION PARAMETERS}
\label{app:simsettings}

\begin{table}[htb]
\caption{\label{Tab:mbpythia}Parameters used in \pythia Tune A simulation.} 
\begin{ruledtabular} \begin{tabular}{ccc}
Parameter  & Setting & Description\\
\hline
MSEL & 1 & Turn on all QCD processes\\
PARP(67) & 4.0 & Set hard scattering scale $\mu^2$\\
PARP(82) & 2.0 & Turn off $p_{T}$ for multiparticle interactions\\
PARP(84) & 0.4 & Radius of core Gaussian matter\\
PARP(85) & 0.9 & Probability that two gluons are \\
&&produced with colors connected\\
& &  to the nearest neighbors\\
PARP(86) & 0.95 & Probability that two gluons \\
&&are produced with PARP(85)\\
& &  conditions or closed loop\\
PARP(89) & 1800 & Reference energy scale of the turn-off $p_T$\\
PARP(90) & 0.25 & Energy dependence of the turn-off $p_T$\\
PARP(91) & 1.5 & Primordial $k_T$ Gaussian width\\
CKIN(3) & 1.5 & Lower cutoff on $\hat{p}_{\perp}$\\
MSTP(51) & 7 & CTEQ 5L, leading order PDF\\	
\end{tabular} \end{ruledtabular}
\end{table}

\begin{table}[htb]
\caption{ \label{Tab:dysetup} Parameters used in \pythia Drell-Yan simulations.} 
\begin{ruledtabular} \begin{tabular}{ccc}
Parameter  & Setting & Description\\
\hline
MSEL  & 0          & Select subprocesses manually       \\
MSTP(43) &3     & Select Drell-Yan process \\
  &   & Complete $Z^{0}/\gamma^{*}$ 
structure\\
MSUB(1) &1      & Turn on $q+\bar{q}$ $\rightarrow$ $Z^{0}/\gamma^{*}$ $\rightarrow$ $\mu^{+}\mu^{-}$\\
MSTP(91) &1     & Gaussian primordial $k_{T}$\\  
PARP(91) &$1.1$  & Gaussian width of $k_{T}$ in \gevc\\ 
MSTP(33)& 1     & Inclusion of k-factors in\\
 & &       hard cross sections\\
MSTP(32) &4      & Use $Q^{2} = \hat{s}^{2}$    \\
CKIN(1)& 0.5      & Lower cutoff on $\hat{m}=\sqrt{\hat{s}}$\\
CKIN(2)& -1        & Upper cutoff on $\hat{m}=\sqrt{\hat{s}}$\\
CKIN(3) &0.0      & Lower cutoff on $\hat{p}_{\perp}$\\
CKIN(4)& -1        & Upper cutoff on $\hat{p}_{\perp}$\\
MSTP(51)& 7     & CTEQ 5L, leading order PDF\\	
\end{tabular} \end{ruledtabular}
\end{table}


\section{DATA TABLES}
\label{app:datatables}

\begin{table*}[htb]
\caption{\label{Tab:ccdphi}The differential yield $dN/d\phi$ of unlike-sign muon pairs from charm with 
mass 1.5--2.5 \gevcc in the ideal muon arm acceptance, as a function of the pair azimuthal opening angle.} 
\begin{ruledtabular} \begin{tabular}{cccccc}
$|\phi_{\mu,1}$--$\phi_{\mu,2}|$ & $dN/d\phi$ & stat. error& sys. error (type B) & sys. error (type C) \\
{[rad]} & $10^{-9}\times$[rad$^{-1}$] & $10^{-9}\times$[rad$^{-1}$]  & $10^{-9}\times$[rad$^{-1}$] & $10^{-9}\times$[rad$^{-1}$] \\
\hline
0--$\frac{\pi}{15}$ & -0.136 & 0.199 & $^{+3.3\times 10^{-2}}_{-4.3\times 10^{-2}}$ & $1.6\times 10^{-2}$ \\
$\frac{\pi}{15}$--$\frac{2\pi}{15}$ & -1.43 $\times 10^{-2}$ & 0.130 & $^{+4.87 \times 10^{-2}}_{-3.22\times 10^{-2}}$ & 1.7 $\times 10^{-3}$\\
$\frac{2\pi}{15}$--$\frac{3\pi}{15}$ & 0.456 & 0.237 & $^{+0.165}_{-0.174}$ & 5.5$\times 10^{-2}$ \\
$\frac{3\pi}{15}$--$\frac{4\pi}{15}$ & 0.238 & 0.280 & $^{+0.334}_{-0.298}$ & 2.9$\times 10^{-2}$ \\
$\frac{4\pi}{15}$--$\frac{5\pi}{15}$ & 1.08 & 0.41 & $^{+0.57}_{-0.51}$ & 0.13 \\
$\frac{5\pi}{15}$--$\frac{6\pi}{15}$ & 0.443 & 0.579 & $^{+0.792}_{-0.685}$ & 5.3$\times 10^{-2}$ \\
$\frac{6\pi}{15}$--$\frac{7\pi}{15}$ & 3.34 & 0.71 & $^{+0.94}_{-0.85}$ & 0.40 \\
$\frac{7\pi}{15}$--$\frac{8\pi}{15}$ & 5.02 & 0.87 & $^{+0.95}_{-0.85}$ & 0.60 \\
$\frac{8\pi}{15}$--$\frac{9\pi}{15}$ & 7.09 & 0.93 & $^{+0.83}_{-0.77}$ & 0.85 \\
$\frac{9\pi}{15}$--$\frac{10\pi}{15}$ & 7.97 & 0.97 & $^{+0.77}_{-0.73}$ & 0.96 \\
$\frac{10\pi}{15}$--$\frac{11\pi}{15}$ & 6.69 & 1.05 & $^{+0.82}_{-0.80}$ & 0.80 \\
$\frac{11\pi}{15}$--$\frac{12\pi}{15}$ & 7.70 & 1.11 & $^{+0.98}_{-0.96}$ & 0.92 \\
$\frac{12\pi}{15}$--$\frac{13\pi}{15}$ & 10.2 & 1.1 & $^{+1.3}_{-1.3}$ & 1.2 \\
$\frac{13\pi}{15}$--$\frac{14\pi}{15}$ & 7.95 & 1.09 & $^{+1.28}_{-1.27}$ & 0.95 \\
$\frac{14\pi}{15}$--$\pi$ & 6.15 & 1.21 & $^{+1.13}_{-1.12}$ & 0.74 \\
\end{tabular} \end{ruledtabular}

\caption{\label{Tab:bbdphi}The differential yield $dN/d\phi$ of 
like-sign muon pairs from bottom with mass 3.5--10.0 \gevcc in the ideal 
muon arm acceptance, as a function of the pair azimuthal opening angle.}
\begin{ruledtabular} \begin{tabular}{cccccc}
$|\phi_{\mu,1}$--$\phi_{\mu,2}|$ & $dN/d\phi$ & stat. error& sys. error (type B) & sys. error (type C) \\
{[rad]} & $10^{-9}\times$[rad$^{-1}$] & $10^{-9}\times$[rad$^{-1}$]  & $10^{-9}\times$[rad$^{-1}$] & $10^{-9}\times$[rad$^{-1}$] \\
\hline
$\frac{4\pi}{12}$--$\frac{5\pi}{12}$ & $8.36\times10^{-2}$ 	&	$3.47\times10^{-2}$		& $^{+9.2\times10^{-3}}_{-9.2\times10^{-3}}$ & $1.0\times10^{-2}$ \\
$\frac{5\pi}{12}$--$\frac{6\pi}{12}$ & $5.74\times10^{-2}$ 	&	$2.72\times10^{-2}$		& $^{+3.5\times10^{-3}}_{-3.6\times10^{-3}}$ & $6.9\times10^{-3}$ \\
$\frac{6\pi}{12}$--$\frac{7\pi}{12}$ & 0.274 				&	$6.6\times10^{-2}$		& $^{+1.1\times10^{-2}}_{-1.2\times10^{-2}}$ & $3.3\times10^{-2}$ \\
$\frac{7\pi}{12}$--$\frac{8\pi}{12}$ & 0.531 				&	$9.6\times10^{-2}$		& $^{+2.3\times10^{-2}}_{-2.8\times10^{-2}}$ & $6.4\times10^{-2}$ \\
$\frac{8\pi}{12}$--$\frac{9\pi}{12}$ & 1.13 				&		0.14				& $^{+5\times10^{-2}}_{-9\times10^{-2}}$ & 0.14 \\
$\frac{9\pi}{12}$--$\frac{10\pi}{12}$ & 1.51				&		0.18				& $^{+9\times10^{-2}}_{-0.18}$ & 0.18 \\
$\frac{10\pi}{12}$--$\frac{11\pi}{12}$ & 1.87 				&		0.22				& $^{+0.16}_{-0.30}$ & 0.22 \\
$\frac{11\pi}{12}$--$\pi$ & 1.94 							&		0.24				& $^{+0.21}_{-0.41}$ & 0.23 \\
\end{tabular} \end{ruledtabular}

\caption{\label{Tab:ccpt}The differential yield $dN/dp_{T}$ of 
unlike-sign muon pairs from bottom with mass 1.5--2.5 \gevcc in the 
ideal muon arm acceptance, as a function of the pair transverse 
momentum.}
\begin{ruledtabular} \begin{tabular}{cccccc}
$p_{T}$ & $dN/dp_{T}$ & stat. error& sys. error (type B) & sys. error (type C) \\
{[GeV/$c$]} & $10^{-9}\times$[$c$/GeV] & $10^{-9}\times$[$c$/GeV]  & $10^{-9}\times$[$c$/GeV] & $10^{-9}\times$[$c$/GeV] \\
\hline
0--0.2 		& 1.02 					& 0.73 					& $^{+0.59}_{-0.59}$ 								& 0.12 \\
0.2--0.4 	& 3.97 					& 1.05 					& $^{+1.32}_{-1.31}$ 								& 0.48 \\
0.4--0.6 	& 8.16 					& 1.17 					& $^{+1.37}_{-1.36}$ 								& 0.98 \\
0.6--0.8 	& 8.91 					& 1.19 					& $^{+1.22}_{-1.21}$ 								& 1.07 \\
0.9-1.0 	& 5.89 					& 1.08 					& $^{+0.92}_{-0.91}$ 								& 0.71 \\
1.0--1.2 	& 6.31 					& 1.00 					& $^{+0.75}_{-0.73}$ 								& 0.76 \\
1.2--1.4 	& 5.58 					& 0.90 					& $^{+0.66}_{-0.64}$ 								& 0.67 \\
1.4--1.6 	& 6.91 					& 0.84 					& $^{+0.62}_{-0.60}$ 								& 0.83 \\
1.6--1.8 	& 5.15 					& 0.75 					& $^{+0.58}_{-0.53}$ 								& 0.62 \\
1.8--2.0 	& 3.46 					& 0.67 					& $^{+0.53}_{-0.49}$ 								& 0.42 \\
2.0--2.4 	& 1.90					& 0.41 					& $^{+0.50}_{-0.43}$ 								& 0.23 \\
2.4--2.8 	& 0.761 				& 0.309 				& $^{+0.408}_{-0.370}$ 								& $9.1\times10^{-2}$ \\
2.8--3.2 	& $-5.97\times10^{-2}$	& 0.239 				& $^{+0.353}_{-0.298}$ 								& $7.2\times10^{-3}$ \\
3.2--3.6 	& $5.02\times10^{-2}$ 	& 0.203 				& $^{+0.259}_{-0.242}$ 								& $6.0\times10^{-3}$ \\
3.6--4.4 	& 0.206 				& 0.102 				& $^{+0.135}_{-0.118}$ 								& $2.5\times10^{-2}$ \\
4.4--5.2 	& $8.18\times10^{-2}$ 	& $8.03\times10^{-2}$ 	& $^{+6.77\times10^{-2}}_{-5.64\times10^{-2}}$ 		& $9.8\times10^{-3}$ \\	
\end{tabular} \end{ruledtabular}
\end{table*}

\begin{table*}[htb]
\caption{\label{Tab:bbpt}The differential yield $dN/dp_{T}$ of like-sign 
muon pairs from bottom with mass 3.5--10.0 \gevcc in the ideal muon arm 
acceptance, as a function of the pair transverse momentum.}
\begin{ruledtabular} \begin{tabular}{cccccc}
$p_{T}$ & $dN/dp_{T}$ & stat. error& sys. error (type B) & sys. error (type C) \\
{[GeV/$c$]} & $10^{-9}\times$[$c$/GeV] & $10^{-9}\times$[$c$/GeV]  & $10^{-9}\times$[$c$/GeV] & $10^{-9}\times$[$c$/GeV] \\
\hline
0--0.5 		& 0.199 				& $5.7\times10^{-2}$ 	& $^{+2.7\times10^{-2}}_{-5.0\times10^{-2}}$ 			& $2.4\times10^{-2}$			\\
0.5--1.0	& 0.576 				& $9.2\times10^{-2}$ 	& $^{+5.6\times10^{-2}}_{-0.122 			}$			& $6.9\times10^{-2}$			\\
1.0--1.5	& 0.754 				& $9.9\times10^{-2}$ 	& $^{+6.1\times10^{-2}}_{-0.126 			}$			& $9.0\times10^{-2}$			\\
1.5--2.0 	& 0.777 				& $9.5\times10^{-2}$ 	& $^{+5.0\times10^{-2}}_{-9.6\times10^{-2}}$ 			& $9.3\times10^{-2}$			\\
2.0--2.5 	& 0.536 				& $7.8\times10^{-2}$ 	& $^{+3.3\times10^{-2}}_{-5.8\times10^{-2}}$ 			& $6.4\times10^{-2}$			\\
2.5--3.0	& 0.376 				& $6.5\times10^{-2}$ 	& $^{+2.0\times10^{-2}}_{-3.1\times10^{-2}}$ 			& $4.5\times10^{-2}$			\\
3.0--3.5	& 0.230					& $4.9\times10^{-2}$ 	& $^{+1.1\times10^{-2}}_{-1.8\times10^{-2}}$ 			& $2.8\times10^{-2}$			\\
3.5--4.0 	& 0.199 				& $4.3\times10^{-2}$ 	& $^{+8\times10^{-3}}_{-1.0\times10^{-2}}$	 			& $2.4\times10^{-2}$			\\
4.0--4.5 	& $9.05\times10^{-2}$ 	& $2.93\times10^{-2}$	& $^{+3.8\times10^{-3}}_{-4.9\times10^{-3}}$ 			& $1.09\times10^{-2}$			\\			
4.5--5.0	& $2.37\times10^{-2}$ 	& $1.75\times10^{-2}$	& $^{+1.4\times10^{-3}}_{-1.4\times10^{-3}}$ 			& $2.9\times10^{-3}$			\\			
\end{tabular} \end{ruledtabular}

\caption{\label{Tab:dy}The differential Drell-Yan cross section 
$\frac{d^{2}\sigma}{dmdy}$ as a function of the muon pair mass, where 
the muon pair rapidity $|y_{\mu\mu}|$ is between 1.2 and 2.2.}
\begin{ruledtabular} \begin{tabular}{cccccc}
$m_{\mu\mu}$ & $\frac{d^{2}\sigma}{dmdy}$ & stat. error& sys. error (type B) & sys. error (type C) \\
{[GeV/$c^{2}$]} & [pb $c^{2}$/GeV] & [pb $c^{2}$/GeV]  & [pb $c^{2}$/GeV]  & [pb $c^{2}$/GeV] \\
\hline
5 	& 114 & 16 & $^{+24}_{-19}$ & 14 \\
5.4 & 95.6 & 13.3 & $^{+16.8}_{-13.4}$ & 11.5 \\
5.8 & 67.8 & 10.7 & $^{+11.6}_{-9.1}$ & 8.1 \\
6.25 & 41.3 & 7.5 & $^{+7.6}_{-5.9}$ & 5.0 \\
6.75 & 38.1 & 6.5 & $^{+4.6}_{-3.6}$ & 4.6 \\
7.3 & 25.3 & 4.8 & $^{+2.9}_{-2.3}$ & 3.0 \\
7.9 & 19.4 & 4.1 & $^{+1.7}_{-1.4}$ & 2.3 \\
12.1 & 4.94 & 1.33 & $^{+0.19}_{-0.19}$ & 0.59 \\
14 & 0.823 & 0.686 & $^{+3.9\times10^{-2}}_{-3.4\times10^{-2}}$ & $9.9\times10^{-2}$ \\
\end{tabular} \end{ruledtabular}

\caption{\label{Tab:dypt}The differential Drell-Yan cross section 
$\frac{1}{2\pi p_{T}}\frac{d^{2}\sigma}{dydp_{T}}$ as a function of the 
muon pair transverse momentum, where the muon pair mass $m_{\mu\mu}$ is 
between 4.8 and 8.2 \gevcc and the muon pair rapidity $|y_{\mu\mu}|$ is 
between 1.2 and 2.2.}
\begin{ruledtabular} \begin{tabular}{cccccc}
$p_{T}$  & $\frac{1}{2\pi p_{T}}\frac{d^{2}\sigma}{dydp_{T}}$  & stat. error & sys. error (type B) & sys. error (type C)\\
{[GeV/$c$]} & [pb ($c$/GeV)$^{2}$] & [pb ($c$/GeV)$^{2}$] & [pb ($c$/GeV)$^{2}$] &  [pb ($c$/GeV)$^{2}$]\\
\hline
0.25 & 12.0 & 3.4 & $^{+1.9}_{-1.6}$ & 1.4\\
0.75 & 13.1 & 2.0 & $^{+1.8}_{-1.4}$ & 1.6\\
1.25 & 7.48 & 1.30 & $^{+1.42}_{-1.14}$ & 0.90\\
1.75 & 6.22 & 0.93 & $^{+1.00}_{-0.81}$ & 0.75\\
2.25 & 2.48 & 0.55 & $^{+0.59}_{-0.46}$ & 0.30\\
2.75 & 1.22 & 0.39 & $^{+0.37}_{-0.29}$ & 0.15\\
3.25 & 0.408 & 0.239 & $^{+0.206}_{-0.157}$ & $4.9\times 10^{-2}$\\
3.75 & 0.688 & 0.198 & $^{+0.123}_{-0.103}$ & $8.3\times 10^{-2}$\\
4.25 & 0.627 & 0.164 & $^{+8.4\times 10^{-2}}_{-7.1\times 10^{-2}}$ & $7.5\times 10^{-2}$\\
4.75 & $9.29 \times 10^{-2}$ & $8.84 \times 10^{-2}$ & $^{+3.30 \times 10^{-2}}_{-2.59 \times 10^{-2}}$ & $1.12 \times 10^{-2}$\\
5.25 & $9.47 \times 10^{-2}$ & $6.61 \times 10^{-2}$ & $^{+2.28 \times 10^{-2}}_{-1.98 \times 10^{-2}}$ & $1.14 \times 10^{-2}$\\
5.75 & 0.127 & $6.6\times 10^{-2}$ & $^{+1.9\times 10^{-2}}_{-1.9\times 10^{-2}}$ & $1.5\times 10^{-2}$\\
\end{tabular} \end{ruledtabular}
\end{table*}


\clearpage


 
%
 
\end{document}